%% file: gpg-COMNET.tex
\documentclass{comnet-mod}

\bibpunct{[}{]}{,}{n}{,}{;}     
\usepackage[utf8]{inputenc}
\usepackage{graphicx}
\usepackage{xcolor}

\usepackage{float}
\usepackage{multicol}
\usepackage{multirow}
\usepackage{times}
\usepackage{latexsym}
\usepackage{array,booktabs,ragged2e}
\newcolumntype{R}[1]{>{\RaggedLeft\arraybackslash}p{#1}}
\usepackage{amsmath, amssymb, mathrsfs}
\usepackage{url}                
\usepackage{ifpdf}              
\usepackage{pdfpages}           
\usepackage{algorithm}
\usepackage{algpseudocode}
\usepackage{graphicx}
\usepackage[font=small,labelfont=bf]{caption}
\usepackage{subcaption}
\algnewcommand{\LineComment}[1]{\State \parbox[t]{\dimexpr\linewidth-\algorithmicindent}{ \emph{#1}}}
\algnewcommand\algorithmicforeach{\textbf{for each}}
\algdef{S}[FOR]{ForEach}[1]{\algorithmicforeach\ #1\ \algorithmicdo}

\usepackage{comment}

\renewcommand{\thesection}{\arabic{section}}
\renewcommand{\thesubsection}{\thesection.\arabic{subsection}}
\renewcommand{\thesubsubsection}{\thesubsection.\arabic{subsubsection}}

\usepackage{todonotes}

\newcommand{\tup}[1]{\left\langle #1 \right\rangle}

\newcommand{\degseq}{\vec{\bf d}}
\newcommand{\tdegseq}{\vec{\bf t}}

\newcommand{\degdist}{\vec{\bf P}}

\graphicspath{
    {/}
    {Images/}
}
\newcommand{\minitab}[2][l]{\begin{tabular}{#1}#2\end{tabular}}

\usepackage{hyperref}
\AtBeginDocument{
  \label{CorrectFirstPageLabel}
  \def\fpage{\pageref{CorrectFirstPageLabel}}
}

\usepackage{amssymb}
\usepackage{wrapfig}
\usepackage{todonotes}
\usepackage{comment}
\usepackage{balance}
\usepackage{paralist}
\usepackage{verbatim}

\begin{document}

\title{A generative graph model for electrical infrastructure networks}

\shorttitle{A generative graph model for electrical infrastructure networks} 
\shortauthorlist{S.G. Aksoy, E. Purvine, E. Cotilla-Sanchez, M. Halappanavar} 

\author{
\name{Sinan G. Aksoy$^*$}
\address{Pacific Northwest National Laboratory, Richland, WA 99354\email{$^*$Corresponding author: sinan.aksoy@pnnl.gov}}
\name{Emilie Purvine}
\address{Pacific Northwest National Laboratory, Seattle, WA 98109}
\name{Eduardo Cotilla-Sanchez}
\address{Oregon State University, Corvallis, OR 97331}
\and
\name{Mahantesh Halappanavar}
\address{Pacific Northwest National Laboratory, Richland, WA 99354}}

\maketitle

\begin{abstract}
{We propose a generative graph model for electrical infrastructure networks that accounts for heterogeneity in both node and edge type. To inform the model design, we analyze the properties of power grid graphs derived from the U.S.~Eastern Interconnection, Texas Interconnection, and Poland transmission system power grids. Across these datasets, we find subgraphs induced by nodes of the same voltage level exhibit shared structural properties atypical to small-world networks, including low local clustering, large diameter, and large average distance. On the other hand, we find subgraphs induced by transformer edges linking nodes of different voltage types contain a more limited structure, consisting mainly of small, disjoint star graphs. The goal of our proposed model is to match both these inter and intra-network properties by proceeding in two phases: 
the first phase adapts the Chung-Lu random graph model, taking desired vertex degrees and desired diameter as inputs, while the second phase of the model is based on a simpler random star graph generation process. We test the model's performance by comparing its output across many runs to the aforementioned real data. In nearly all categories tested, we find our model is more accurate in reproducing the unusual mixture of properties apparent in the data than the Chung-Lu model. We also include graph visualization comparisons, {a brief analysis of edge-deletion resiliency, and guidelines for artificially generating the model inputs in the absence of real data}. 
}
{power grid graph model, Chung-Lu model, network-of-networks.}
\end{abstract}

\input{gpg-introduction}

\input{gpg-preliminaries}

\input{gpg-real-data}

\input{gpg-model}

\input{gpg-results}

\input{gpg-synthetic}

\input{gpg-conclusion}

\paragraph{Funding.} This work was supported in part by the Applied Mathematics Program of the Office of Advanced Scientific Computing Research within the Office of Science of the U.S. Department of Energy (DOE) through the Multifaceted Mathematics for Complex Energy Systems (M2ACS) project. 
\paragraph{Acknowledgement.} {The authors would like to thank Josh Tobin and Stephen Young for helpful discussions, as well as an anonymous referee for suggestions which improved the manuscript}.

\bibliographystyle{comnet}
\bibliography{powergridGen}
%








\input{gpg-appendices}

\end{document}

%% file: gpg-introduction.tex
\section{Introduction}


Graph theoretic approaches to modeling the electric power grid aim to leverage tools in network science to analyze its topological structure. This research area spans a variety of applications, including vulnerability analysis \cite{albert2004structural,hines2010vul}, controlled islanding \cite{AMRAEE2017135, xu2010controlled}, locational marginal pricing \cite{cheverez2009admissible}, and the location of sensors \cite{anderson2012graph}. In its most basic form, {the physical infrastructure of the grid} is represented as a graph, where the vertices represent generators, substations, or loads, while edges represent power transmission lines or a voltage transformer.


A fundamental obstacle in applying network science to the power grid stems from the restricted access and proprietary nature of power grid data. To better facilitate the application of network science tools to the power grid, researchers need generative graph models to create synthetic power grid graphs. Ideally, these models require few, compact inputs (either measured from real data or generated artificially) and produce graphs exhibiting meaningful structural properties that can be quantitatively tuned according to user input. In this way, generative graph models serve as {\it null models} for testing hypotheses and algorithms at different scales. In this work, we investigate structural properties of power grid graphs, and then apply those observations to design a generative graph model. However, before salient features worth modeling can be selected, one must determine a suitable underlying data structure for the model.

In regard to this issue, Halappanavar et ~al. \cite{Halappanavar2015} argue that the inherently heterogeneous nature of the real-world power grid necessitates graph models which, at a minimum, account for different node and edge types. Accordingly, they argue it may be inappropriate to apply graph algorithms or metrics to the entire power grid graph in a way that treats vertices or edges within a network as homogenous in type. As an alternative, the authors propose using a more nuanced power grid graph model where each vertex has a nominal voltage rating, and each edge is either a power transmission line (connecting two vertices of the same voltage) or transformer edge (connecting vertices of different voltage levels). In this approach, a power grid graph is then viewed as a ``network-of-networks," composed of vertex or edge homogenous subgraphs. That is, power grid graphs consist of a collection of same-voltage subgraphs for each voltage level, which are connected to each other via transformer edges. In this framework, network science tools can then be used to study and model the structure of the same-voltage subgraphs themselves, as well as how these subgraphs connect to each other through transformer edges.

In this work, we develop this viewpoint by designing a generative random graph model for the power grid that accounts for node and edge type. Our model aims to match, according to user-specified input, meaningful structural properties encountered in real power grid graph data. The paper is organized as follows: in Section \ref{sec:prelims}, we describe the graph theoretic basics underlying the model, which largely mirrors that which is presented in \cite{Halappanavar2015}. Then, in Section \ref{sec:real_data}, we investigate characteristics of real-world power grid graphs. We analyze both the same-voltage subgraphs, as well as the transformer edges between them, in the U.S.~Eastern Interconnection, Texas Interconnection, and Poland transmission networks\footnote{The U.S. power grid data used in this paper were obtained through the U.S.~Critical Energy Infrastructure Information (CEII) request process. Poland open-source data comes from MATPOWER:~\url{http://github.com/MATPOWER/matpower}.}. {As we will soon explain, we find that the same-voltage subgraphs exhibit atypical structure, such as low clustering, high diameter and high average distance}. Given that many existing graph generation models are unable to capture this combination of properties, we propose a new graph model, described in Section \ref{sec:model}, based on the well-known Chung-Lu model, which we call the Chung-Lu Chain model. We also describe our process for generating the transformer edges, as well as how both phases of the model may be used in conjunction to output the aggregate graph. In Section \ref{sec:results}, we test the model's performance by comparing its output over many runs against the real data. This comparison not only includes a number of graph metrics, but also graph visualizations and resiliency to single edge failure analysis. {In Section \ref{sec:synthetic}, we provide guidelines for artificially generating the model inputs.} We conclude and briefly mention avenues for future work in Section \ref{sec:conc}.

\subsection{Prior Work}
Before proceeding, we describe prior work on graph modeling of the electrical power grid. While in this paper we focus both on modeling the vertex and edge-homogenous subgraphs, as well as the aggregate graph, Hines et.~al.~\cite{hines2010topological} consider the aggregate power grid graph associated with the IEEE 300-bus test case and the U.S.~Eastern Interconnection power grid. They compare the structure of these aggregate graphs with comparably sized Erd\H{o}s-R\'{e}nyi, preferential attachment, and small-world graphs and found these models lack utility as stand-alone models for the power grid since they differ substantially in degree distribution, clustering, diameter and assortativity. As an alternative, they proposed a modification of the random geometric graph model called the ``minimum-distance" model, which they show outperforms the aforementioned models on the selected criteria. In \cite{WANG2009114} Wang, Scaglione and Thomas also analyze the structural properties of power grid graphs and propose their own generative model. They take a hierarchical approach to generating synthetic power grid graphs based on geographic zones; in contrast, our approach decomposes the network into vertex and edge-homogenous subgraphs based only on voltage rating and requires no geographic knowledge of the underlying power grid. 
Lastly, building off of work in \cite{birchfield2016}, Birchfield et~al.~\cite{birch17} put forth an extensive list of 18 validation metrics for assessing the realism of synthetic power grid data. Some of these metrics, such as the number of substations per given voltage range and the ratio of transmission lines to substations of a given voltage level, concern basic information that is directly built into our generative model. Other metrics mentioned, such as generator dispatch percentage and reactive power limits, require additional information that while not included in our base generative model, may be later appended to its output data. 

{Although our work here focuses on modeling the physical connections in electrical transmission networks rather than the dynamics of power flow, prior work has established ways in which grid dynamics are influenced by underlying graph structure. For example, \cite{mureddu2016islanding} analyzes power grid resiliency under link failures, using structural graph metrics, such as betweenness centrality, to determine which edges to delete. Relatedly, \cite{pagani2015complex} studies grid vulnerability based on various graph edge and vertex deletion scenarios; they conclude that while their graph-based approaches ``are not meant to be a substitute for current reliability assessment methods...graph topology plays an important role in the robustness of the network at all power levels, including the medium and low voltage levels." Graph topology has also been linked to dynamics via spectral analyses of graphs: for example, \cite{fioriti2012spectral} investigates the relationship between Laplacian graph eigenvalues on synchronizability and epidemic thresholding, \cite{deka2015structural} considers adjacency graph eigenvalues and adversarial attacks, and \cite{pinar2010optimization} analyzes the correspondence between the Jacobian matrix of power flow and the Laplacian matrix of the graph. The aforementioned topics detail only a few examples -- for a more complete survey examining the scope, limitations, and directions of graph theoretic approaches towards power grid modelling, readers are referred to \cite{PAGANI20132688}. In short, while graph models clearly are not holistic models of the operational grid, their study is not only interesting in its own right, but also a valuable aspect of power grid data linked to dynamics.}




%% file: gpg-preliminaries.tex
\section{Preliminaries} \label{sec:prelims}

\subsection{Graph theory basics}\label{sec:basics}
A \emph{graph} $G=(V,E)$ is a set of vertices, $V$ with $|V|=n$, and set of edges, $E\subseteq V\times V$, where each edge is an unordered pair of distinct vertices.
Two vertices and $i,j \in V$ are {\it adjacent} if $\{i,j\} \in E$, and we call $i$ and $j$ {\em neighbors}.
The {\em degree} of a vertex $i$, denoted $d_i$, is the number of vertices adjacent to $i$.
The \emph{degree sequence} of a graph, denoted $\degseq = \tup{d_1, d_2, \ldots, d_n}$, is the list of degrees for each vertex in the graph.

A path of length $k$ between vertices $i$ and $j$ is a sequence of $k+1$ distinct vertices, $i=i_0,\dots,i_k=j$, where $i_\ell$ is adjacent to $i_{\ell+1}$ for $\ell=0,\dots,k-1$.
If $i=j$, then we call $i=i_0,\dots,i_k=j$ a {\em cycle} of length $k$, and a cycle of length 3 is called a {\it triangle}.
The {\it distance} between two vertices is the length of the shortest path between them, and is denoted $d(i,j)$.

A subset $S\subseteq V$ is {\it connected} if there is a path between all pairs of vertices in $S$.
If $C \subseteq V$ is a maximally connected set (i.e., if there is no $S \supsetneq C$ that is also connected), we call $C$ a {\em connected component} of $G$.
A {\it tree} is a connected graph with no cycles.
The degree 1 vertices in a tree are called {\it leaves}. 
A {\it $k$-star} is a tree consisting of $k$ leaves connected to one degree $k$ vertex, called the {\it center}.

The subgraph induced by vertex set $S\subseteq V$ is the graph with vertex set $S$ and whose edge set consists of all edges in $E$ with both endpoints in $S$.
Similarly, the subgraph induced by edge set $T\subseteq E$ is the graph with edge set $T$, and vertex set consisting of all endpoints of edges in $T$.
We denote these by $G[S]$ and $G[T]$ respectively.
It will be clear from context whether we mean vertices or edges.

\subsection{Graph metrics}
{In applications of graph theory, one is often concerned with global quantities associated with the graph as a whole, or local quantities associated with a vertex or edge.}
We refer to these as \emph{metrics}, and describe a few that are important in the context of power networks.

The \emph{degree distribution} of a graph, $\degdist = \tup{P(1), P(2), \ldots, P(\Delta)}$, is the sequence of the number of vertices for each possible degree, i.e., $P(i) =$ the number of vertices, $v$, with $d_v = i$, and $\Delta$ is the maximum degree in the graph.
{This can be obtained from, but is not to be confused with, the degree sequence defined in Section \ref{sec:basics}.
Two numerical measures for comparing the similarity between two degree distributions are the {\it Kolmogorov-Smirnov (KS) statistic} and {\it Relative Hausdorff (RH) measure} \cite{Simpson15}. 
While the KS statistic is bounded between 0 and 1, the RH measure can exceed 1. Readers may refer to \cite{Simpson15, matulef2017sampling} for further discussions about these measures.
}

Two basic measures of distance in a graph are diameter and average distance.
The \emph{average distance} in a connected graph $G$ is the average distance between all pairs of vertices.
The \emph{diameter} of a connected graph $G$, denoted $\delta$, is the maximum distance over all pairs of vertices.
\[ \textnormal{avgDist}(G) = \frac{1}{|V|(|V|-1)}\sum_{i \in V} \sum_{\stackrel{j \in V}{j\neq i}} d(i,j), \qquad \delta(G) = \max_{\stackrel{i, j \in V}{i\neq j}} d(i,j) \]
If a graph contains multiple connected components, we define the diameter as the maximum over the diameters of all connected components.
Similarly, the average distance does not take into account pairs of vertices in two different connected components since there is no path between them.

The \emph{local clustering coefficient} (LCC) of a vertex is used to measure how tightly connected the neighborhood of a vertex is.
It is defined as
\[
lcc(i)=\frac{\# \mbox{ triangles incident to vertex } i}{{d_i \choose 2}}.
\]
Note that the LCC of a degree 1 vertex is undefined.
The local clustering coefficient of a graph is the average LCC over all vertices for which the LCC is defined.


\subsection{The power grid as a graph}

\begin{figure}[t!]
\centering
\begin{subfigure}[b]{0.3\textwidth}
\includegraphics[width = \linewidth]{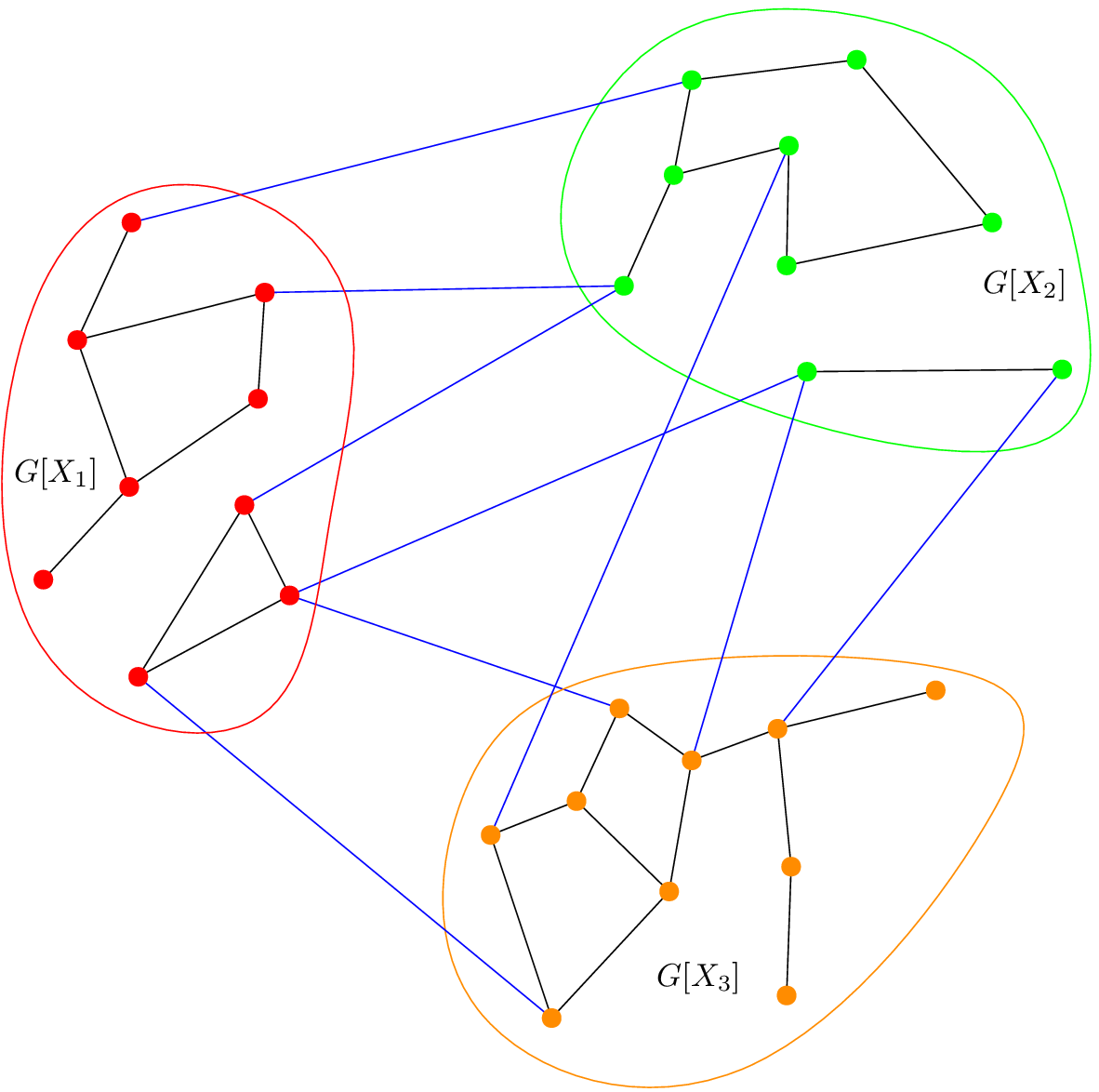}
\caption{} \label{fig:decomposition_model}
\end{subfigure}
\qquad
\qquad
\begin{subfigure}[b]{0.3\textwidth}
\scriptsize
\begin{tabular}{ |l| m{3.8cm} |}
  \hline
{\bf Notation} & {\bf Description} \\
  \hline \hline
$G$               & Entire power grid graph                                             \\
$n$               & Number of vertices in $G$                                           \\\hline
$G[X]$            & Subgraph induced by vertices of voltage level $X$                   \\
$\degseq^X$       & Degree sequence of voltage $X$ vertices in $G[X]$                             \\\hline
$T$               & Set of all transformer edges in $G$                                 \\
$T[X,Y]$          & Set of transformer edges between vertices of voltage $X$ and $Y$    \\
$G[S]$            & Subgraph induced by the set of transformer edges $S\subseteq T$     \\
$\tdegseq[X,Y]$   & Transformer degree sequence of voltage $X$ vertices with respect to voltage $Y$ vertices, i.e., $t[X,Y]_i=|\{j \in Y: \{i,j\}\in T[X,Y] \}|$ \\ 
\hline
\end{tabular}
\caption{} \label{tab:notation}
\end{subfigure}
\caption{A sample power grid graph $G$ with three voltage levels.
  Vertices in $G[X_1]$ are red, in $G[X_2]$ are green, and in $G[X_3]$ are orange.
  The transformer edges between two different voltage levels are blue, {constituting $T$}.}
\end{figure}

We associate a graph with a power grid in the same manner as described in \cite{Halappanavar2015}.
Namely, power grid graphs are undirected graphs in which each vertex additionally has a ``type" according to its voltage level.
More formally, a power grid graph is $G=(V,E,f,\mathcal{X})$, where $V$ is a set of vertices representing power stations, buses, generators, loads, etc., $E$ is a set of edges representing power lines connecting the vertices, $\mathcal{X}$ is a set of possible voltage levels, and $f:V \to \mathcal{X}$ assigns a voltage level to each vertex.
We further call an edge $\{i,j\}$ a {\it transformer} edge if $f(i)\not=f(j)$.
{We emphasize all edges are undirected as these graphs are modeling the physical connections in the transmission network. Of course, actual power-flows may be directional, vary in magnitude, etc, but such properties depend on the particular dynamics and operational state of the grid.}

As introduced in \cite{Halappanavar2015}, a power grid graph with multiple voltage levels can be decomposed as the union of same-voltage subgraphs and transformer edges.
More precisely, the subgraph induced by voltage level $X\in \mathcal{X}$ consists of all vertices of voltage $X$ and all edges between them.
With slight abuse of notation, we denote this by $G[X]$, where $X$ denotes the set of vertices of voltage level $X$.
Though $X$ is not a set of vertices, it can be uniquely identified with the set of vertices for which $f(i)=X$.
Thus, transformer edges are those between two vertices in different same-voltage subgraphs.
{In this way, the entire power grid graph consists of all vertices of all voltage levels, with edge set given by the set union of all same-voltage subgraph edges and all transformer edges.
}

In table in Figure \ref{tab:notation} presents the basic notation we use to refer to power grid graphs.
Hereafter, whenever the voltage level $X$ is clear from context, we will drop the superscript (e.g., write $\vec{\bf d}$ instead of $\vec{\bf d}^X$).
Figure \ref{fig:decomposition_model} shows an example of a power grid graph with three voltage levels, $\mathcal{X} = \{X_1, X_2, X_3\}$.
The vertices for the three different same-voltage subgraphs are shown in red, green, and orange, and the transformer edges are blue.
Notice that each same-voltage subgraph may have multiple connected components. {See below\footnote{{If we circularly label the (red) vertices from 1 to 9 in $G[X_1]$ clockwise, starting with the topmost vertex, then $\degseq^{X_1}=\tup{1,2,2,2,2,2,1,3,3}$, $\tdegseq[X_1,X_2]=\tup{1,1,0,1,1,0,0,0,0}$, and $\tdegseq[X_1,X_3]=\tup{0,0,0,0,1,1,0,0,0}$}.} for examples of degree and transformer degree sequences in Figure \ref{fig:decomposition_model}}.

%% file: gpg-real-data.tex
\section{Characteristics of power grid graphs}\label{sec:real_data}
This section summarizes properties of data from three real-world power grid graphs:~U.S.~Eastern Interconnection, Texas Interconnection, and the Poland transmission system.
In each case, we compute the degree distribution, average distance, diameter, and average local clustering coefficient for all of the same-voltage subgraphs and also consider properties of the transformer edges graph.
We show these results, summarize the trends, and provide a justification as to why power grid graphs might have these common properties.

\subsection{Same-voltage subgraphs}

Table \ref{tab:same_voltage_measures} reports the vertex and edge count, diameter, average distance, and average local clustering coefficient computed on each of the same-voltage subgraphs found in the Eastern, Texas, and Polish networks.
There are two major points elucidated by the table.
{First, the average distance and diameter values are roughly on the order of $\sqrt{n}$, a relationship also independently suggested by Young et~al.~\cite{stephen} and further explored in Section \ref{sec:synP1}. As we discuss in Section \ref{sec:dist_random_graphs}, most random graph models produce graphs with much smaller diameter and average distance, typically on the order of $\log(n)$.}
Second, the clustering coefficient is always very small.
Except in the case of Poland's 400 kV subgraph, the smallest graph in the group, the average LCC is less than 0.10, and typically much less.
This means that for most vertices in the same-voltage subgraphs, their neighborhoods contain very few triangles.
This does not mean that cycles are rare, only that the cycles must be longer than three edges.
In fact, since power grids must be highly resilient to failure of single edges (transmission lines), there are many cycles.
However, because each transmission line costs money and time to build, the extra redundancy given by a triangle is often outweighed by the cost of commissioning.

\vspace{2mm}

\begin{minipage}{\textwidth}
\begin{minipage}[c]{0.55\textwidth}
\centering
\scriptsize
\begin{center}
\begin{tabular}{ |m{1.6cm}|m{0.15cm}|m{0.15cm}|m{0.15cm}|m{0.3cm}|m{0.15cm}| }
  \hline
  \multicolumn{6}{|c|}{\bf{Largest Component}} \\
  \hline
&$|V|$&$|E|$&Diam& Av. Dist.& Local CC  \\ \hline
Eastern: 138 kV & 12997 & 15752 & 386 & 135.117 & 0.049 \\
Eastern: 230 kV & 5691 & 6831 & 211 & 73.503 & 0.033\\
Eastern: 345 kV & 1066 &1360 & 78 & 25.038 & 0.073 \\
Eastern: 500 kV  & 370 & 377 & 48 & 18.945 & 0.092 \\
Eastern: 765 kV & 79 &112 & 16 & 6.204 & 0.037 \\ \hline
Texas: 69 kV & 1169 & 1257 & 134 & 47.978 & 0.007 \\
Texas: 138 kV & 2768 & 3272 & 84 & 32.466 & 0.019 \\
Texas: 345 kV & 208 & 290 & 22 & 8.320 & 0.062 \\ \hline
Poland: 110 kV & 2024 & 2302 & 92 & 37.661 & 0.008 \\
Poland: 220 kV& 135 & 174 & 20 & 7.899 & 0.032 \\
Poland: 400 kV & 50 & 58 & 17 & 6.484 & 0.141\\
\hline
\end{tabular}
\end{center}
\captionof{table}{Vertex and edge count, diameter, average distance, and local clustering
coefficient for Eastern, Texas, and Polish transmission networks.}\label{tab:same_voltage_measures}
\end{minipage}
\hfill
\begin{minipage}[c]{0.42\textwidth}
\centering
\includegraphics[scale=0.21]{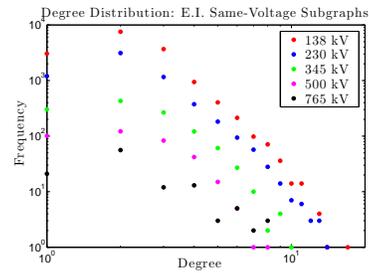}
\vspace{5mm}
\captionof{figure}{{Degree distributions for same-} \\ {voltage subgraphs in the Eastern} \\ {transmission network.}}\label{fig:deg_EI}
\end{minipage}
\end{minipage}
\vspace{1mm}

Figure \ref{fig:deg_EI} illustrates the degree distributions, on a log-log scale, for all five same-voltage subgraphs of the Eastern Interconnection.
Degree distributions from each of the other networks show similar trends but are not pictured here.
{The shape of these degree distributions indicates a preponderance of degree 1 and 2 vertices, and relatively few high-degree vertices}. 
{In general, degree distributions are a fundamental property of complex networks and are ubiquitously studied in the literature, particularly for social networks and the connectivity structure of the Internet \cite{barabasi03networks}.}
However, a key difference between many such real-world networks and the power grid is that the power grid networks represent physical infrastructure.
Transmission lines are physical connections between generators, buses, or loads; in contrast, in a social network or the Internet the link may be symbolic and cost may not be strongly influenced by a physical distance.



\subsection{Transformer edges graph}\label{sec:real_transformer}

Recall that an edge is considered part of the transformer subgraph if its endpoints have different voltage levels.
The real transformer subgraphs for the Eastern, Texas, and Polish networks tend to have many small connected components, almost all of which are star graphs.
Recall that a star graph consists of a single vertex of degree $k$ connected to $k$ vertices of degree 1.
Table \ref{tab:EI_transformer_data_pairs} presents the number of connected components alongside the number of non-star components for the voltage pairwise transformer subgraphs of the U.S.~Eastern Interconnection.
For example, the first row indicates that the transformer subgraph consisting only of edges between the 138 kV vertices and the 230 kV vertices of the U.S.~Eastern Interconnection has 970 connected components, and only five are not stars.
Additionally, Table \ref{tab:EI_transformer_data} contains information about component sizes and counts for the entire transformer subgraph of the U.S.~Eastern Interconnection.
The information for the other networks studied is similar but omitted for the sake of brevity.
The main conclusion to draw from these tables is the vast majority of the connected components in the transformer subgraph are $k$-stars. 
Comparing Table $\ref{tab:EI_transformer_data_pairs}$ with Table $\ref{tab:EI_transformer_data}$, it is apparent the majority of the (already few) non-stars are formed when considering all of the transformer edges together, rather than in the individual subgraphs between pairwise voltage components.
From a total of 1,772 connected components in the transformer subgraph, only 21 are not stars, and in all of the pairwise voltage level graphs there are only 6 non-star components.
This observation is used to build our generative model of the transformer subgraph in Section \ref{sec:model_phase2}.
Rather than a generic random graph model, we use a more specialized algorithm to produce random stars in order to more closely match the real graphs we wish to mimic.

The occurrence of star components may be explained by the nature of the edges themselves.
Transforming voltage from one level to another occurs at an electrical substation.
These substations can be large or small; when they are large, they may send transformed power out in multiple directions, yielding the star pattern in the graph.
Additionally, though not evident in this table, we observe that the non-star components in the transformer subgraph are almost never cycles.
This also makes sense from an engineering perspective: when transforming from voltage $X$ to voltage $Y$, it is typically more economically prudent to push power through a relatively long distance after stepping up in voltage.

\begin{table}[h]
\scriptsize
\centering
\begin{subtable}{.5\textwidth}
\centering
\begin{tabular}{ |p{1cm}|R{1cm}|R{1.37cm}| }
  \hline
  \multicolumn{3}{|c|}{\bf{E.I. Transformer Subgraphs, $T[X,Y]$}} \\
  \hline
Graph & \# components &  \# non-star components     \\
     \hline
  $T[138,230]$ & 970 &  5     \\
  $T[138,345]$  & 493 &  1  \\
  $T[138,500]$  & 31 & 0  \\
  $T[138,765]$  & 15 & 0  \\
  $T[230,345]$  & 151  & 0 \\
  $T[230,500]$  & 169 & 0   \\
 $T[230,765]$  & 11  & 0 \\
  $T[345,500]$  & 9  & 0 \\
  $T[345,765]$  & 33 & 0   \\
 $T[500,765]$  & 5  & 0 \\
  \hline
\end{tabular}
\caption{}\label{tab:EI_transformer_data_pairs}
\end{subtable}
\begin{subtable}{.5\textwidth}
\centering
\begin{tabular}{ |R{1.1cm}|R{1cm}|R{1.37cm}| }
  \hline
  \multicolumn{3}{|c|}{\bf{E.I. Transformer Graph, $T$}} \\
  \hline
Component size & \# components &  \# non-star components     \\
\hline
  2 & 1482 &  0     \\
  3 & 235 &  6  \\
  4 & 34 & 7  \\
  5 & 16 & 7  \\
  6 & 2  & 0 \\
  8 & 2 & 1   \\
 18 & 1  & 0 \\
  \hline
\end{tabular}
\caption{}\label{tab:EI_transformer_data}
\end{subtable}
\caption{The number of non-star components of different sizes in the transformer graph between each pair of voltage levels (\ref{tab:EI_transformer_data_pairs}) and the aggregate transformer graph (\ref{tab:EI_transformer_data}) in the U.S.~Eastern Interconnection.}
\end{table}

%% file: gpg-model.tex
\section{A generative model}\label{sec:model}

In the previous section, we identified a number of characteristics that consistently appear in same-voltage subgraphs, as well as features common to the transformer edges subgraph.
{The same-voltage subgraphs, which account for most edges in a power grid graph, have similarly shaped degree distributions, large diameter and average distance relative to the number of vertices, and low clustering coefficients.}
The transformer edges graph also has a distinctive, albeit much less rich, structure: in our data, we find that $G[T]$ consists almost entirely of disjoint, small-degree stars.
We now turn our attention to random graph models.
First we describe distance properties of well-known random graph models, which precludes their use in the power grid setting.
We then introduce our new model which has a favorable distance property for the power grid application.

\subsection{Distances in random graph models}\label{sec:dist_random_graphs}

Small diameter and average distance are often cited as a ubiquitous property of real-world networks.
Indeed, in addition to high clustering coefficients, a key defining property of small-world networks is that the distance between two randomly selected vertices is proportional to the logarithm of the number of vertices \cite{Watts1998}.
The study of diameter and average distance in random graph models has a rich and extensive history.
In particular, depending on the random model considered and assumptions placed on its inputs, there are a plethora of results about the asymptotic behavior of diameter and average distance.
In both regards, the same-voltage subgraphs of power grid graphs stand in stark contrast.
Using standard asymptotic notation,\footnote{A function $f(n)=O(g(n))$ if for all sufficiently large values of $n$, there is a positive constant $c$ such that $|f(n)|\leq c \cdot |g(n)|$.
We say $f(n)=\Theta(g(n))$ if both $f(n)=O(g(n))$ and $g(n)=O(f(n))$, and say$f(n)=o(g(n))$ if $\lim_{n\to\infty}\frac{f(n)}{g(n)}=0$.}
Table \ref{tab:asymDiam}, lists some of these for well-known random graph models under various mild assumptions.
In this table, $n$ denotes the number of vertices in the graph.

\begin{table}[h]
\scriptsize
\centering
\begin{tabular}{ |c|l|l|l| }
\hline
{\bf Model}                                     & {\bf Diameter}                                  & {\bf Av. Distance}                                 & {\bf Assumption}  \\ \hline  \hline
\multirow{4}{*}{\minitab[c]{Erd\H{o}s-R\'{e}nyi\\$G(n,p)$}} & \multirow{4}{*}{$\Theta(\log{n})$}  & \multirow{4}{*}{$(1+o(1))\left(\frac{\log{n}}{\log{np}}\right)$}  & Av. dist: $np\geq c>1$, \\
                                  &                                           &                                              & \phantom{Av. dist: }$\frac{\log{n}}{\log{np}}\to \infty$ \\
                                  &                                           &                                              & Diam: $np=c>1$ \ \cite{Chung2002}  \\ \hline
\multirow{2}{*}{Chung-Lu, $G(\degseq)$}  & \multirow{2}{*}{$\Theta\left(\frac{\log{n}}{\log{\tilde{d}}}\right)$} & \multirow{2}{*}{ $(1+o(1))\left(\frac{\log{n}}{\log{\tilde{d}}}\right)$} & {\it admissible} deg. seq. \\
                                  &                                           &                                              & See \cite{Chung2002} \\\hline
Pref. Attachment with             & $\Theta(\log{n})$                         & $O(\log{n})$                            & $\beta=3\quad$ \cite{Bollobas2005,pittel94} \\
power law exponent $\beta$        & $\Theta(\log{n})$                  & $\Theta(\log{n})$                          & $\beta>3\quad$ \cite{Dommers2010}  \\ \hline
Stochastic Kronecker              & $O(1)$                                    & $O(1)$                                       & See \cite{Mahdian2010}  \\
\hline
\end{tabular}
\vspace{1mm}
  \caption{Diameters and average distances in random graph models.}
\label{tab:asymDiam}
\end{table}

Recall our observation that both the diameter and average distance are much closer to $\sqrt{n}$ than to $\log(n)$ in the real power grid same-voltage subgraphs.
Consequently, a direct application of these well-known models will not suffice for generating graphs with large diameter and average distance without introducing additional modifications.
Our proposed model for the same-voltage subgraphs of the power grid does precisely this.
Motivated by the Chung-Lu (CL) model's effectiveness in {matching degree distributions} \cite{Pinar2012}, its documented tendency to produce graphs with low clustering coefficients \cite{Kolda2014}, its simplicity, and its efficient implementations \cite{miller2011efficient,Winlaw2015}, we design a heavily adapted version which we call the {\it Chung-Lu Chain} (CLC) model.
As we will soon explain in detail, while the Chung-Lu model takes a desired degree sequence as its sole input, our CLC model additionally takes desired diameter as a second input.

Our generative model for transformer edges aims to replicate the disjoint star structure observed in power grid graphs while also matching desired transformer degrees between each pair of voltages.
Accordingly, our model for $G[T]$ is based on random star generation.
While both models function independently, they may be used together to output the entire power-grid graph $G$.
Before describing each phase formally, we first provide a brief overview of the model.

\subsection{Model overview}
Our generative model for the power grid proceeds in two ``phases" and is described formally by the pseudocode in Algorithms $\ref{alg:setup}-\ref{alg:entire}$ in Appendix \ref{apen:algs}.
In the first phase, we model each same-voltage subgraph, $G[X]$, separately.
For each $X$, this phase takes two inputs: $\degseq^X$, the desired degree sequence, and $\delta^X$, the desired diameter.
These inputs are initially used in a preprocessing stage, described in the function \textsc{Setup}$(\degseq, \delta)$ in Algorithm \ref{alg:setup}, which groups vertices and records information to be used when generating the output graph.

Using the vertex grouping information along with an updated degree sequence, both recorded in preprocessing, the actual graph generation procedure for Phase 1 is described in the function \textsc{CLC}$(\degseq', {\bf v}, D, S)$ in Algorithm \ref{alg:phase1}.
Here, $\degseq', {\bf v}, D, S$ are the output from \textsc{Setup}$(\degseq, \delta)$.
Loosely speaking, this model creates a number of smaller Chung-Lu random graphs which are linked together in a special ``chain-like" structure to achieve the desired diameter.
For this reason, we call the Phase 1 model the Chung-Lu Chain model.

The second phase of our model creates the transformer edges graph, $T[X,Y]$, between every pair of voltage levels, $X$ and $Y$, separately.
For each $X$, $Y$ voltage pair, we only need the desired transformer degrees of each voltage level with respect to each other, $\tdegseq[X,Y]$ and $\tdegseq[Y,X]$, as inputs.
Based on our finding (detailed in Section \ref{sec:real_transformer}) that the transformer edges of a power grid graph consist almost entirely of disjoint stars, the graph generation process is accordingly based on random star generation.
The procedure is described in function \textsc{Stars}$(\tdegseq[X,Y], \tdegseq[Y,X])$ in Algorithm \ref{alg:phase2}.

We note that Phase 1 and Phase 2 operate independently in the sense that none of the inputs required for Phase 1 are required for Phase 2, and no information about the graphs returned by Phase 1 is necessary to run Phase 2.
This allows the user to control how inputs for Phase 1 (the desired degrees for the same-voltage subgraphs) may be correlated with the inputs to Phase 2 (the desired transformer degrees for a pair of same-voltage subgraphs).
In addition to this increased flexibility, using separate phases for the same-voltage and transformer edge subgraphs is a practical necessity, as these graph structures differ radically and thus benefit from individually tailored generation processes.
Ultimately, the entire power grid graph $G$ may be generated by applying Phase 1 to each voltage level, applying Phase 2 to each pair of voltage levels, and taking the union of all returned edge sets.
The function \textsc{CLCStars} in Algorithm \ref{alg:entire} describes how to use both phases to generate the entire power grid graph on $k$ voltage levels.


\subsection{Phase 1: the Chung-Lu Chain model}

Our random graph model for the same-voltage subgraphs consists of two algorithms: (1) a preprocessing stage and (2) a graph generation stage.
Given desired vertex degrees and desired graph diameter, the preprocessing is done in Algorithm $\ref{alg:setup}$ to randomly partition vertices into \emph{boxes} which are key to matching the large distances observed in power grid graphs.
Among the vertices within each box, one vertex is marked as a \emph{diameter path} vertex and in some boxes another is marked as a \emph{subdiameter path} vertex.
These selections involve subtle considerations to ensure that, in the graph generation stage of Algorithm $\ref{alg:phase1}$, the output graph more accurately matches the desired properties specified by the inputs.
These are explained in Appendix \ref{apen:algRem}.

\begin{figure}[t!]
\centering
\begin{subfigure}[b]{0.46\textwidth}
\includegraphics[width = \linewidth]{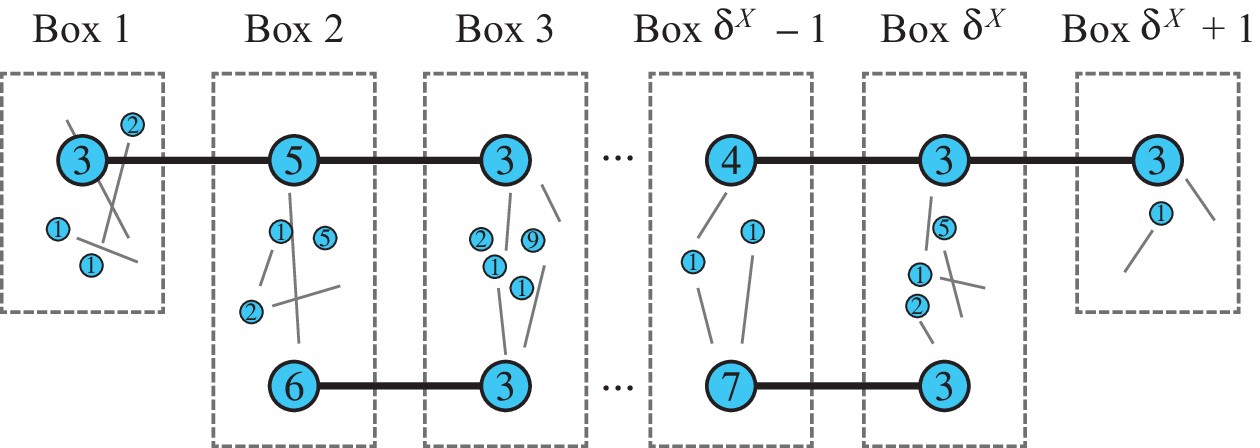}
\caption{} \label{fig:Phase1}
\end{subfigure}
\qquad
\qquad
\begin{subfigure}[b]{0.3\textwidth}
\includegraphics[width = \linewidth]{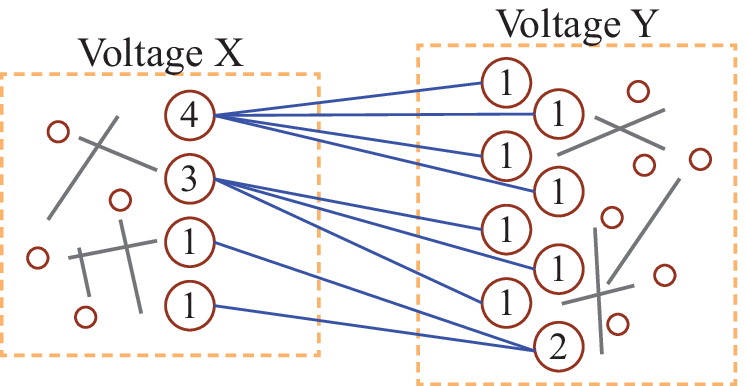}
\caption{} \label{fig:phase2}
\end{subfigure}
\caption{{\bf (a)} Output of Algorithms \ref{alg:setup}-\ref{alg:phase1}.
Algorithm \ref{alg:setup} partitions vertices into boxes, identifies a single diameter path vertex in each box and subdiameter path vertex in some boxes.
Algorithm \ref{alg:phase1} creates all edges.
Each vertex is labeled with its desired degree.
The larger vertices are the diameter path (top) and subdiameter path (bottom) vertices.
{\bf (b)} Sample transformer graph between a voltage $X$ subgraph and voltage $Y$ subgraph.
  Small unlabeled circles are vertices in each subgraph with no transformer degree, numbered vertices are labeled with their desired transformer degree.
  Blue edges are transformer edges, gray are edges formed in Phase 1 within each same-voltage subgraph.}
\end{figure}

In Algorithm $\ref{alg:phase1}$, a separate Chung-Lu random graph is generated on the vertices in each of the boxes and a deterministic diameter path is generated on the vertices chosen in Algorithm $\ref{alg:setup}$.
We note that since the vertices in each box will be modeled as a separate random graph, additional edges between vertices in different boxes are not possible, and hence the diameter of a realization of the CLC model must be at least as large as the length of this diameter path.
Furthermore, a deterministic ``subdiameter path" is also created on the chosen vertices.
The purpose of this path is to facilitate a richer connectivity structure consistent with the data by allowing for the creation of alternate paths and cycles which connect vertices assigned to different boxes.
The existence of such paths is implied by long cycles observed in visualizations of power grid graphs, as well as by computations of edge connectivity.
Figure \ref{fig:Phase1} shows a cartoon illustration of Phase 1.


\subsection{Phase 2: inserting transformer edges}\label{sec:model_phase2}

Algorithm \ref{alg:phase2} generates the transformer edges between two subgraphs of different voltage levels $X$ and $Y$.
The desired transformer degree of each vertex in the subgraphs of voltage $X$ and $Y$ are provided as input.
Following our observation in Section \ref{sec:real_transformer} that the transformer subgraphs appear to consist largely of star graphs, Algorithm 2 aims to match these desired transformer degrees by generating the transformer edges graph as a collection of random, disjoint star graphs.
To achieve this, vertices with nonzero transformer degree are partitioned into two sets:~vertices with transformer degree at least 2 and those with transformer degree 1.
Vertices of degree at least $k\geq2$ are intended to be the centers of a $k$-star, whereas vertices of degree 1 are intended as the leaf vertices which connect to the center of the star.
The stars are generated by randomly selecting a vertex from voltage $X$ with transformer degree $k\geq 2$ and connecting this vertex with $k$ randomly selected vertices from voltage $Y$ that have desired transformer degree 1.
The selected vertices are then removed from their respective sets, and another vertex is selected until the degree $k\geq 2$ vertices from voltage $X$ have been exhausted.
The same procedure is then applied for the centers of $k$ stars centered at voltage $Y$ vertices.
Finally, the remaining degree 1 vertices from each voltage level are randomly matched with each other.
Figure \ref{fig:phase2} illustrates an example output of this procedure.
In the example, the voltage $X$ transformer degrees are $\tup{4, 3, 1, 1}$ and voltage $Y$ transformer degrees are $\tup{2, 1, 1, 1, 1, 1, 1, 1}$.
In both voltage $X$ and $Y$, there are vertices which do not participate in the transformer graph (depicted by the smaller blank circles) .


Lastly, we note that Algorithm $\ref{alg:phase2}$ accounts for a special case in which the desired transformer degrees of all vertices cannot be realized as a collection of disjoint stars via the aforementioned process.
Namely, if, at any point, there are insufficiently many degree-1 vertices from the remaining pool to generate the current $k$-star, the degree $k$ vertex is removed from its pool and placed in a ``leftover bin".
The final part of the algorithm then generates a random bipartite Chung-Lu graph \cite{Aksoy2016} on the vertices in these leftover bins.
In Appendix \ref{apen:algRem}, we describe a sufficient condition on the desired transformer degree sequences of two voltage levels which guarantees this special case will not occur.
We note this condition is met by every pair of voltage levels in all our power grid graph data.
Furthermore, satisfying this condition also guarantees that the actual transformer degrees of vertices in the output graph of Algorithm $\ref{alg:phase2}$ will match the input desired degrees {\it exactly}.

%% file: gpg-results.tex
\section{Comparison of model output to real data}\label{sec:results}

After discussing observations within the real data (Section \ref{sec:real_data}), and presenting our Chung-Lu Chain (CLC) and star generation algorithms (Section \ref{sec:model}), we can now summarize the results of the algorithm in comparison to the real-world data.
In particular, we first present results for Phase 1, the same-voltage subgraphs, then for the transformer edges, and finally for the aggregate graph, taking the union of the same-voltage subgraphs and the transformer edges.
The measures we focus on matching are: number of vertices and edges in the largest component, diameter, average distance, average local clustering coefficient, and degree distribution.
We discuss the measures separately in the following subsections.

As an aside, the CLC model also performed more favorably than Chung-Lu for a number of other metrics tested, including degree correlation statistics like the assortativity coefficient, as well as geometric notions of connectivity like the spectral gap of the normalized Laplacian matrix (which is intimately related to ``algebraic connectivity").
For brevity, we limit our presentation here to the aforementioned metrics due to their relevance to the aims of our model. For the interested reader, data on assortativity and the spectral gap are available in Appendix \ref{apen:eig}.



\subsection{Same-voltage subgraphs} \label{sec:same_voltage_comp}
\begin{figure}[h!]
\centering
\begin{subfigure}[b]{0.3\textwidth}
\includegraphics[width = \linewidth]{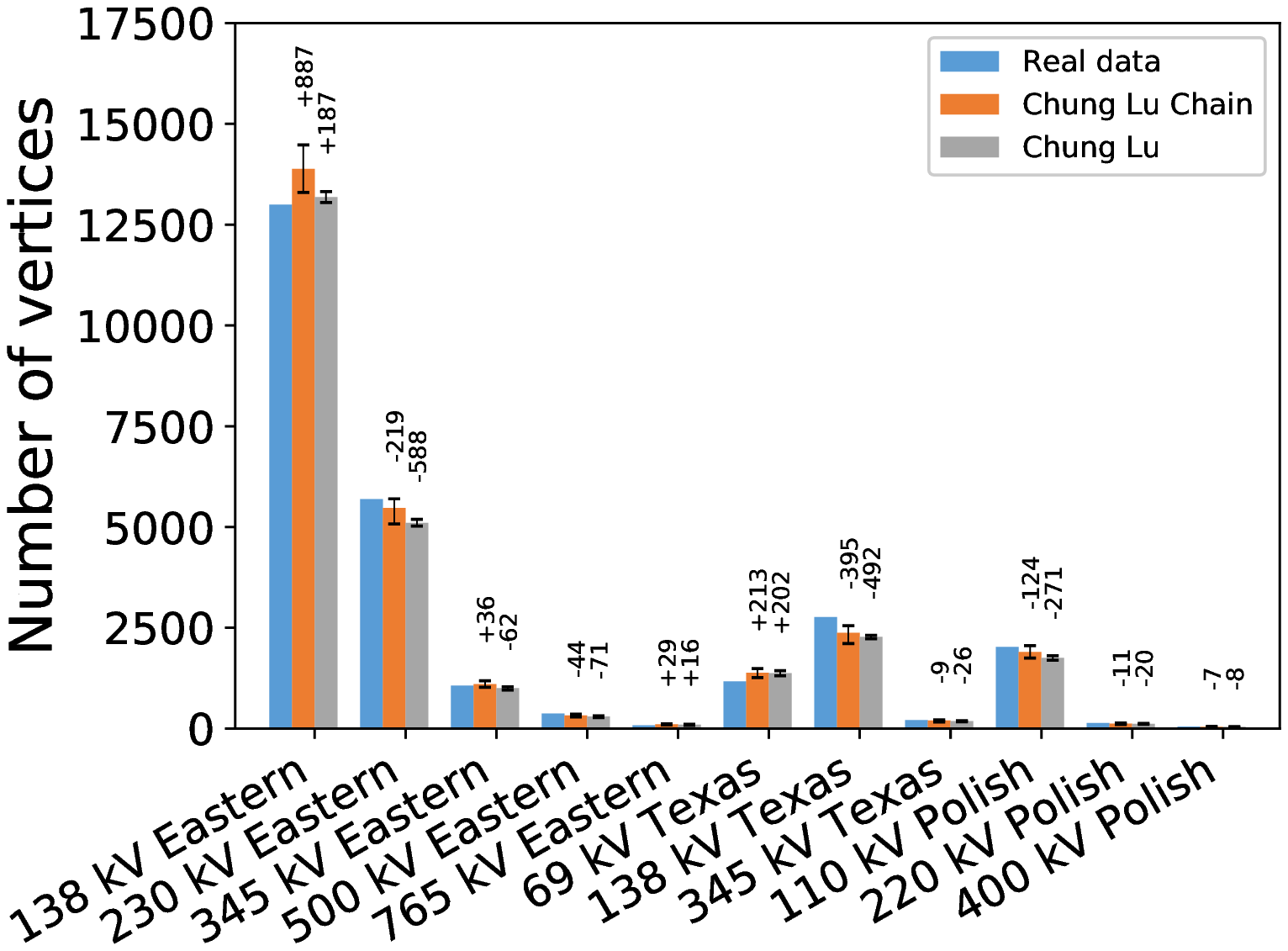}
\caption{}\label{fig:NumVertCompare}
\end{subfigure}
\begin{subfigure}[b]{0.3\textwidth}
\includegraphics[width = \linewidth]{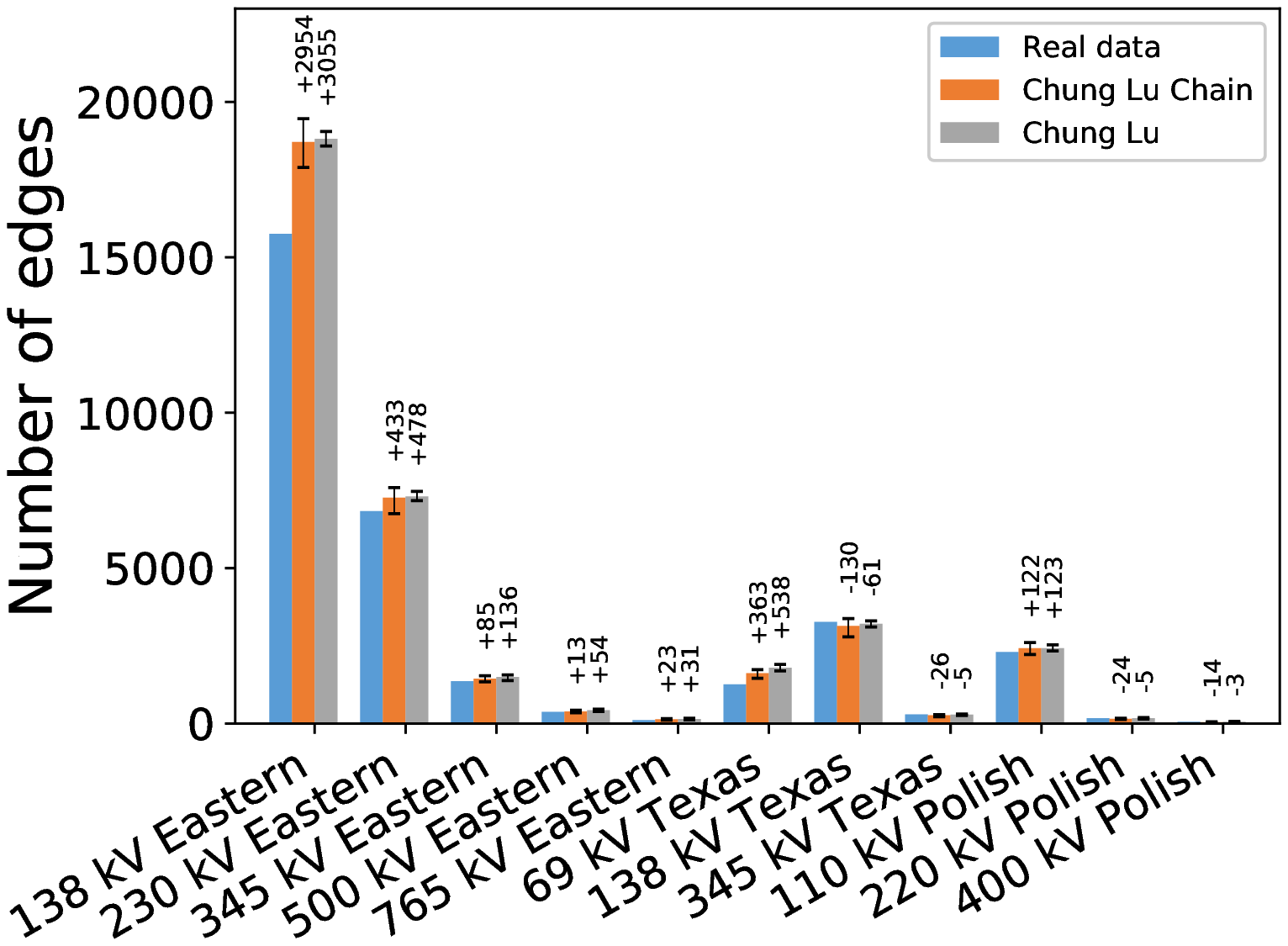}
\caption{}\label{fig:NumEdgeCompare}
\end{subfigure}
\begin{subfigure}[b]{0.3\textwidth}
\includegraphics[width = \linewidth]{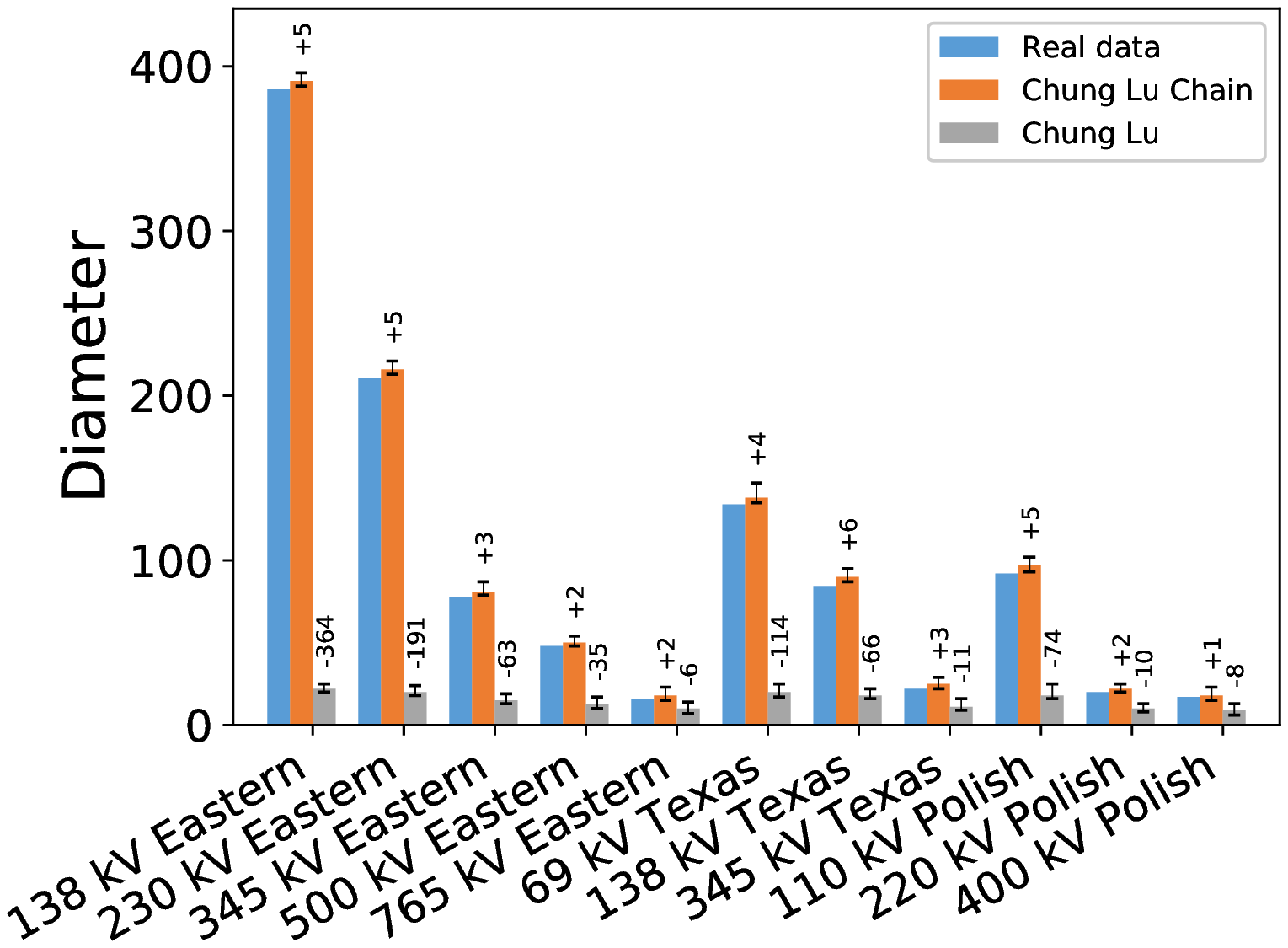}
\caption{}\label{fig:DiamCompare}
\end{subfigure}
\\
\begin{subfigure}[b]{0.3\textwidth}
\includegraphics[width = \linewidth]{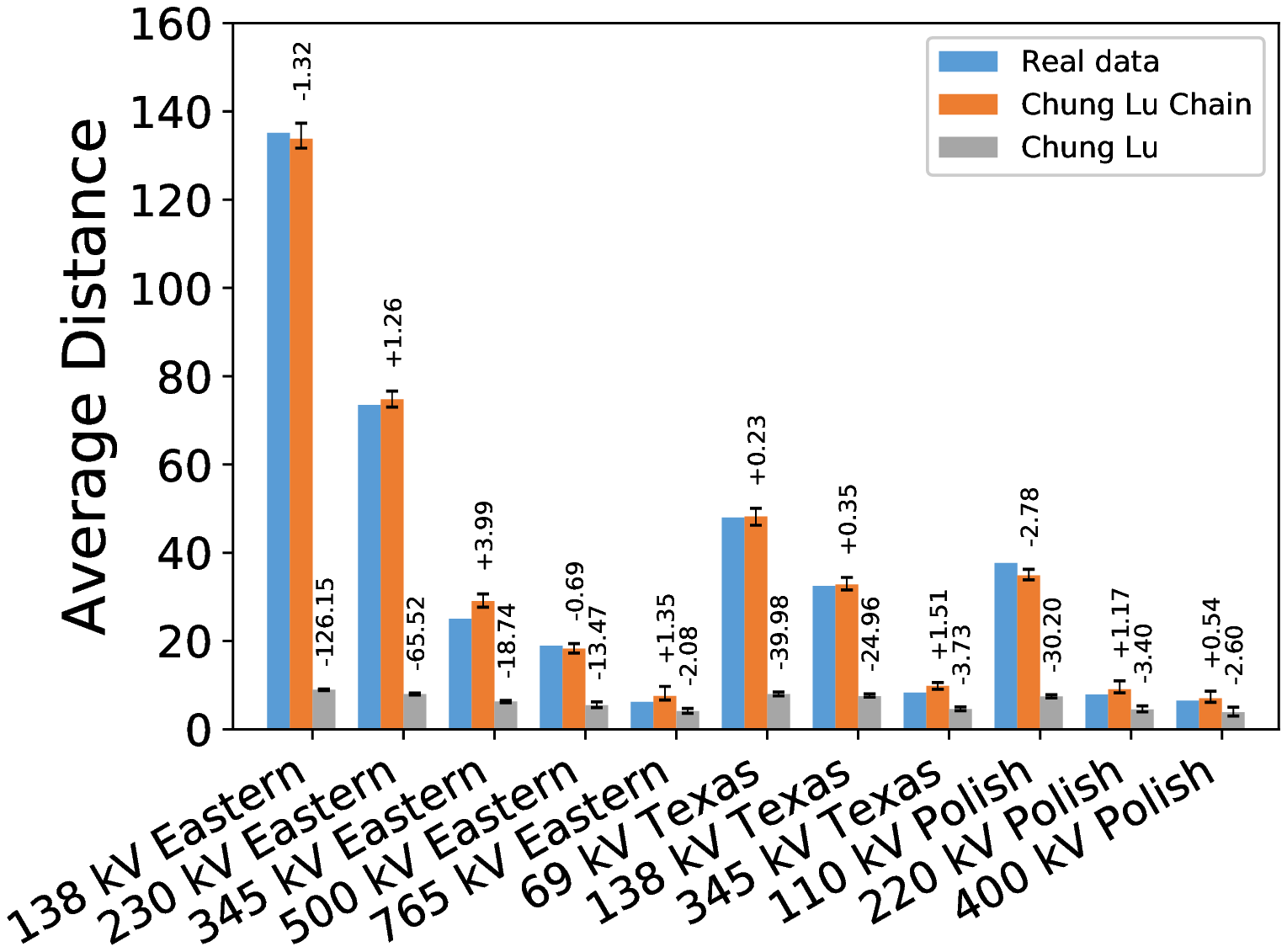}
\caption{}\label{fig:AvDistCompare}
\end{subfigure}
\begin{subfigure}[b]{0.3\textwidth}
\includegraphics[width = \linewidth]{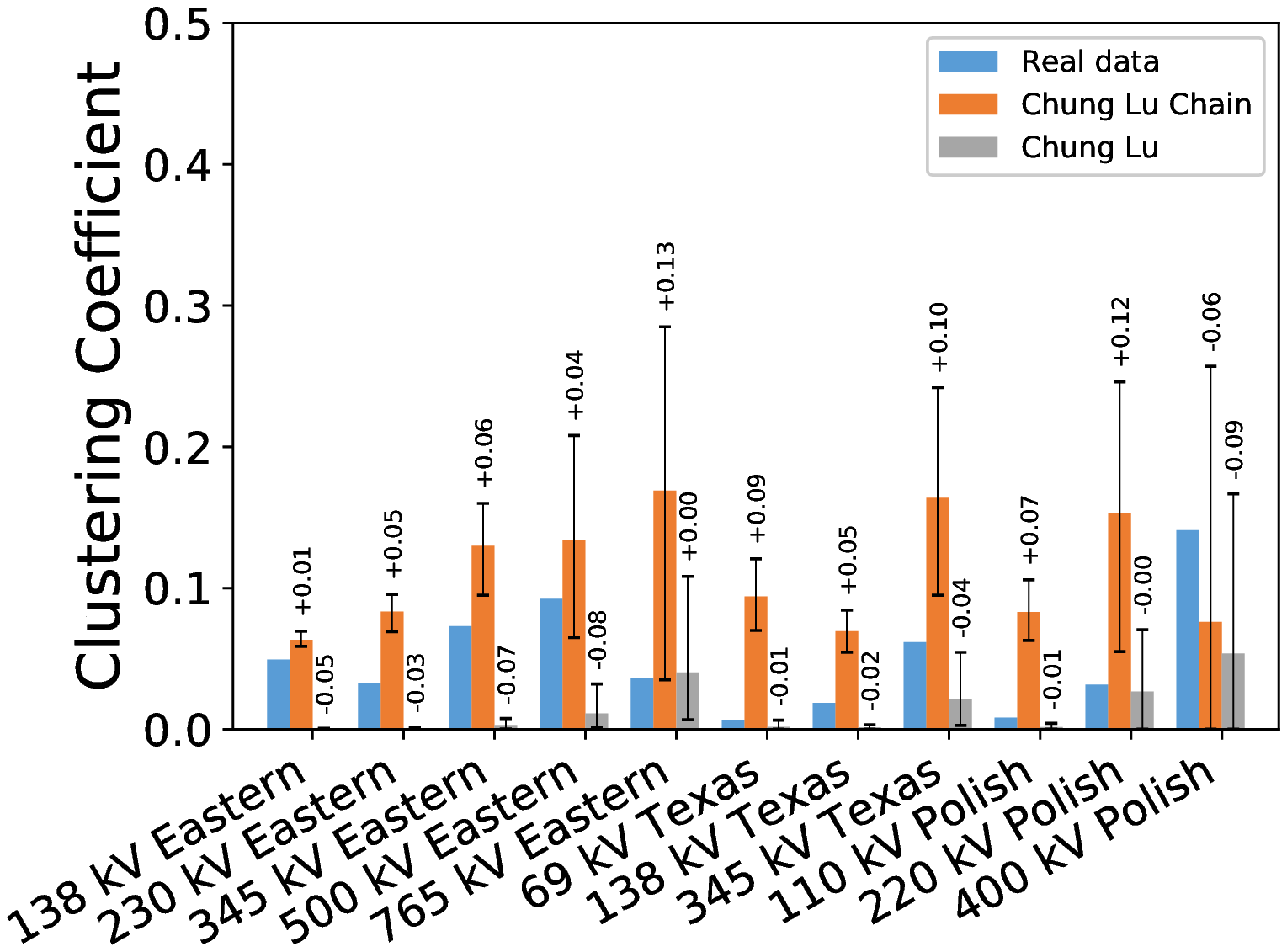}
\caption{}\label{fig:AvLCCCompare01}
\end{subfigure}
\caption{Bar charts comparing measures in the same-voltage subgraphs.}
\label{fig:subBars}
\end{figure}

The bar charts in Figures \ref{fig:NumVertCompare}--\ref{fig:AvLCCCompare01} show the relevant measures on same-voltage subgraphs compared across the real data (blue), our new CLC model (orange), and the Chung-Lu model (gray).
Each same-voltage subgraph is treated separately and they are listed along the bottom of the chart.
Both random models are run 100 times with the necessary input data being measured from the corresponding real subgraph, and the measures are collected for each run.
For both Chung-Lu and CLC, the bar height is the average of those 100 trials, {the raw residual values are reported above each bar}, and error bars indicate the minimum and maximum values {observed over the 100 trials.} 

We must point out that the real same-voltage subgraphs may not be connected, but tend to contain a single large connected component and many very small connected components.
Moreover, Chung-Lu and CLC are not guaranteed to return connected graphs.
However, in \cite{Chung2006}, Chung and Lu prove that if expected average degree is strictly greater than 1 (a mild condition certainly met by our data), the Chung-Lu model will almost surely output a graph with a giant component.
Experimentally, this is confirmed for both the Chung-Lu and CLC models.
Therefore, for all cases including the real data, we only report data corresponding to the largest connected component of each graph.

\paragraph{Number of vertices and edges}
Figures \ref{fig:NumVertCompare}--\ref{fig:NumEdgeCompare} compare the number of vertices and edges in the largest component across all test cases and the two generative models.
Chung-Lu and CLC are very similar to each other in vertex and edge count, and additionally tend to match the real data well.
The worst matching appears to be the number of edges for the largest same-voltage subgraph, the Eastern 138 kV subgraph.
But even in this case both are within 20\% on the average with CLC performing slightly better than Chung-Lu.
It is worth emphasizing that only (non-isolated) vertices in the largest component contribute to reported total vertex counts in both the original graph data and the random graph model data.
The Chung-Lu model is never guaranteed to match largest component size nor total number of non-isolated vertices in the entire graph since any vertex may become isolated with nonzero probability.
To ensure that CLC better matches non-isolated vertex counts and largest component sizes in the original data, we use prior theoretical results \cite{Chung2006} on the Chung-Lu model to adjust the inputted degree sequence of the CLC model (see Appendix \ref{apen:algRem} for details).


%

\paragraph{Diameter and average distance}
These two measures, with comparisons in Figures \ref{fig:DiamCompare} and \ref{fig:AvDistCompare}, show stark differences between the CLC model and the Chung-Lu model.
In all test cases and trials, the CLC model generates a graph which matches the real diameter and average distance significantly more accurately than Chung-Lu.
This stems from using the diameter and subdiameter paths in the CLC model.
As discussed in Section \ref{sec:dist_random_graphs} the Chung-Lu model outputs a graph that under mild assumptions (see {\it admissible degree sequence} in \cite{Chung2006}) has diameter and average distance roughly $\log(|V|)$ with high probability, whereas we observe the diameter and average distance in the real data are on the order of $\sqrt{|V|}$.
Moreover, because diameter is an input parameter to our CLC model, which can match graphs with both small and large diameter with minor lower and upper limits.

While average distance is not an explicit input of the model, we nonetheless are able to match it well by virtue of the data exhibiting a consistent diameter to average distance ratio, which our model replicates.
That is, across all same-voltage subgraphs in our data, the diameter tends to be about 3 times as large as the average distance (more precisely, the mean ratio is 2.71, ranging from 2.44 to 3.12).
Furthermore, since the Chung-Lu Chain model stitches together many small graphs (each with small diameter) in a path-like structure (see Figure $\ref{fig:Phase1}$), the diameter to average distance ratio of the output graphs will be approximately the same as that of the path graph on $n$ vertices, $P_n$.
A routine computation shows that $P_n$ has a diameter to average distance ratio of $\frac{3(n-1)}{n+1}\approx 3$.
In this way, our model utilizes the dependency between diameter and average distance featured in the data so that, in matching diameter correctly, we also match average distance without having to invoke it as a separate parameter.


%

\paragraph{Average local clustering coefficient}
The comparison plot for average local clustering coefficient is in Figure \ref{fig:AvLCCCompare01}.
{For clarity, we plot the local clustering coefficient on a $[0,0.5]$ $y$-axis, although possible clustering coefficient values range between 0 and 1.}
It is oft-noted that the Chung-Lu model tends to produce graphs of low clustering coefficients, whereas many real-world networks exhibit much larger clustering coefficients \cite{Chung2006, Kolda2014}.
However, as we have observed, power grid graphs stand apart in this regard:~their clustering coefficients appear to be small, typically below 0.1 in our data with only one exception (Polish 400 kV).
Therefore, what is sometimes cited as a unrealistic property of the Chung-Lu model is in fact quite appropriate for power grid graphs.
Since our model is just a set of Chung-Lu graphs connected via diameter and subdiameter paths it is expected, and experimentally confirmed, that CLC will also provide a small clustering coefficient.

Both theoretical and experimental research \cite{Bollobas2005, Ostroumova2013, vdHofstad2017} show that, under a variety of assumptions, the LCC of scale-free power law graphs tends toward zero or some positive constant as the network size approaches infinity.
{Accordingly, it is not surprising that when modeling real-world networks, those of smaller size may have relatively larger LCC compared to those of larger networks.}
Since the CLC model ``stitches together" smaller Chung-Lu random graphs, one might expect both that (1) the LCC of the CLC model exceeds that of the Chung-Lu model run on the same degree distribution; and (2) the LCC is larger for CLC graphs in which the expected number of vertices per box is smaller when compared to Chung-Lu Chain graphs in which the expected number of vertices per box is larger.
Such expectations are consistent with the data in Figure \ref{fig:AvLCCCompare01}, which shows: (1) the CLC model generates graphs with larger clustering coefficients than the corresponding Chung-Lu model on all datasets, and (2) the CLC model has larger clustering coefficients on smaller graphs, like the Eastern 765 kV or Texas 345 kV, than on larger graphs like the Eastern 138 kV.
While local clustering coefficient is not a directly tunable parameter of the CLC model, our data shows the model produces ``small" clustering coefficients, within about 0.1 of the original data and often closer for larger subgraphs.

\begin{figure}[t!]
\centering
\includegraphics[scale=0.14]{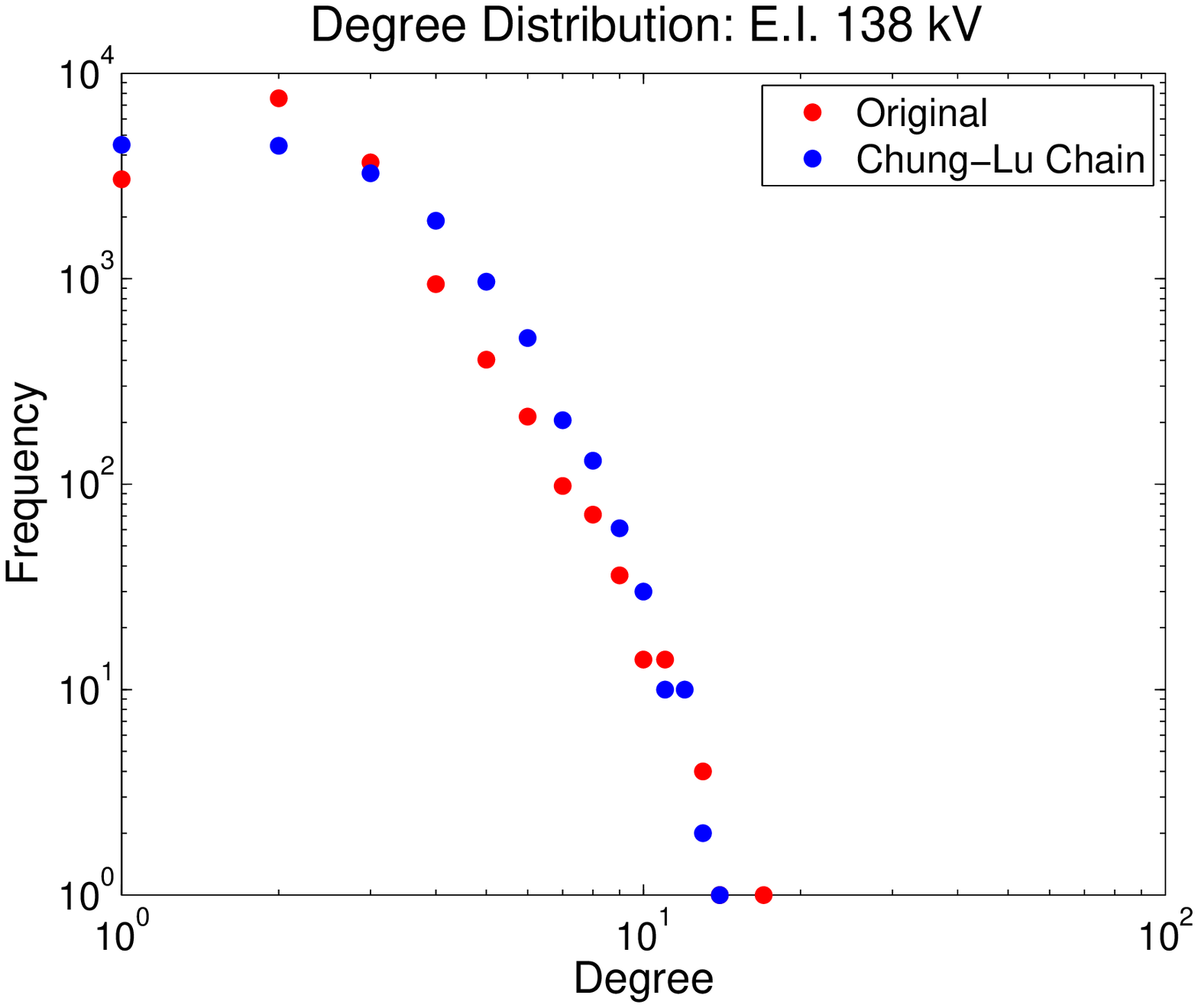}
\includegraphics[scale=0.14]{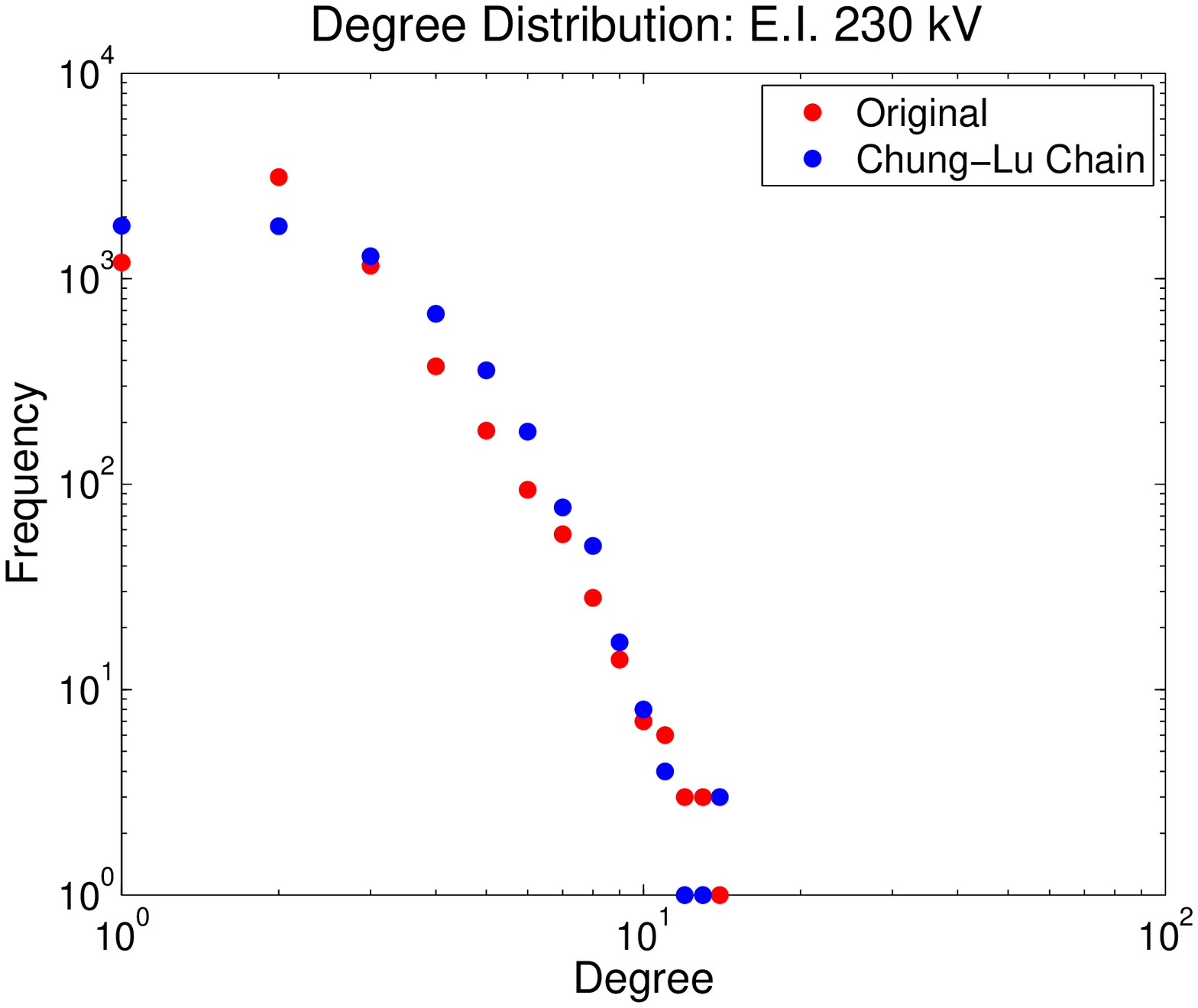}
\includegraphics[scale=0.14]{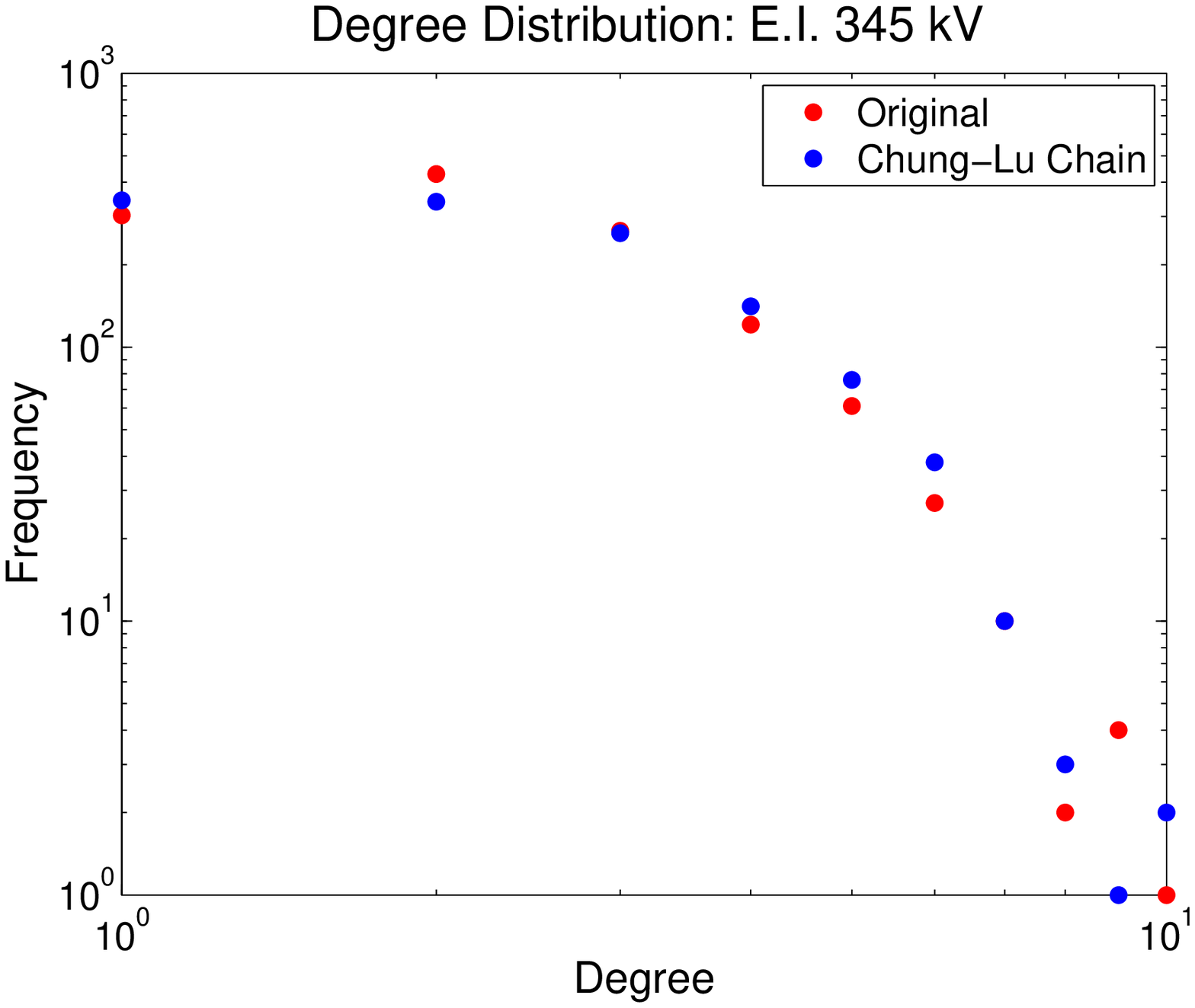}
\includegraphics[scale=0.14]{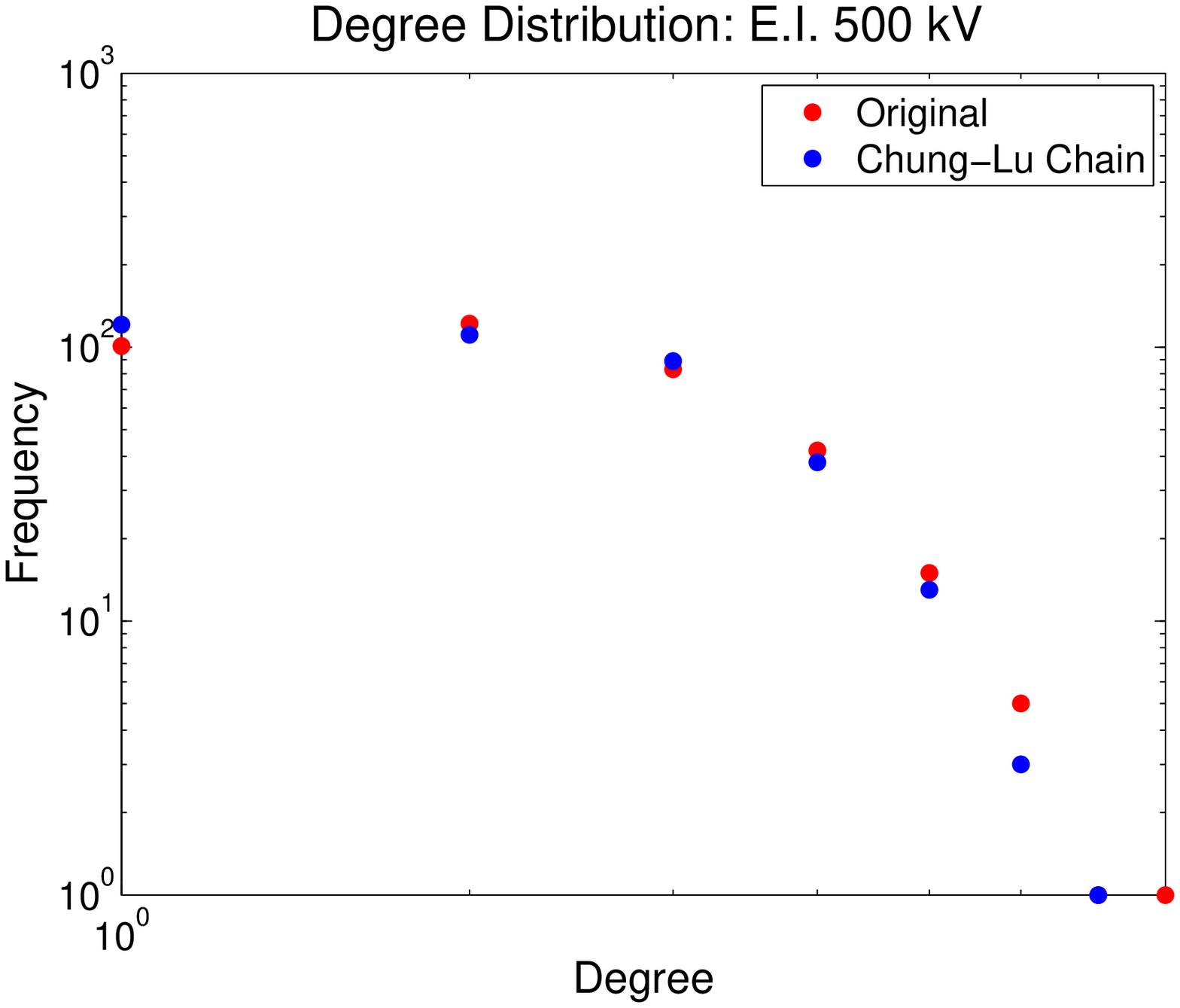}
\includegraphics[scale=0.14]{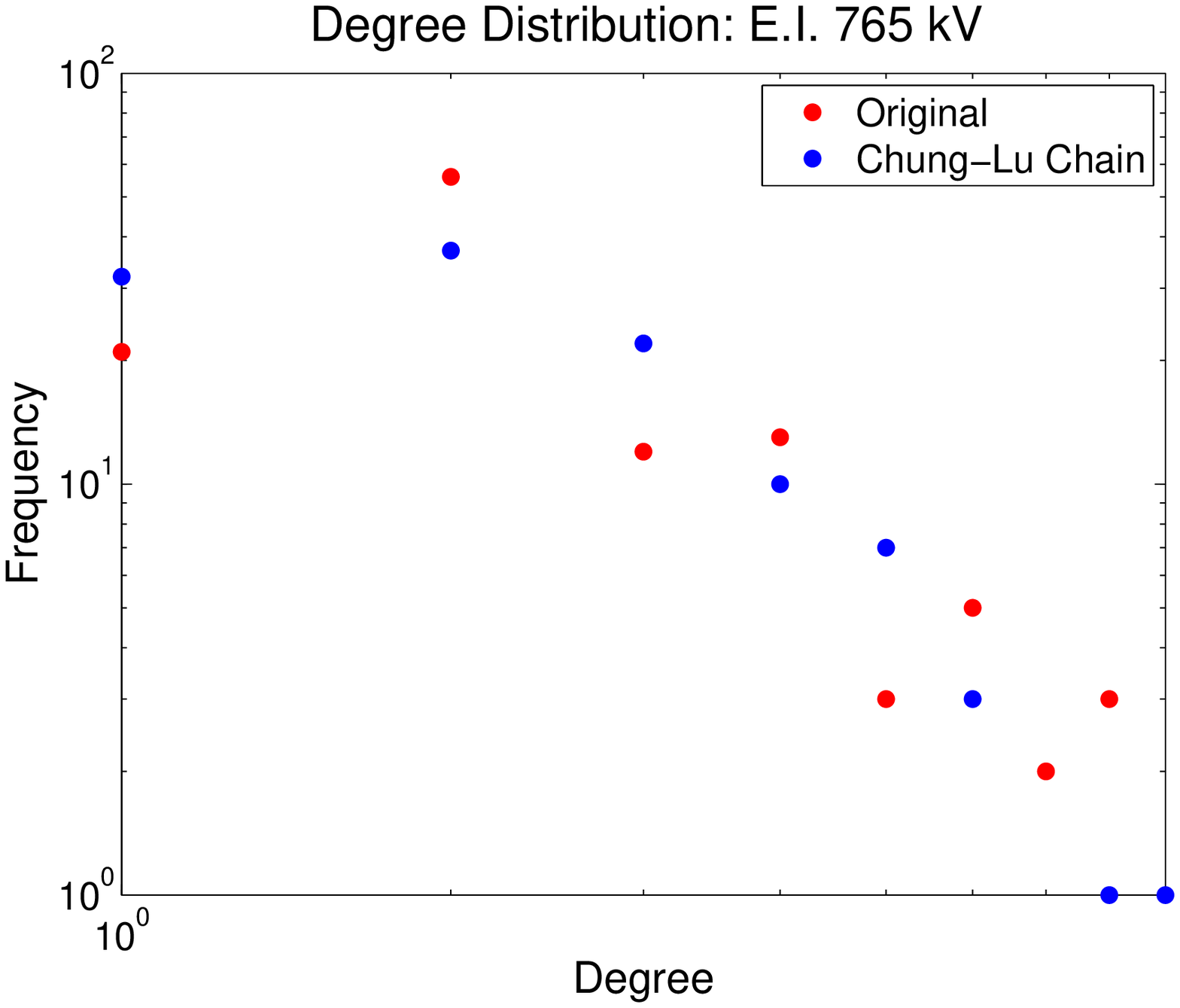} \\
\vspace{2mm}
\scriptsize
\begin{tabular}{ |l|p{0.1cm}|p{0.1cm}|}
  \hline
  \multicolumn{3}{|c|}{\bf{Eastern: DD Comparison Metrics}} \\
  \hline
   & $RH$ & $KS$  \\ \hline
138 kV & 0.21 & 0.13  \\ 
230 kV & 0.18 & 0.12  \\ 
345 kV & 0.10 & 0.05  \\ 
500 kV & 0.13 & 0.04 \\ 
765 kV & 0.27 & 0.12 \\  \hline
\end{tabular}
\quad
\begin{tabular}{ |p{0.75cm}|p{0.1cm}|p{0.1cm}|}
  \hline
  \multicolumn{3}{|c|}{\bf{Texas: DD Comparison Metrics}} \\
  \hline
   & $RH$ & $KS$  \\ \hline
69 kV & 0.24 & 0.10  \\ 
138 kV & 0.21 & 0.12  \\ 
345 kV & 0.11 & 0.12  \\  \hline
\end{tabular}
\quad
\begin{tabular}{ |p{0.75cm}|p{0.1cm}|p{0.1cm}|}
  \hline
  \multicolumn{3}{|c|}{\bf{Polish: DD Comparison Metrics}} \\
  \hline
   & $RH$ & $KS$  \\ \hline
110 kV & 0.18 & 0.12  \\ 
220 kV & 0.13 & 0.11  \\ 
400 kV & 0.24 & 0.24  \\  \hline
\end{tabular}
\caption{{\it Top row:} Degree distribution, on a log-log scale, for all five same-voltage subgraphs of the Eastern Interconnection. {\it Bottom row:} Degree distribution comparison metrics for the Eastern, Texas, and Polish networks.}
\label{fig:degResults_EI}
\end{figure}

%

\paragraph{Degree distribution}
{The degree distribution is critical to match because of its fundamental nature}.
Both in theory and practice, the Chung-Lu model (which matches desired degrees in expectation) is already well-known to {match degree distributions well}. Thus, we skip this data in the comparison.
However, as the Chung-Lu Chain model is heavily adapted, and contains deterministic diameter and subdiameter paths, it is not immediately clear to what extent Chung-Lu's properties in matching degree distribution extend to the CLC model -- particularly at different scales.
But, in Figure \ref{fig:degResults_EI} we see the degree distribution in the resulting CLC model (blue) approximately matches the real degree distribution (red).
{This similarity can be quantified by the Relative Hausdorff (RH) measure and Kolmogorov-Smirnov (KS) statistic, which are tools for comparing two graph degree distributions or complementary cumulative degree histograms (see \cite{Simpson15, matulef2017sampling}). Below the plots, we present the RH and KS values between the original data versus our model's output for all datasets, each averaged over 100 runs. Smaller values indicate a closer match. Furthermore, while KS is bounded between 0 and 1, RH values can exceed 1.
As a guideline, \cite{stolman2017hyperheadtail} suggests interpreting RH values less than 0.30 as indicative of a ``good" match between two graph's degree distributions, while RH values less than 0.10 are considered ``excellent".}

\begin{figure}
\centering
\begin{subfigure}[b]{0.3\textwidth}
\includegraphics[width=\linewidth]{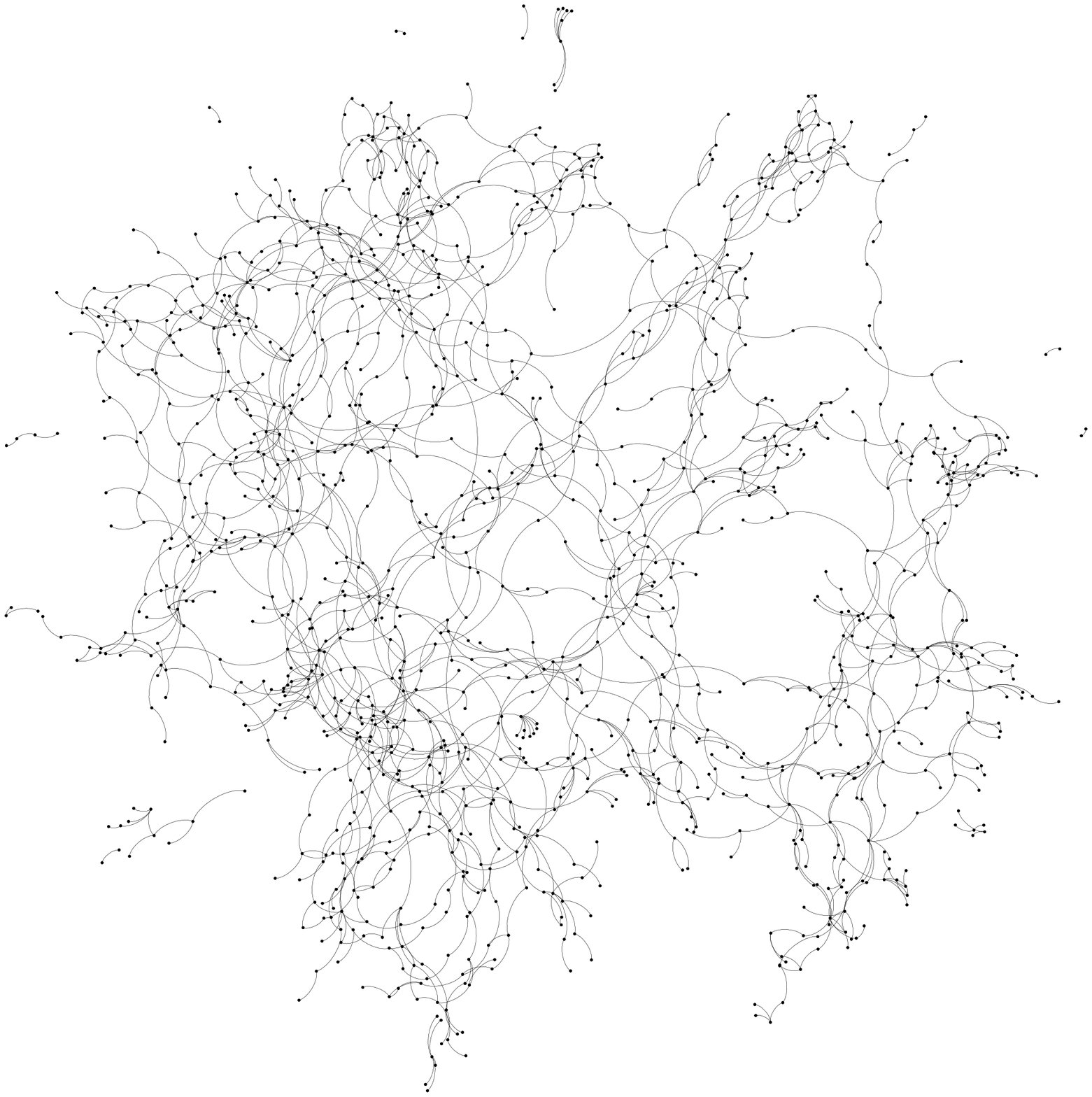}
\caption{E.I. 345 kV (Original)} \label{fig:EVizCompOrig}
\end{subfigure}
\hspace{5mm}
\centering
\begin{subfigure}[b]{0.3\textwidth}
\includegraphics[width=\linewidth]{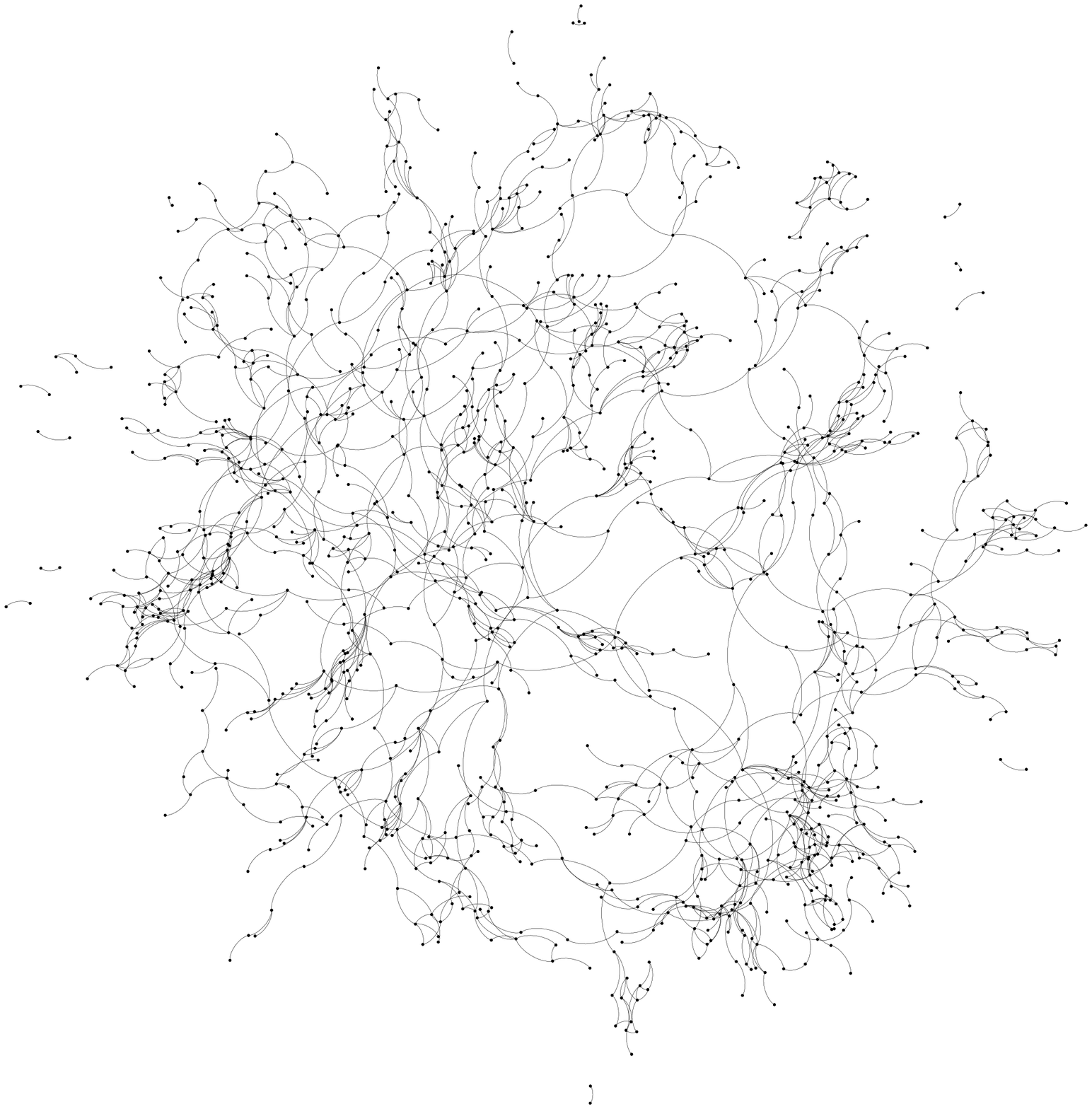}
\caption{E.I. 345 kV (CLC)}\label{fig:EVizCompCLC}
\end{subfigure}
\hspace{5mm}
\begin{subfigure}[b]{0.3\textwidth}
\includegraphics[width=\linewidth]{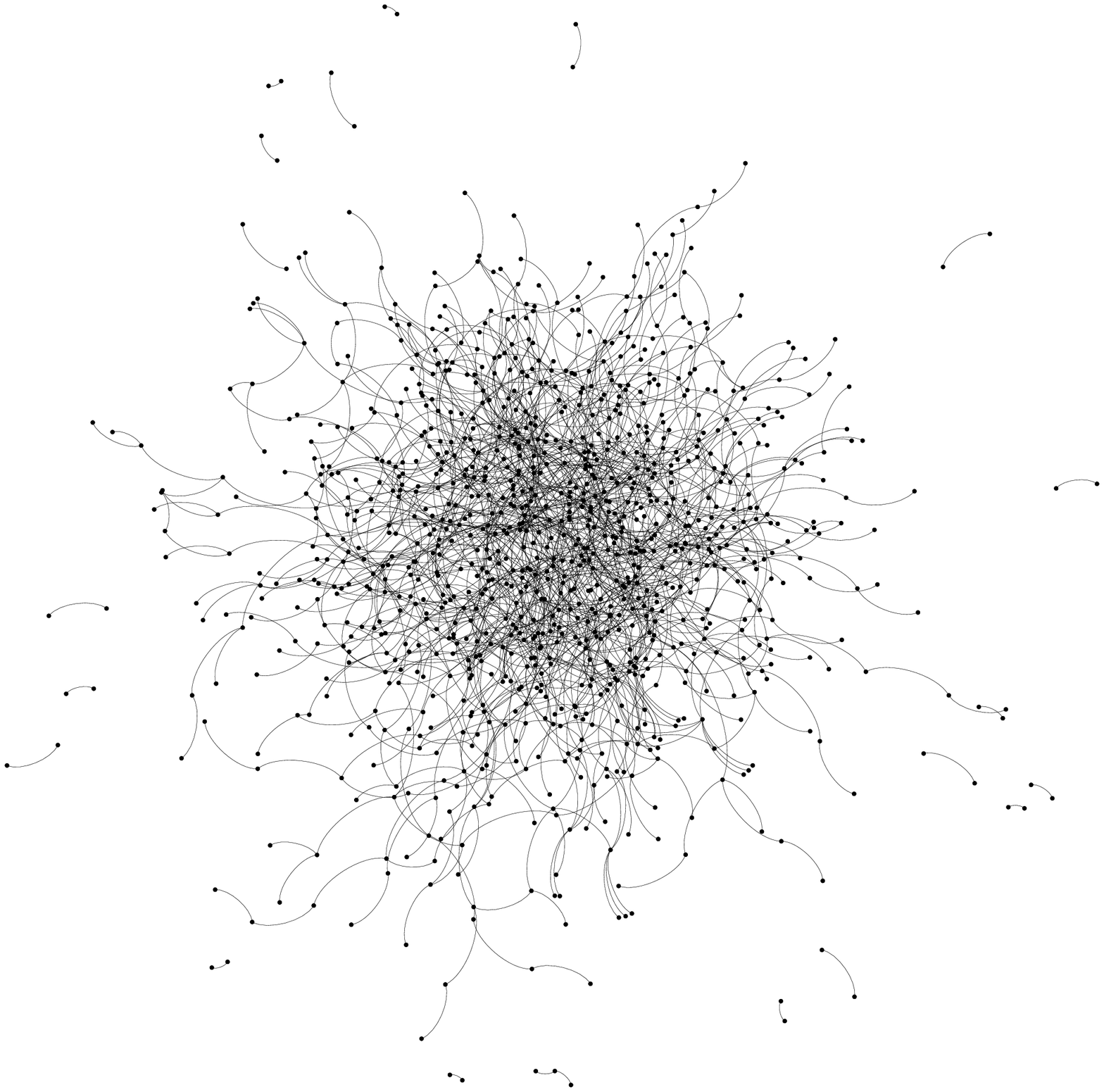}
\caption{E.I. 345 kV (CL)}\label{fig:EVizCompCL}
\end{subfigure}
\\
\begin{subfigure}[b]{0.26\textwidth}
\includegraphics[width=\linewidth]{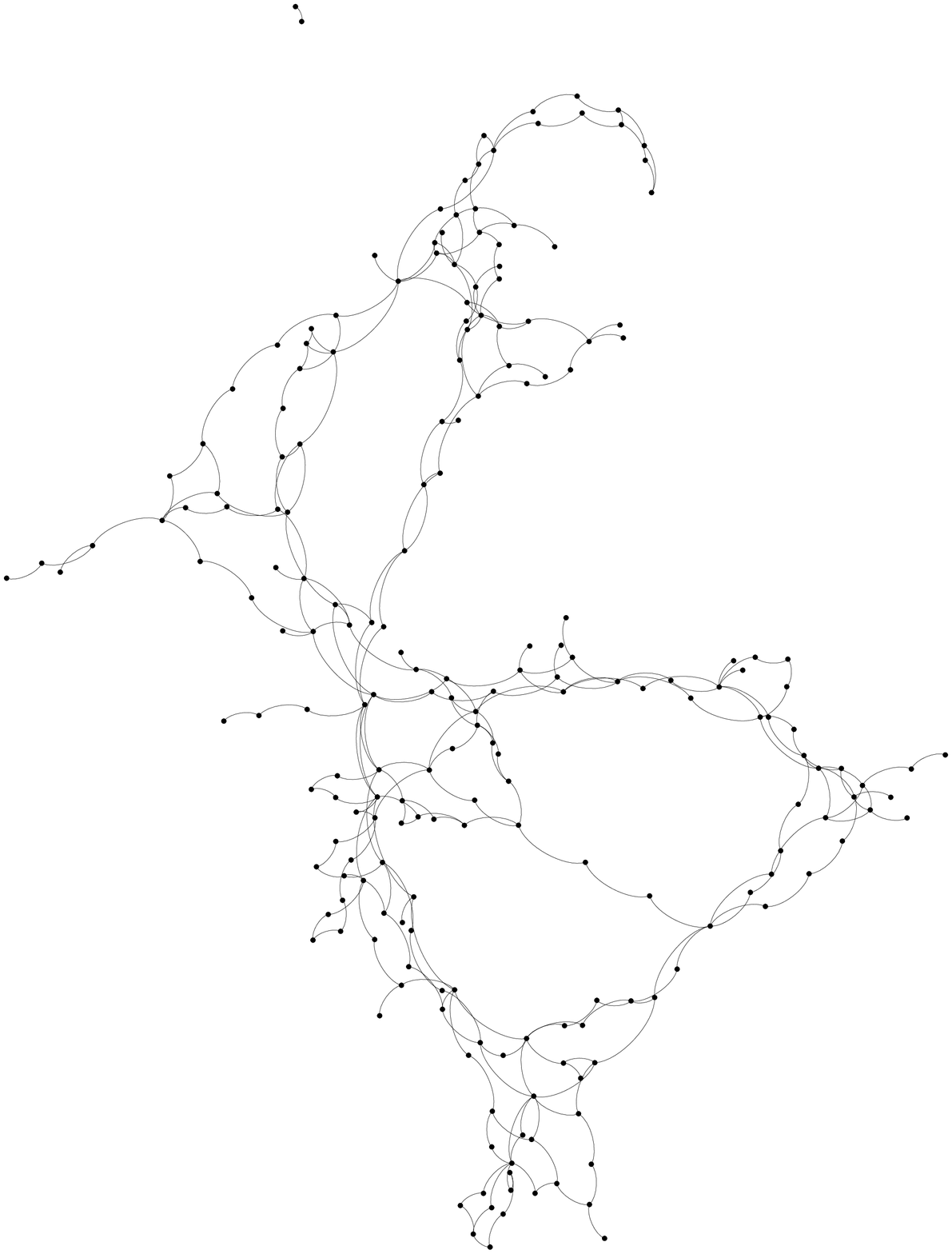}
\caption{T.I. 345 kV (Original)}\label{fig:TVizCompOrig}
\end{subfigure}
\hspace{5mm}
\centering
\begin{subfigure}[b]{0.26\textwidth}
\includegraphics[width=\linewidth]{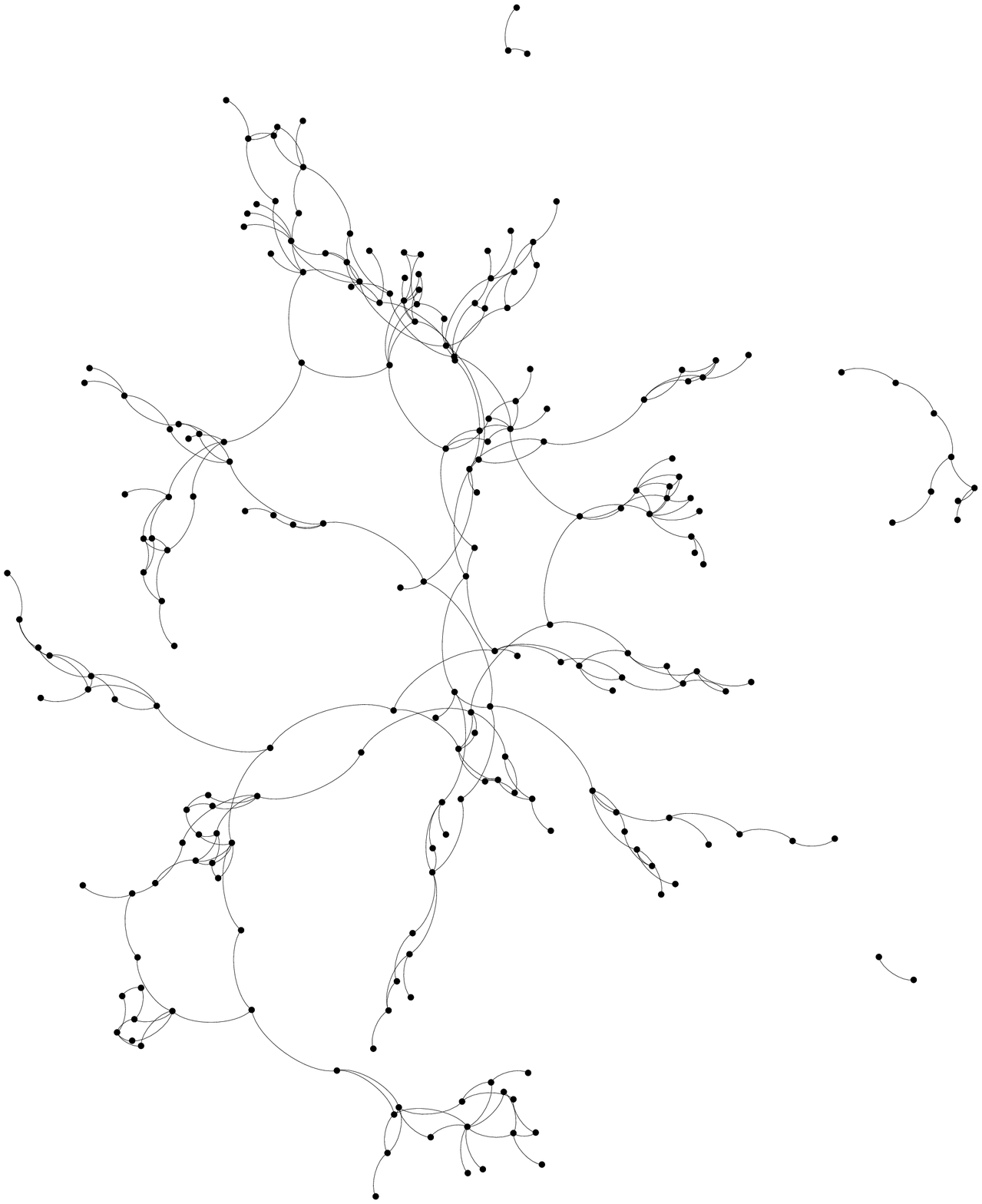}
\caption{T.I. 345 kV (CLC)}\label{fig:TVizCompCLC}
\end{subfigure}
\hspace{5mm}
\begin{subfigure}[b]{0.26\textwidth}
\includegraphics[width=\linewidth]{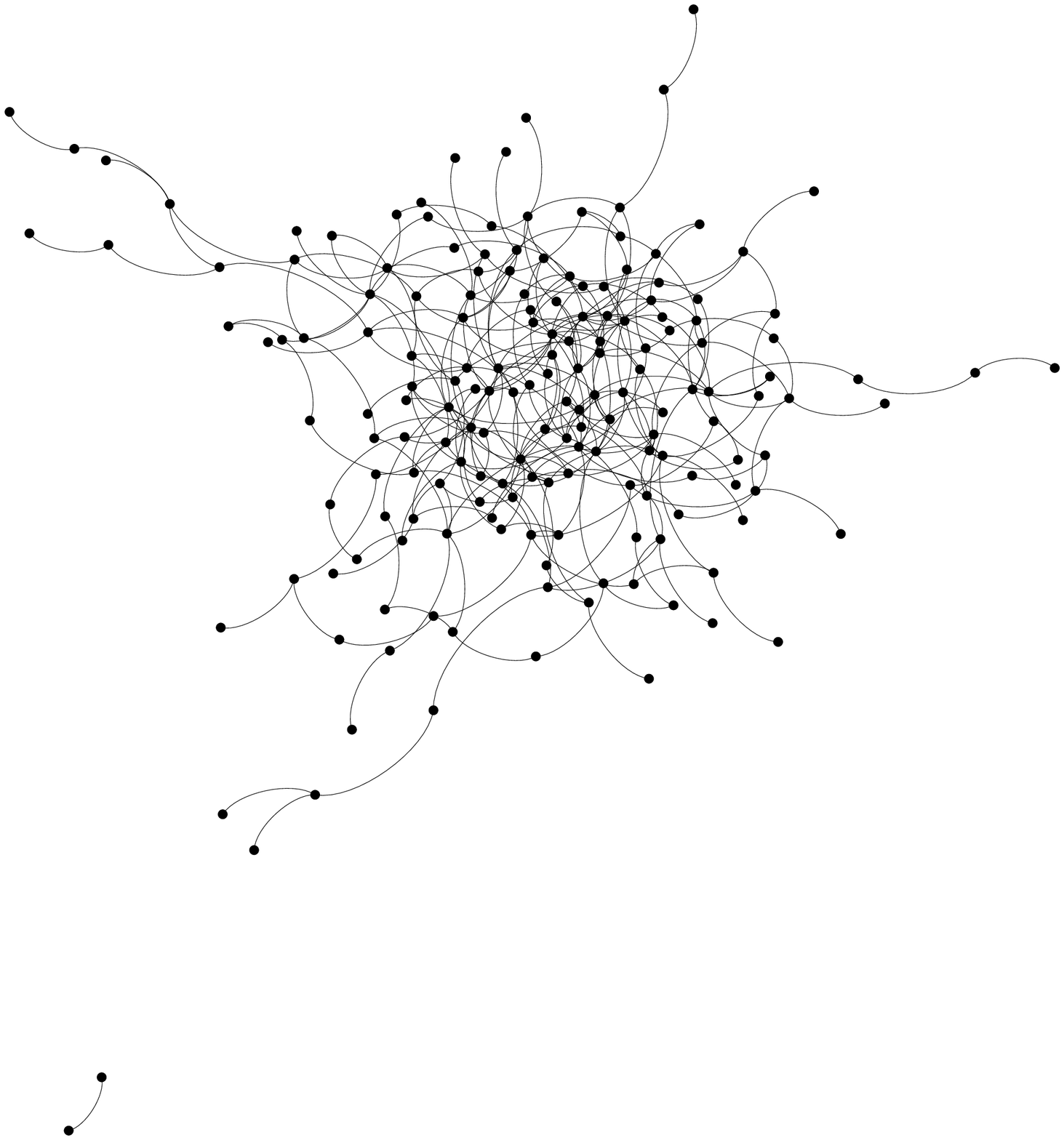}
\caption{T.I. 345 kV (CL)}\label{fig:TVizCompCL}
\end{subfigure}
\\
\begin{subfigure}[b]{0.26\textwidth}
\includegraphics[width=\linewidth]{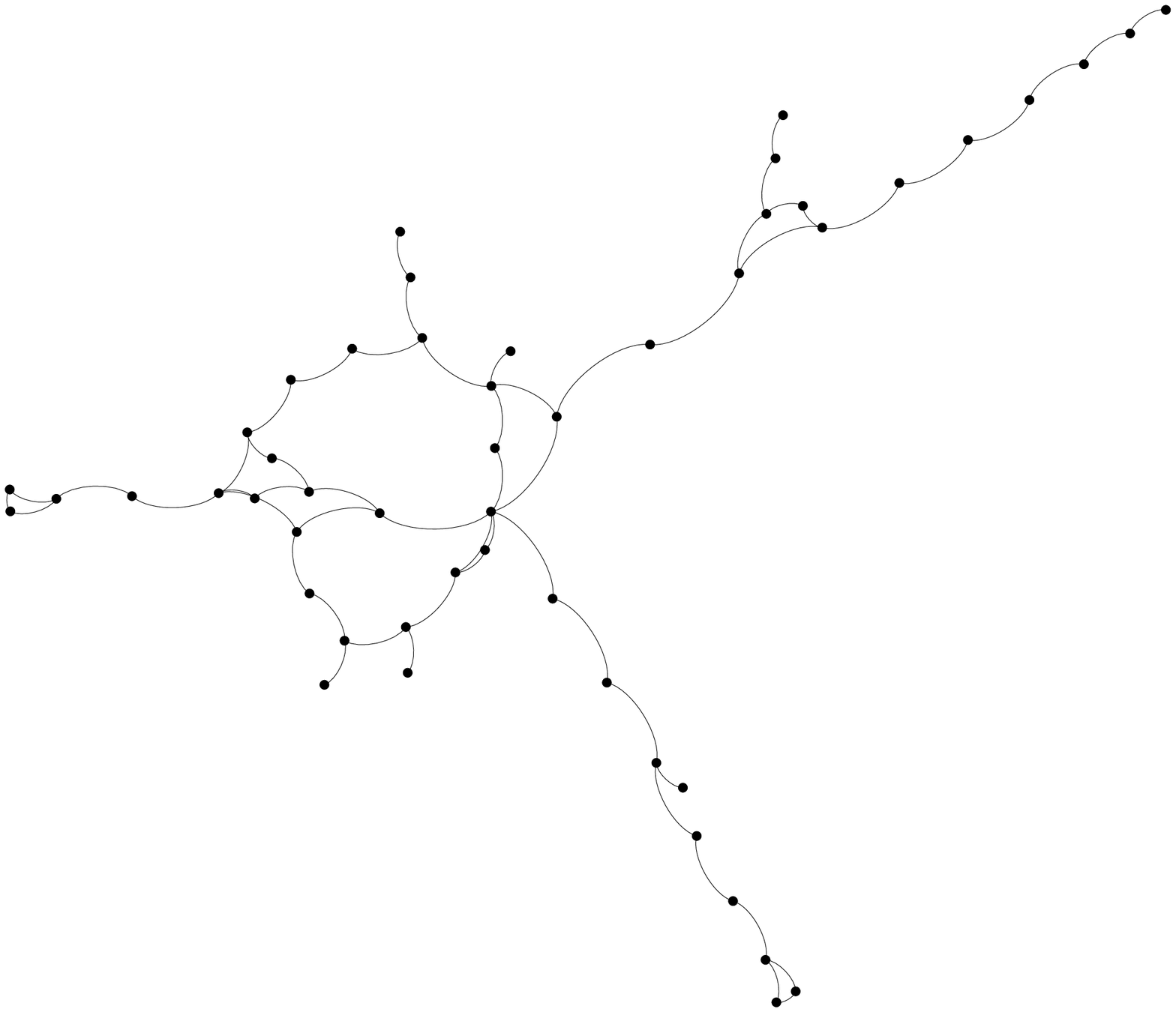}
\caption{Polish 400 kV (Original)}\label{fig:PVizCompOrig}
\end{subfigure}
\hspace{5mm}
\begin{subfigure}[b]{0.26\textwidth}
\includegraphics[width=\linewidth]{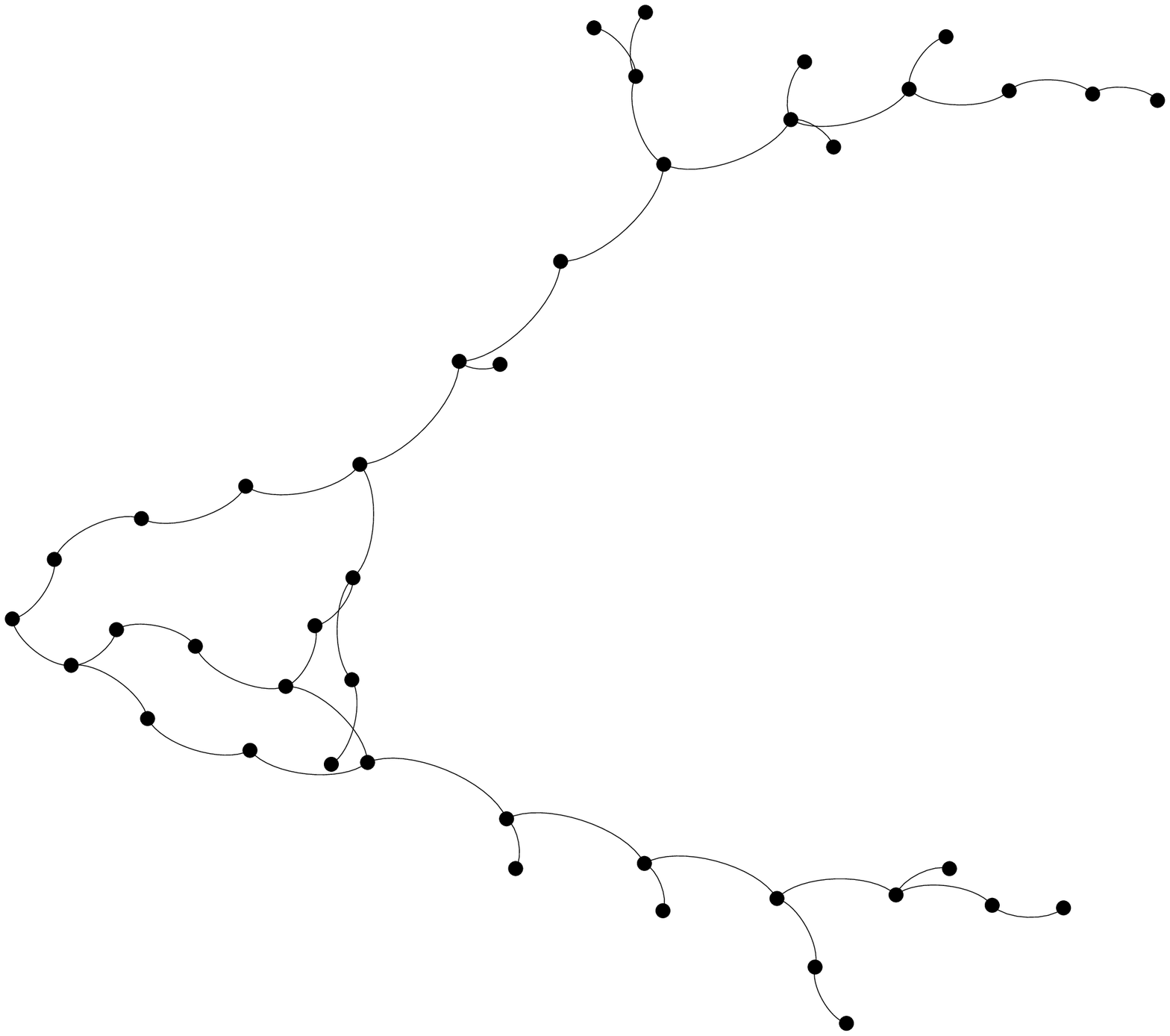}
\caption{Polish 400 kV (CLC)}\label{fig:PVizCompCLC}
\end{subfigure}
\hspace{5mm}
\begin{subfigure}[b]{0.26\textwidth}
\includegraphics[width=\linewidth]{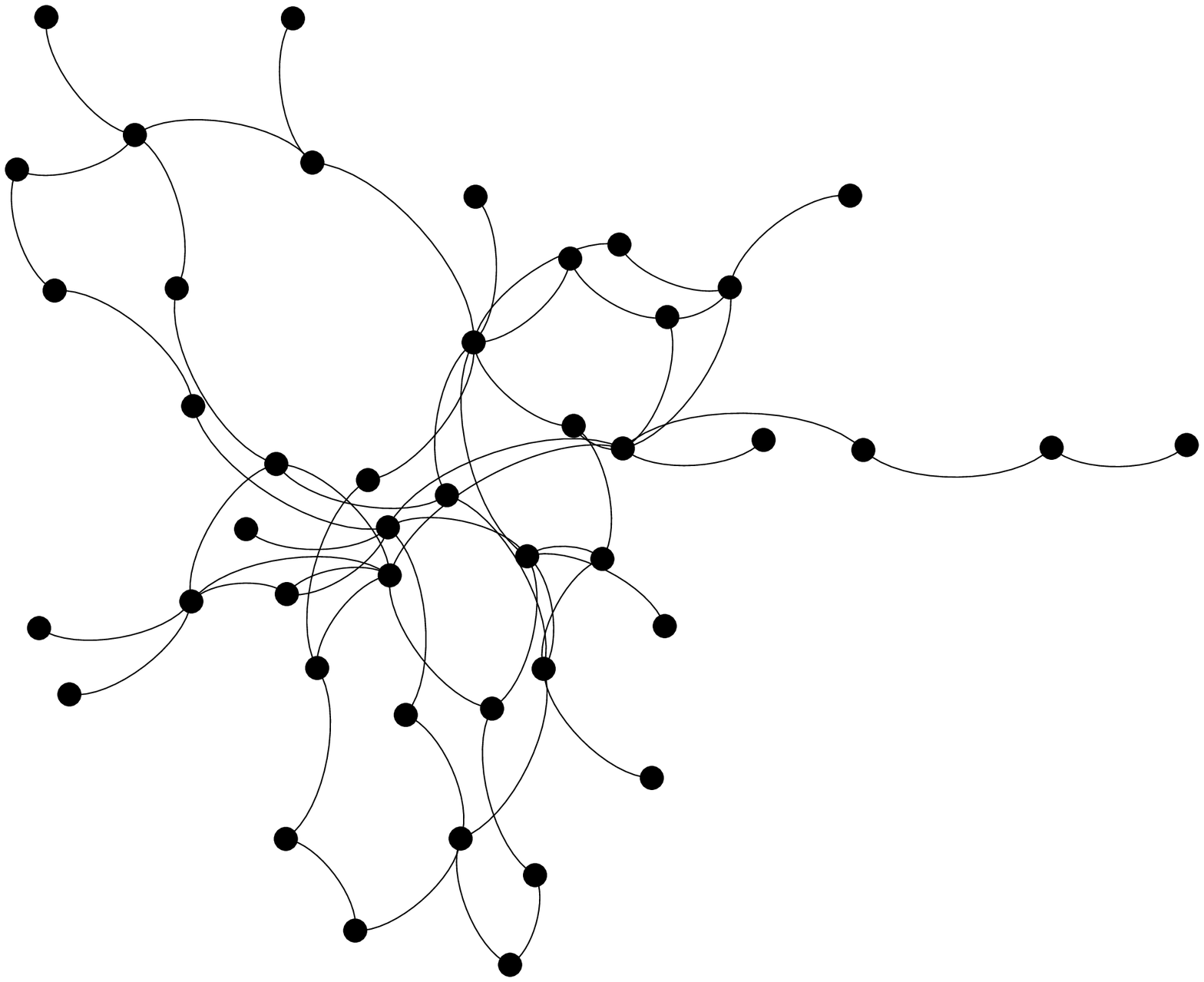}
\caption{Polish 400 kV (CL)}\label{fig:PVizCompCL}
\end{subfigure}
\caption{Visual comparison of graph structure for three different same-voltage subgraphs.}\label{fig:EViz}
\end{figure}

\paragraph{Visualizations}
Finally, we offer visual comparisons of selected same-voltage subgraphs to show that in addition to all of the measures captured, the visual appearance of our randomly generated graphs is similar to that of the real graphs.
This strong similarity in the visualization is consistent with the similar structural properties measured in Section \ref{sec:same_voltage_comp}.
We also include visualizations of corresponding Chung-Lu graphs, which suggest the visualization similarity between our model and the original data is not merely a consequence of having similar degree distributions. 
In all cases, the visualization layouts were done in a similar fashion using the Yifan Hu layout \cite{YifanHu2005} by repelling vertices that are not connected and attracting those that are.
Figures \ref{fig:EVizCompOrig}--\ref{fig:EVizCompCL} depict the Eastern Interconnect 345 kV original, CLC, and CL graphs.
Since these graphs are quite large it's difficult to tell exactly how similar they are.
Next, Figures \ref{fig:TVizCompOrig}-\ref{fig:TVizCompCL} compares smaller systems, the Texas Interconnect 345 kV original, CLC, and CL graphs.
Again, the graphs are fairly large, but there is noticeable similarity in the presence of a few longer cycles, some shorter cycles, and chains hanging off of the edges.
Here, the smaller diameter and average distance of the CL graph relative to the original and CLC graphs are apparent in the visualization. 
Finally, we compare three small graphs, the Polish 400 kV original, CLC, and CL graphs in Figures \ref{fig:PVizCompOrig}-\ref{fig:PVizCompCL}.
In this case, they are small enough to easily compare visually.
Notice the few small cycles on the left of both graphs along with the two long chains going off to the right.
In the original and CLC graphs, these long chains have a few small branches.


\subsection{Transformer subgraphs}
As noted in Section \ref{sec:real_transformer}, the real transformer graphs tend to consist mostly of star graphs, especially when restricted to only one pair of voltage levels.
When all of the transformer edges for a particular network are aggregated, the resulting transformer graph still consists almost entirely of star graphs.
For this reason, we created our model to specifically construct pairwise transformer subgraphs as collections of stars.
Because of this choice of model, and the structure of the transformer subgraphs themselves, it does not make sense to compare the degree distributions, diameter, average distance, or clustering coefficient between the real data and the model.
Degree distributions will always be matched exactly because of how the stars are formed;
in each pairwise transformer subgraph we are guaranteed to have diameter 2 in the model (all stars have diameter 2, the distance from one leaf to another within the same star), average distance is similarly uninteresting;
and clustering coefficient is necessarily zero in star graphs as there are no cycles.
However, this does not mean the Phase 2 graph is structurally identical to the original graph, nor does it mean Phase 2 produces graphs that are structurally identical across different runs of the model.
This is because when we consider the aggregate of the union of all transformer edges in Phase 2, the stars can interact in different ways to create larger structures.
Therefore, the only interesting comparison from the aforementioned list is that between the number of non-star components produced by the model versus observed in the real data.

Table \ref{tab:model_transformer_data} reports the total number of components and non-star components of each size in the transformer subgraph for our three test cases.
We give the values observed in the real data as well as those generated by the model.
In all three example networks, the number of components of a given size are matched closely, always on the same order of magnitude and often within 10\% of the original observed data.
Similarly, the number of non-star components is matched very closely in almost all cases.


\begin{table}
\scriptsize
\begin{center}
\begin{tabular}{|r||r|r||r|r|c|}
\hline
Component & \multicolumn{2}{c||}{\# components} & \multicolumn{2}{c|}{\# non-star} & \\
size      & (Original)    & (Simulated)       & (Original) & (Simulated)           &\\\hline
2         & 81            & 80                & 0          & 0 & \multirow{3}{*}{\rotatebox[origin=c]{270}{Polish}} \\
3         & 38            & 40                & 0          & 0 & \\
4         & 3             &  2                & 2          & 1 & \\ \hline \hline
2         & 341           & 338                & 0         & 0 & \multirow{3}{*}{\rotatebox[origin=c]{270}{Texas}}  \\
3         & 35            & 30                & 1          & 1 & \\
4         & 5             &  8                & 4          & 4 & \\
5         & 2             &  1                & 2          & 2 & \\
7         & 0             &  2                & 0          & 2 & \\ \hline \hline
2         & 1482          & 1467              & 0          & 0 & \multirow{3}{*}{\rotatebox[origin=c]{270}{Eastern}}  \\
3         & 235           & 220               & 6          & 6 & \\
4         & 34            & 48                & 7          & 20 & \\
5         & 16            & 19                & 7          & 10 & \\
6         & 2             &  4                & 0          & 2 & \\
8         & 2             &  1                & 1          & 0 & \\
18        & 1             &  1                & 0          & 0 & \\  \hline
\end{tabular}
\end{center}
\caption{Non-star component counts by size in the transformer subgraphs (Original vs. our simulated Phase 2).}
\label{tab:model_transformer_data}
\end{table}

\subsection{Entire aggregate graph} \label{sec:agg}

Thus far, our models of the same-voltage subgraphs and transformer edges agree well on the test networks.
While our model is explicitly designed to match these vertex and edge-homogenous subgraphs, we now put them together to form the aggregate graph and assess the model's accuracy.
A number of properties, such as degree distribution, number of vertices and edges in the largest component, and local clustering coefficients are closely matched in the aggregate graph largely by virtue of the local matches for each subgraph.
However, while our model is quite accurate in matching diameter and average distance of same-voltage subgraphs, it may be less accurate in matching the {\it heterogeneous diameter} of the aggregate graphs (which is computed from paths which may contain a mixture of transformer and non-transformer edges).
As we will explain, this is because transformer edges can shorten distances between vertices that were previously farther apart in their same-voltage subgraph.
In all cases, we compare our models accuracy against that of the Chung-Lu model applied to the aggregate graph.

In addition to the metrics already discussed, we also consider \emph{resiliency}.
Entire power grids must satisfy $N-1$ security, meaning that if a single power line (edge) or bus (vertex) is removed the grid must still be operational.
With this in mind, we consider the effect of removing a single edge of the aggregate graph.
In particular, for every \emph{non-trivial cut edge} (an edge which, if removed, disconnects more than a single vertex from the rest of the graph) we calculate the size of the largest connected component that remans when it is removed.
We compare the distribution of these sizes for the real data and our model.
We explain the methodology in more detail within that subsection.

\paragraph{Number of vertices \& edges, degree distribution, and local clustering coefficient}

\begin{figure}[t]
\centering
\begin{subfigure}[b]{0.22\textwidth}
\includegraphics[width=\linewidth]{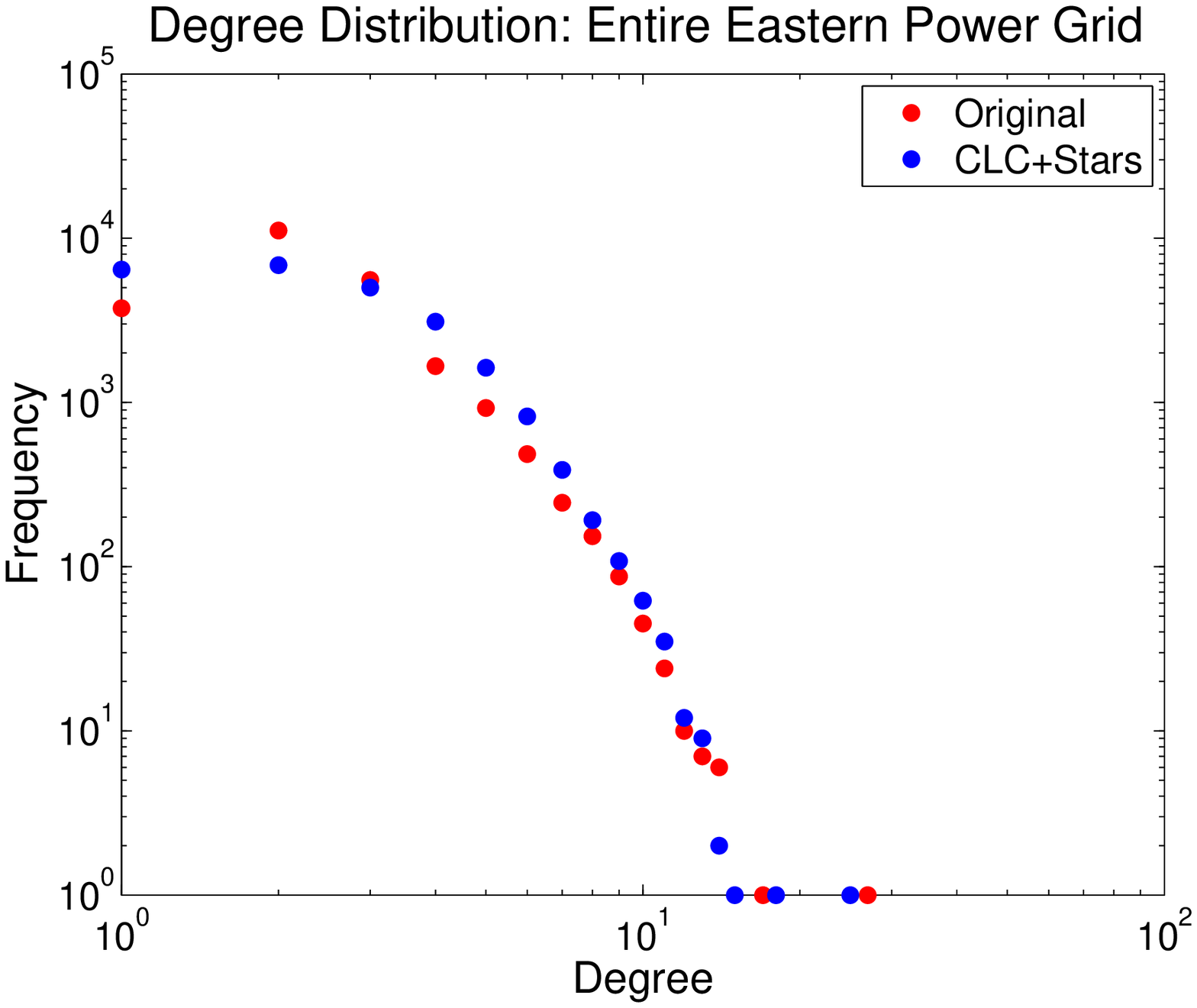}
\caption*{}
\end{subfigure}
\quad
\begin{subfigure}[b]{0.22\textwidth}
\includegraphics[width=\linewidth]{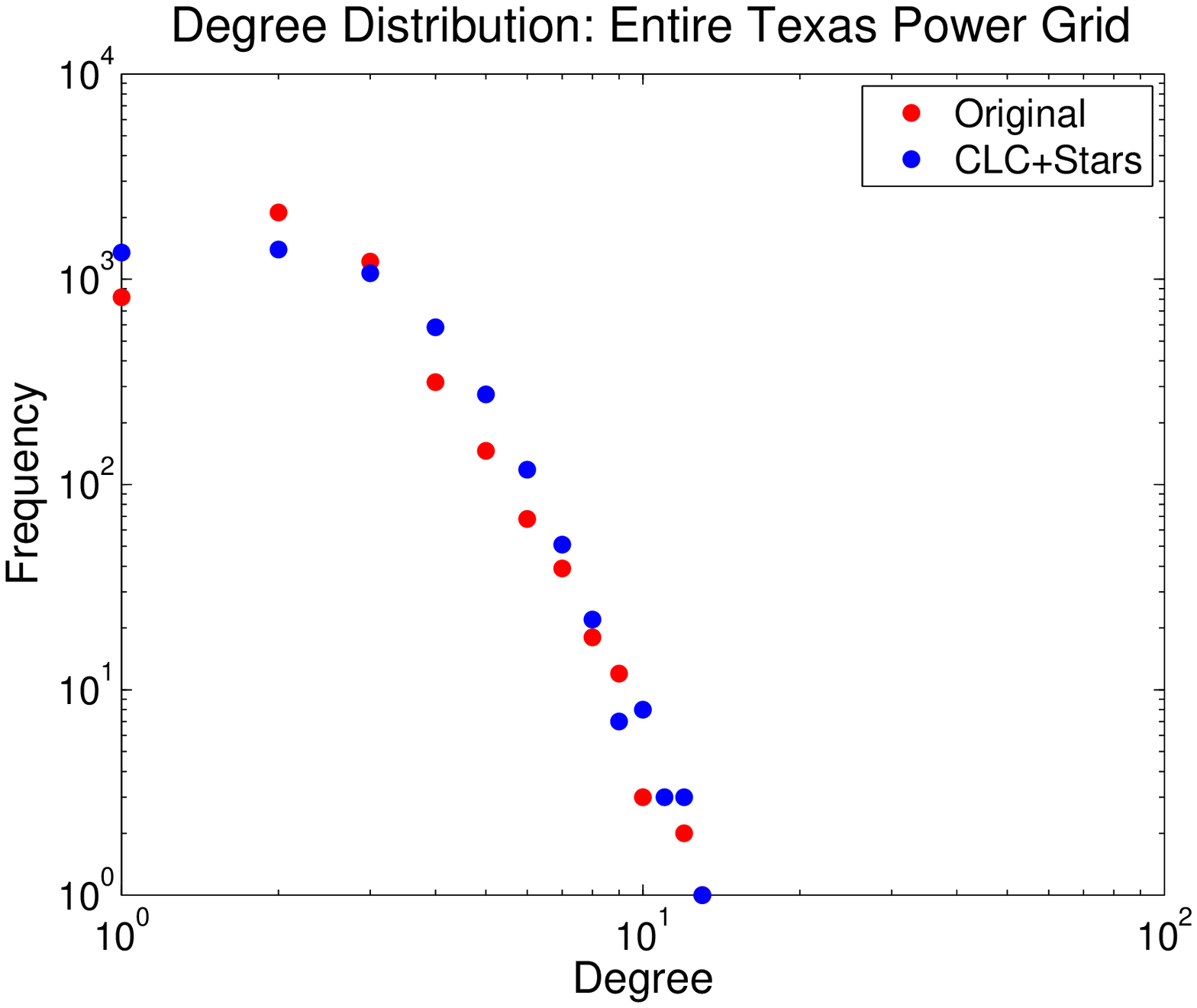}
\caption*{}
\end{subfigure}
\quad
\begin{subfigure}[b]{0.22\textwidth}
\includegraphics[width=\linewidth]{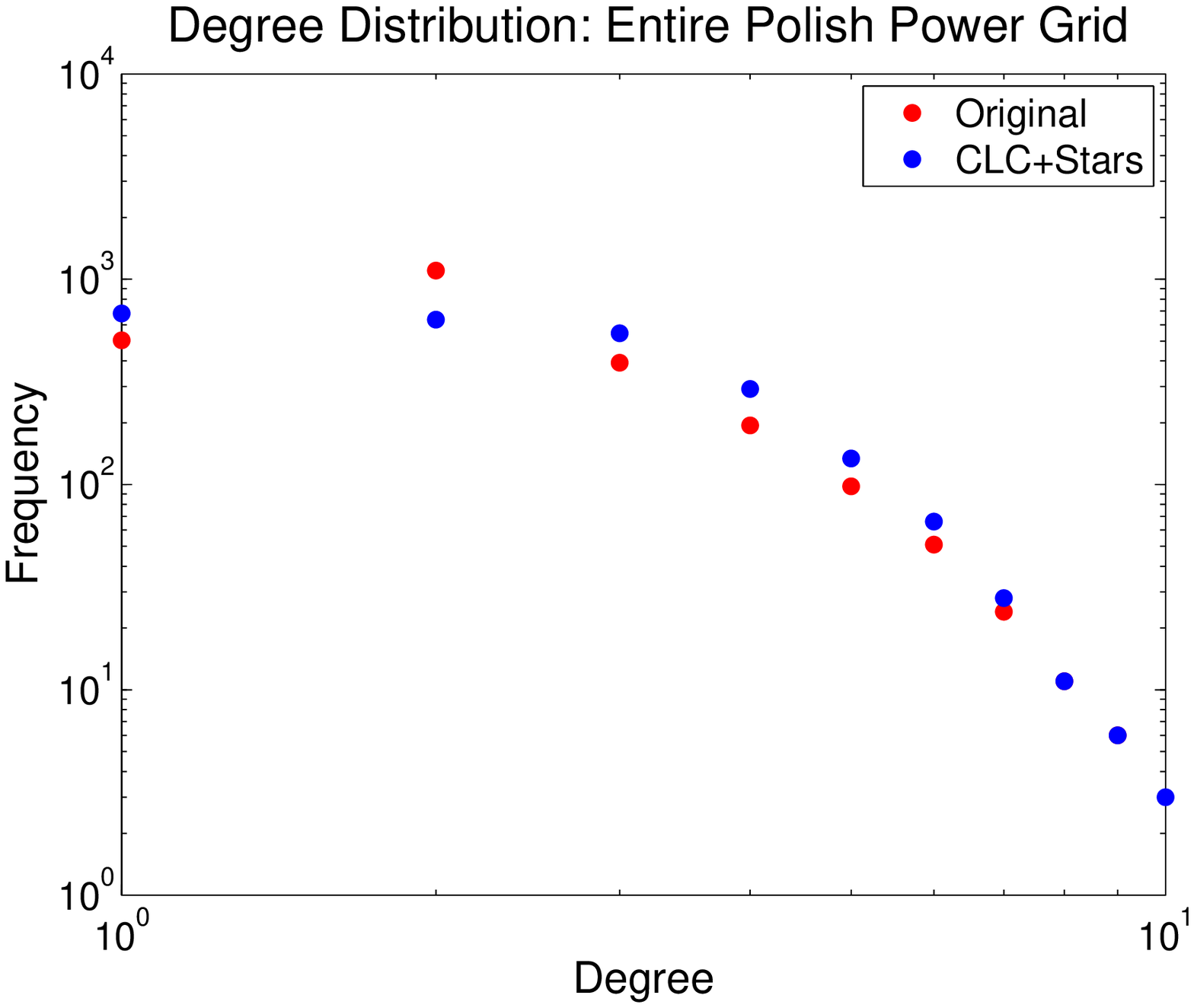}
\caption*{}
\end{subfigure}
\quad
\begin{subfigure}[b]{0.22\textwidth}
\scriptsize
\begin{tabular}{ |p{0.75cm}|p{0.1cm}|p{0.1cm}|}
  \hline
  \multicolumn{3}{|c|}{\bf{DD Comparison Measures}} \\
  \hline
   & $RH$ & $KS$  \\ \hline
Eastern & 0.18 & 0.11  \\ 
Texas & 0.16 & 0.10  \\ 
Polish & 0.14 & 0.11  \\  \hline
\end{tabular}
\vspace{4.7mm}
\caption*{}
\end{subfigure}
\vspace{-4mm}
\caption{Degree distribution, displayed on a log-log scale, for the entire Eastern, Texas, and Polish networks.}
\label{fig:degResultsEntire_all}
\end{figure}

\begin{figure}[h]
\centering
\begin{subfigure}[b]{0.195\textwidth}
\includegraphics[width=\linewidth]{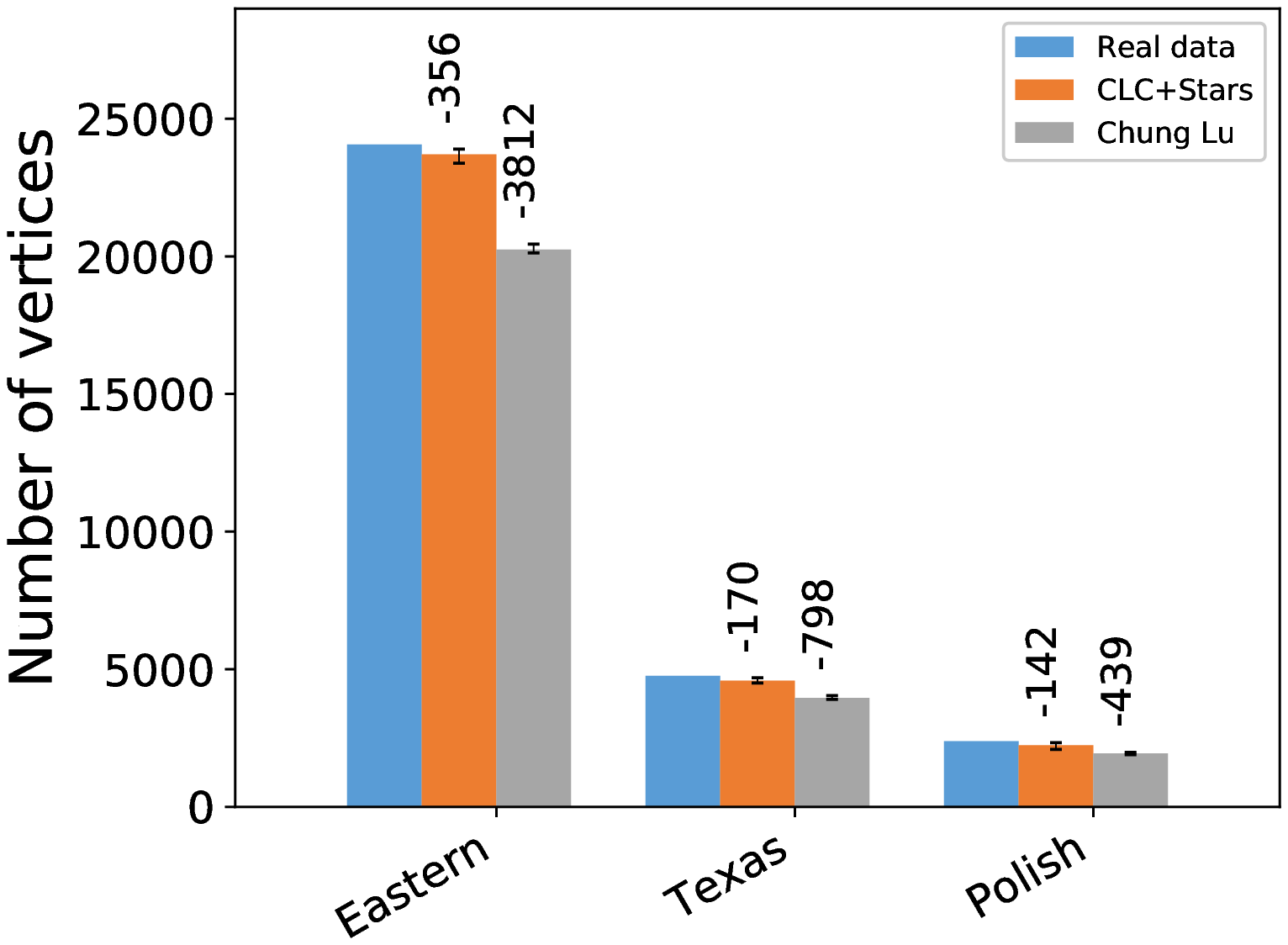}
\caption{}\label{fig:NumVCompareAgg}
\end{subfigure}
\begin{subfigure}[b]{0.195\textwidth}
\includegraphics[width=\linewidth]{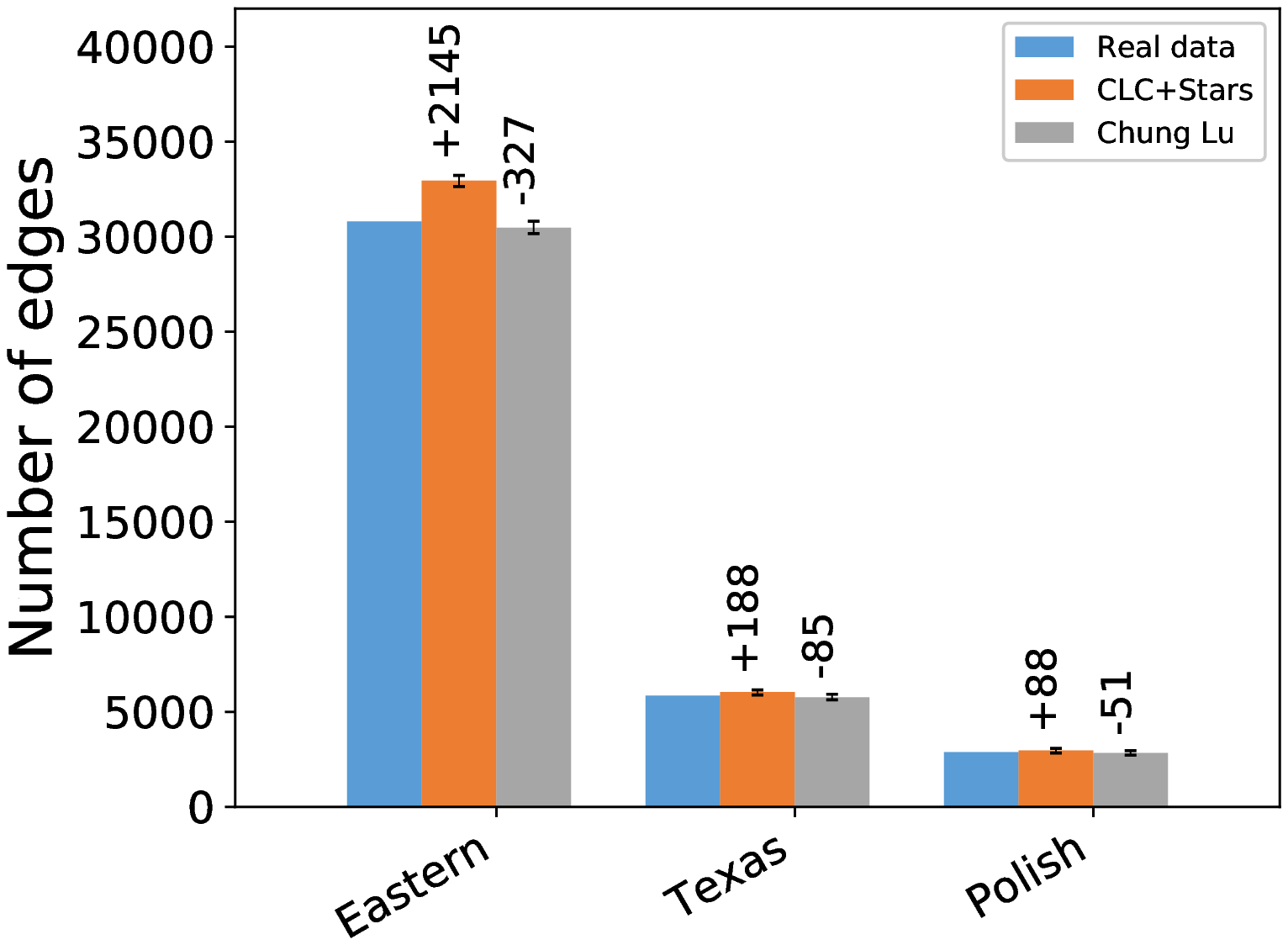}
\caption{}\label{fig:NumECompareAgg}
\end{subfigure}
\begin{subfigure}[b]{0.195\textwidth}
\includegraphics[width=\linewidth]{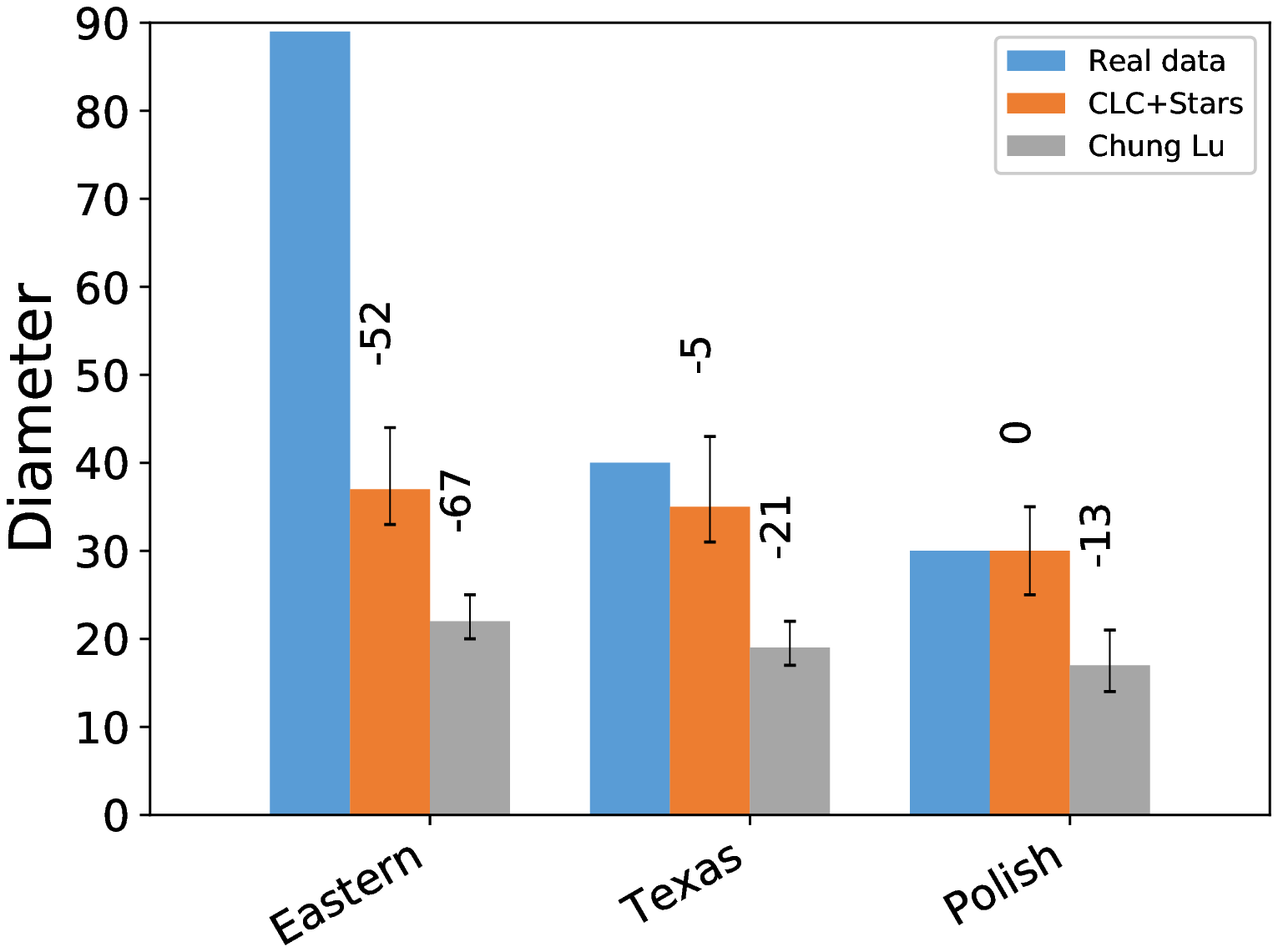}
\caption{}\label{fig:DiamCompareAgg}
\end{subfigure}
\begin{subfigure}[b]{0.195\textwidth}
\includegraphics[width=\linewidth]{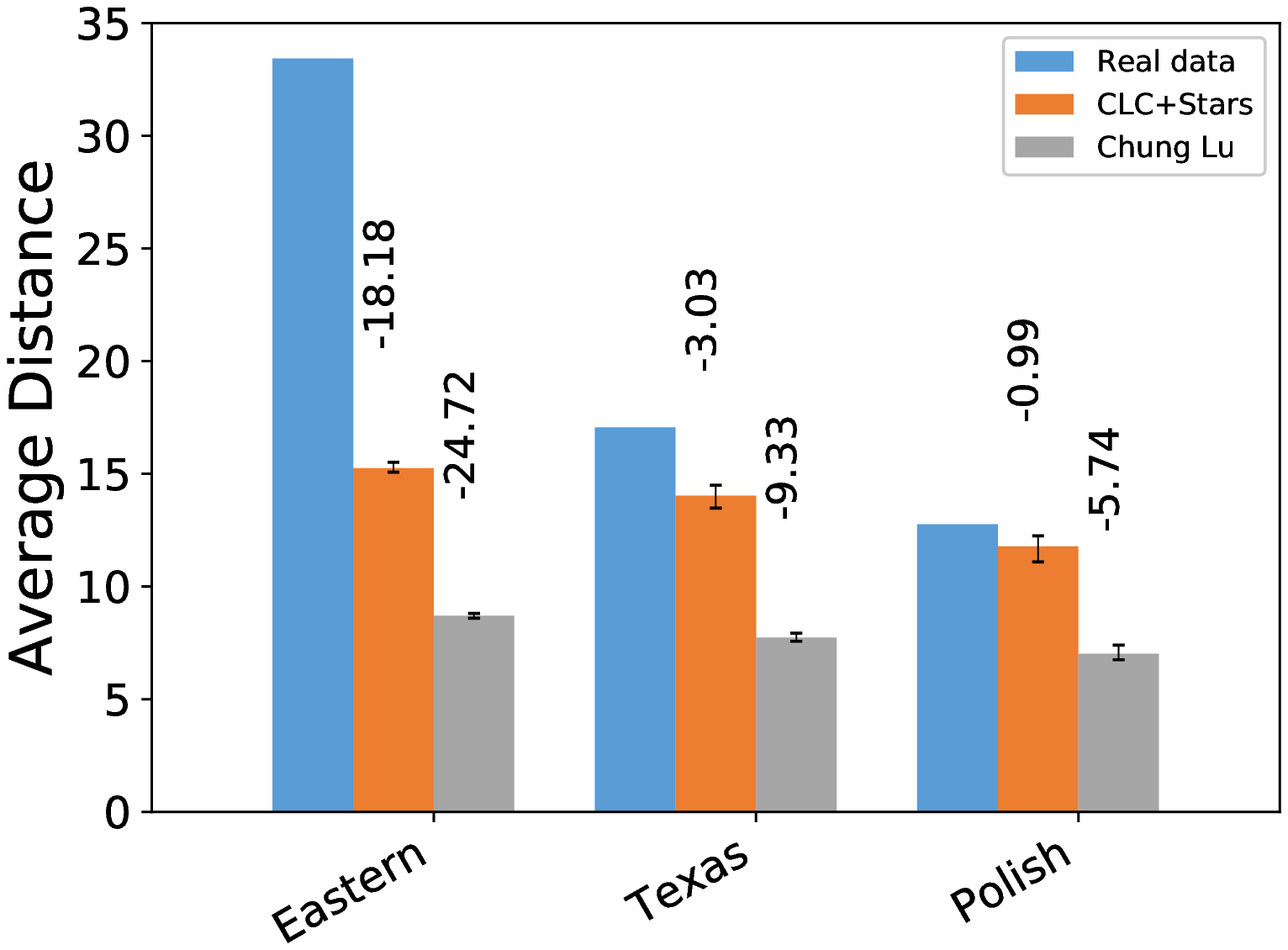}
\caption{}\label{fig:AvDistCompareAgg}
\end{subfigure}
\begin{subfigure}[b]{0.195\textwidth}
\includegraphics[width=\textwidth]{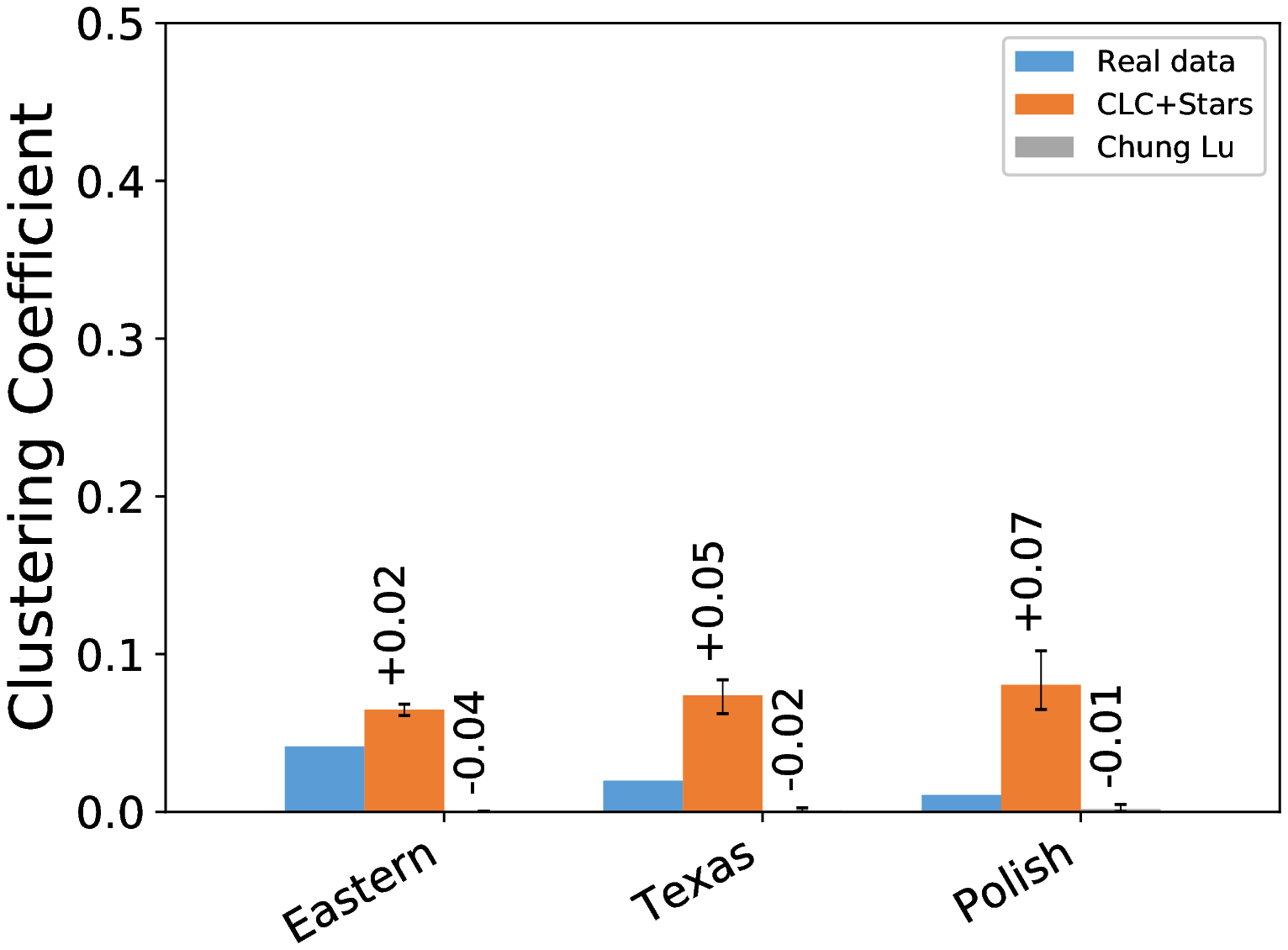}
\caption{}\label{fig:LCCCoompareAgg}
\end{subfigure}
\caption{Bar charts comparing measures in the aggregate graphs.}
\label{fig:aggBars}
\end{figure}

Figures \ref{fig:NumVCompareAgg}--\ref{fig:NumECompareAgg} compare vertex and edge counts for the Eastern, Texas, and Polish networks. Both the CLC+Stars model, as well as the Chung-Lu model, perform comparably in matching the number of vertices and edges in the largest connected component of the aggregate graph. Figure \ref{fig:degResultsEntire_all} compares degree distributions. 
Our CLC+Star model produces degree distributions comparable to the real degree distribution due to the convergence of our local matches:
our same-voltage level CLC model is designed to match the degree within each voltage level and the disjoint star algorithm for the transformer edges matches the transformer degree exactly (under easily met assumptions) or within a slim margin (otherwise).
Figure \ref{fig:LCCCoompareAgg} compares average local clustering coefficients.
As was the case for the same-voltage subgraphs, the average local clustering coefficient for the aggregate graph is low.
We again overestimate the average local clustering coefficient using CLC+Stars, but remain within the same order of magnitude, while Chung-Lu is typically an order of magnitude lower than the real data.


\paragraph{Diameter and average distance}
Figures \ref{fig:DiamCompareAgg}--\ref{fig:AvDistCompareAgg} compare the diameter and average distance in the real data against the CLC+Stars and Chung-Lu models.
Recall that in the same-voltage subgraphs the CLC model achieved more accurate diameter and average distance than the Chung-Lu model.
After stitching those graphs together using the transformer edges, modeled as collections of stars, the accuracy of the CLC+Stars model declines.
While the diameter of the aggregate graph is not a tunable parameter of our model, the diameter results for our model are more accurate than for Chung-Lu.
For the larger network, the Eastern Interconnect, CLC+Stars has the worst performance.
Here, because leaf nodes are chosen randomly in the star-generation process, the insertion of star graphs in Phase 2 created shorter paths between two vertices in a same-voltage subgraph previously farther apart in graph distance.

\paragraph{Resiliency}
Real-world power grids are required to be resilient to certain types of failures.
For example, the loss of a single line or other component which is often referred to as $N-1$ contingency \cite{Kundur2004}. Resilience to this type of failure makes the system $N-1$ secure.
When considering failures of multiple lines or components, this is called $N-x$ contingency, where $x$ is the number of failures allowed.
In the context of a graph, being resilient to $N-1$ contingency means that the removal of any single vertex or edge leaves the graph connected, or, if the graph is disconnected each piece can still function independently (for example, a generator can be found in each component).

\begin{figure}[t]
\centering
\begin{subfigure}[b]{0.24\textwidth}
\includegraphics[width=\linewidth]{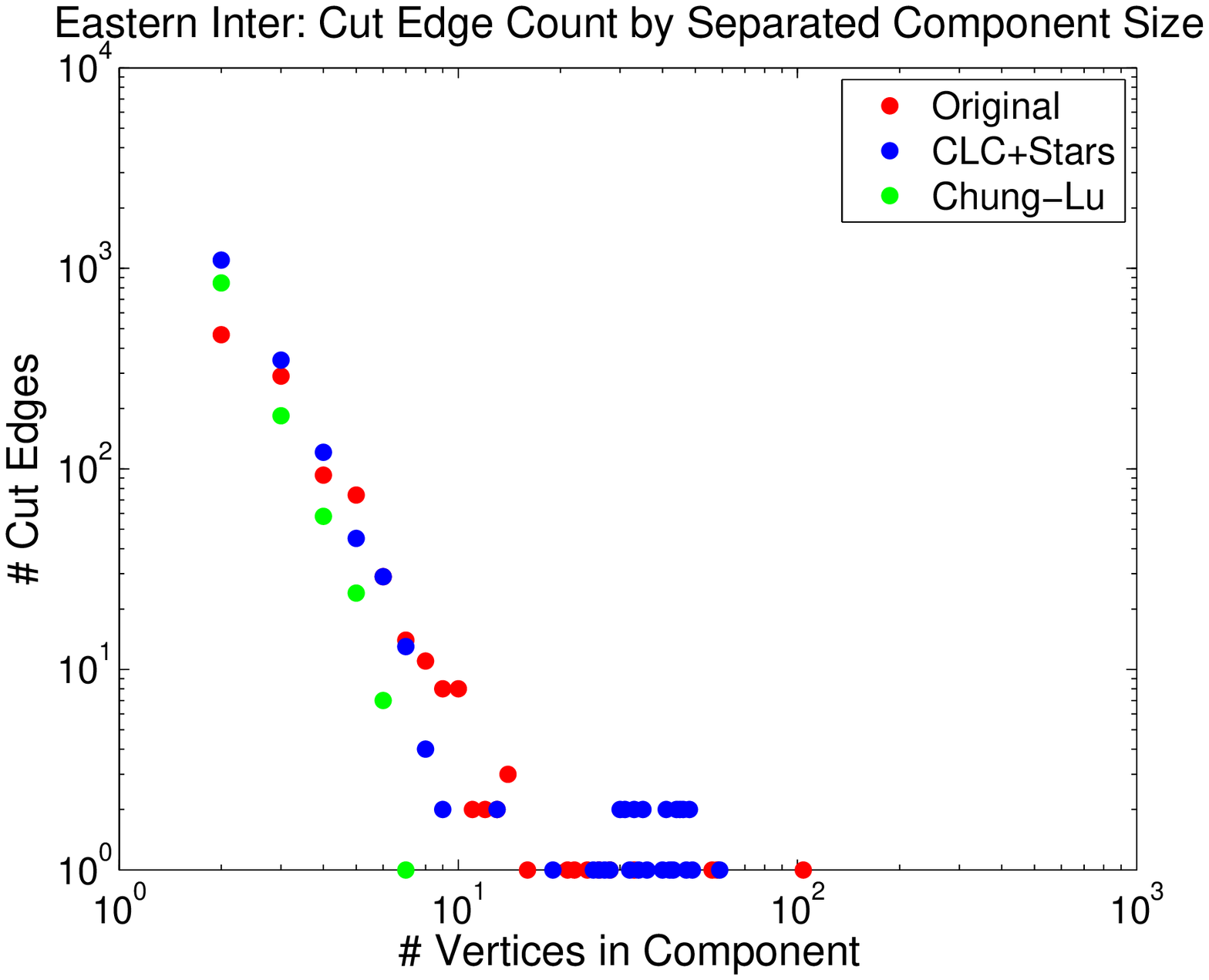}
\caption{}
\label{fig:resiliency_Eastern}
\end{subfigure}
\quad
\begin{subfigure}[b]{0.24\textwidth}
\includegraphics[width=\linewidth]{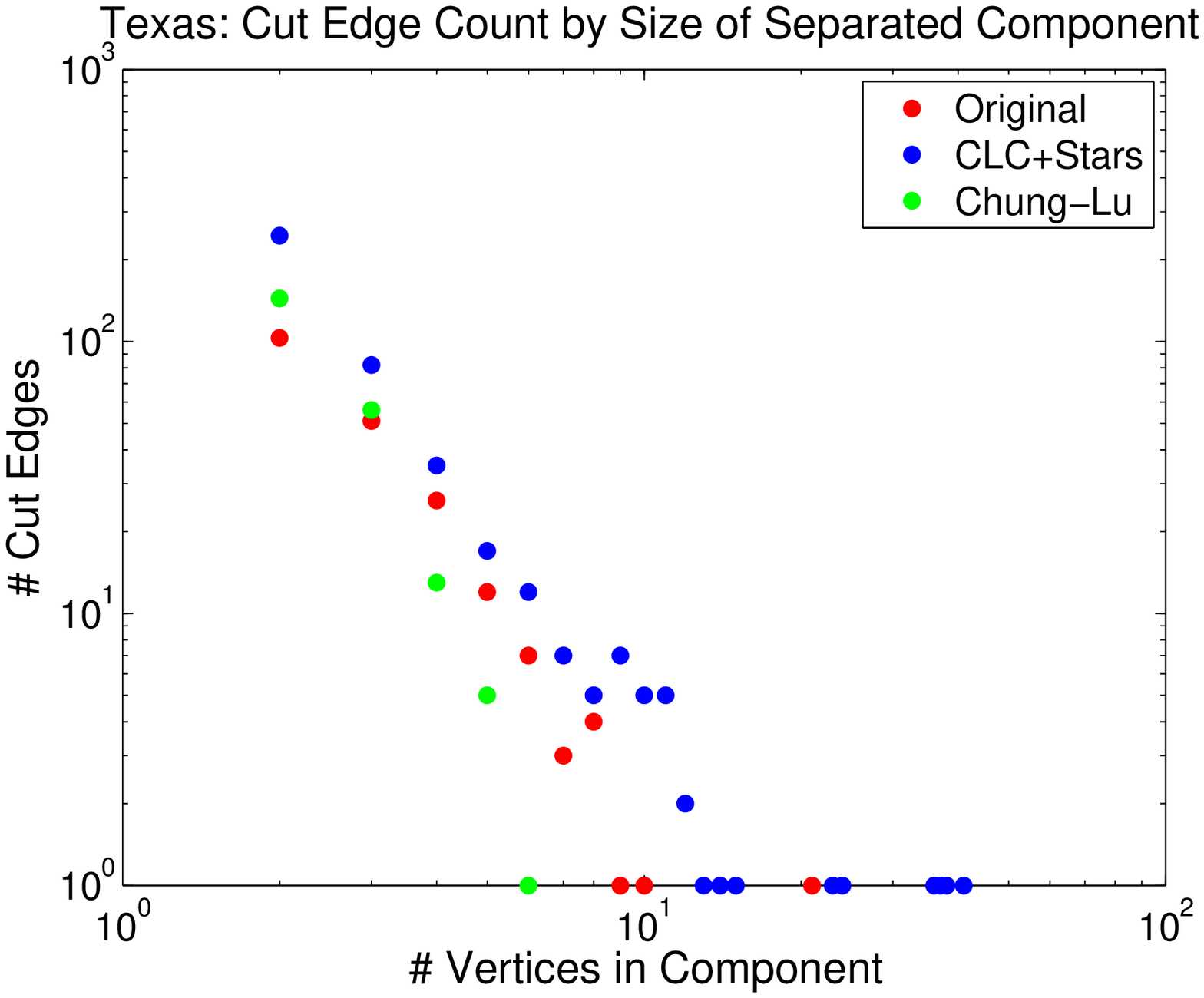}
\caption{}
\label{fig:resiliency_Texas}
\end{subfigure}
\quad
\begin{subfigure}[b]{0.24\textwidth}
\includegraphics[width=\linewidth]{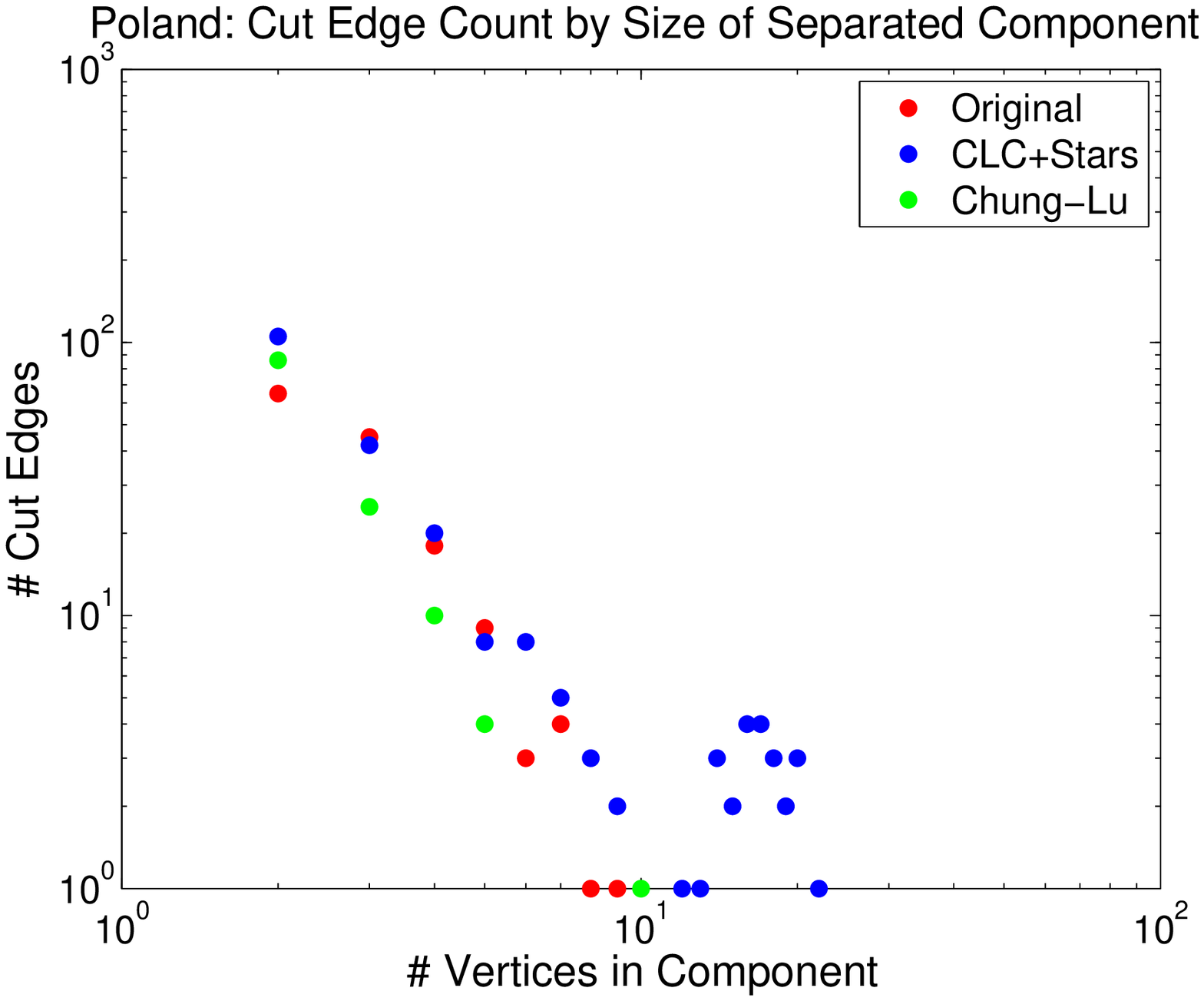}
\caption{}
\label{fig:resiliency_Poland}
\end{subfigure}
\caption{Comparison of the result of removing single cut edges from the Eastern, Texas, and Polish networks.}
\label{fig:cutEdge}
\end{figure}

We test resiliency to the loss of single edges in our CLC+Stars model as well as the Chung-Lu model and compare both to the real data.
Our data lack labels on the vertices to identify generators, so we focus more generally on the sizes of the smallest connected component remaining when a single edge is removed.
We consider \emph{nontrivial cut edges}, those edges whose removal disconnects more than one vertex.
In all networks, both real and synthetic, there are very few such edges -- between 3-5\% in the real data and Chung-Lu model, and 5-7\% in the CLC+Stars model.
Figures \ref{fig:resiliency_Eastern}--\ref{fig:resiliency_Poland} compare sizes of the smallest connected component remaining after removal of single cut edges.
These are frequency plots, meaning that a point at $(x, y)$ indicates that there were $y$ cut-edges for which the smallest connected component remaining when the edge is removed had $x$ vertices.
For example, the blue point in Figure \ref{fig:resiliency_Eastern} at (2, 1000) means that there were 1000 cut edges in the CLC+Stars model of the Eastern Interconnect for which the smallest remaining connected component had only 2 vertices.

In the Eastern and Texas networks, the CLC+Stars model catches the few cut edges that leave a rather large (in comparison) portion of the graph disconnected,  while overestimating these for the Polish network. 
This is likely due to our random stars model for the transformer edges.
If there is a same-voltage connected component that only has one transformer edge linking it to the rest of the network then it will be disconnected when that transformer is cut.
In contrast, the Chung-Lu model tends to underestimate the number of cut edges in general. 
{Measurements of graph resiliency need not be limited to the case of single edge failures. Interested readers should refer to Appendix \ref{apen:eig}, where we consider one such additional measure called the normalized Laplacian spectral gap, a quantity approximating a graphs ``bottleneckeness" or sparsest cut.}


\paragraph{Entire graph visualizations}

The large number of vertices in the entire power grid graphs (e.g., the Eastern Interconnection has 24K vertices and 31K edges) {pose challenges for visualization}. Since depicting every vertex and edge would result in a cluttered and potentially uninformative visualization, it is advantageous to ``reduce" the data while preserving some underlying structure. We present one such compact visualization of the aggregate graph, which we call ``interconnection graphs". {We emphasize these graphs are distinct to, but derived from, the aggregate graphs  studied in Section \ref{sec:agg}}. 

Interconnection graphs represent how transformer edges link connected components from different same-voltage subgraphs.
In an interconnection graph, each connected component within a same-voltage subgraph is collapsed to a single vertex and two such vertices are connected by an edge if there exists a transformer edge linking any vertex in one component to any vertex in the other component.
More formally, if $C_1,\dots,C_k$ denote the connected components of the same-voltage subgraphs of an entire power grid graph $G$ with set of transformer edges $T$, then its interconnection graph $H$ has vertex set $V(H)=\{1, \dots, k\}$ and edge set $E(H)=\{\{i,j\}: \exists \{u,v\} \in T \mbox{ with } u\in C_i, \ v \in C_j\}$.
The weight of a vertex $i\in H$ is the number of vertices in $C_i$ in $G$ and the weight of a edge $\{i,j\}$ is the number of transformer edges between $C_i$ and $C_j$ in $G$.
{In the language of graph theory, interconnection graphs are weighted, attributed {\it quotient graphs} (under the equivalence relation of belonging to the same connected component within a same-voltage subgraph) with attributes given by voltage levels, and additional vertex and edge weightings based on multiplicities.}
With the exception of the edge and vertex weights, an identical construct was considered in \cite{Halappanavar2015}.

\begin{figure}[h!]
\centering
\begin{subfigure}[b]{0.75\textwidth}
\includegraphics[width=\linewidth]{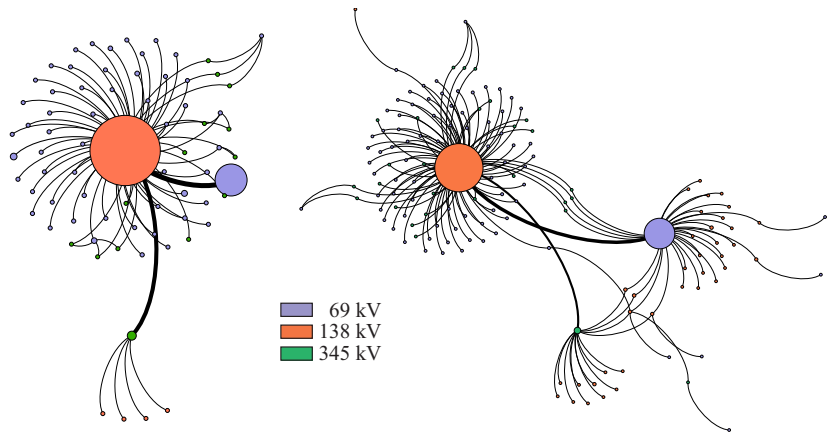}
\caption{Texas real data (left) and CLC+Stars model (right)}
\end{subfigure}
\\
\begin{subfigure}[b]{0.75\textwidth}
\centering
\includegraphics[width=\linewidth]{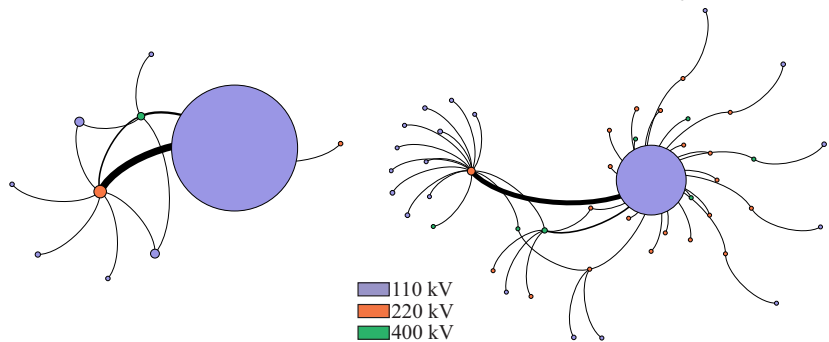}
\caption{Polish real data (left) and CLC+Stars model (right)}
\end{subfigure}
\\
\begin{subfigure}[b]{0.75\textwidth}
\centering
\includegraphics[width=\linewidth]{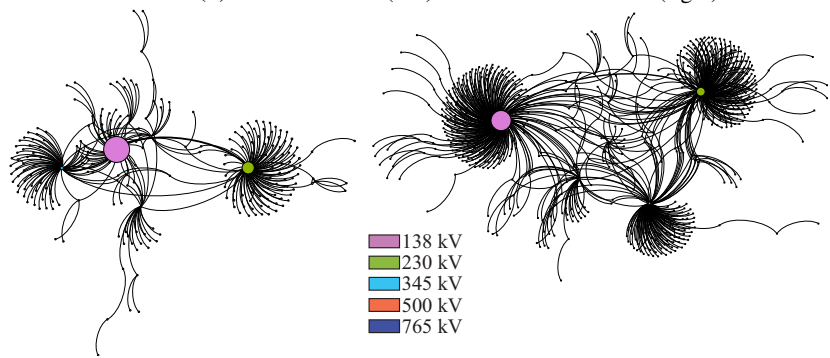}
\caption{Eastern real data (left) and CLC+Stars model (right)}
\end{subfigure}
\caption{Comparison of interconnection graphs for each of the three test networks.}\label{fig:interconnectionVizs}
\end{figure}

Figure \ref{fig:interconnectionVizs} compares visualizations of the interconnection graph derived from the original entire power grid graphs against those derived from our model (i.e. the output of Algorithm \ref{alg:entire}).
The color of a vertex indicates the voltage level of the nodes in its corresponding connected component, while the size of a vertex and thickness of an edge are proportional to their weight as defined above.
Given the relatively small nature of the visualizations, a number of qualitative similarities are readily apparent: first, the fact that nearly every edge connects to the largest vertex of each color type reflects that nearly every transformer edge is incident to the largest component of some voltage level.
In other words, transformer edges between two components are rare when neither component is the largest within its respective same-voltage subgraph.
Furthermore, we also observe that the thickest edges occur between the largest vertices of each color type.
This reflects that the multiplicity of transformer edges between components is greatest when those components are the largest connected components.



%% file: gpg-synthetic.tex
\section{{Guidelines for synthetic generation of model inputs}} \label{sec:synthetic}

{The inputs to our generative model (desired same-voltage subgraph degrees and diameter for Phase 1, and desired transformers degrees for Phase 2) may be easily extracted from given power grid data. In this way, one can fit the model to existing power grid graphs and create ensembles of structurally similar graphs, as we did in Section \ref{sec:results}. 
However, in cases where these inputs are completely unavailable or when users want to scale or vary the inputs, it is desirable to have a framework for artificial input generation. Below, we suggest a simple framework for generating model inputs synthetically, requiring users only specify the number of vertices in each same-voltage subgraph. We then briefly test the model's performance on these synthetic inputs.}

{While the guidelines included in this work are partially informed by our data and the complex networks literature, they are by no means definitive. Given the limited amount of data available, our intention is not to argue rigorously that the approximations used are statistically significant, but to illustrate our model's performance under inputs synthetically generated with these guidelines is comparable to that under inputs extracted from real data. Accordingly, the exact constants we suggest below are less important than the methods used for synthetic generation. Furthermore, that such generalized guidelines provide our model with sufficiently accurate approximations to begin with attests to the network-of-networks approach: if there weren't consistent and clear structural commonalities in the same-voltage subgraphs and transformer subgraphs, the low-complexity guidelines we describe would be ineffective.}

{Ultimately, users are free to decide which assumptions are most appropriate for their purposes. Part of the utility of generative models is the ability to tweak and experiment with model inputs to produce graphs with certain properties that may intentionally deviate from those typical of the network being modeled. In this regard, degree distributions are a flexible model input as there are a variety of efficient methods for fitting and artificially generating differently shaped degree distributions \cite{clauset2009power}. While we construct the guidelines below by sometimes fitting functions to graph parameters extracted from {\it all} our datasets, for brevity we illustrate the model's performance on our largest, the Eastern Interconnection. }

\subsection{{Synthetic Generation of Phase 1 Inputs}} \label{sec:synP1}
\begin{figure}[h!]
\centering
\begin{subfigure}[b]{0.23\textwidth}
\includegraphics[width=\linewidth]{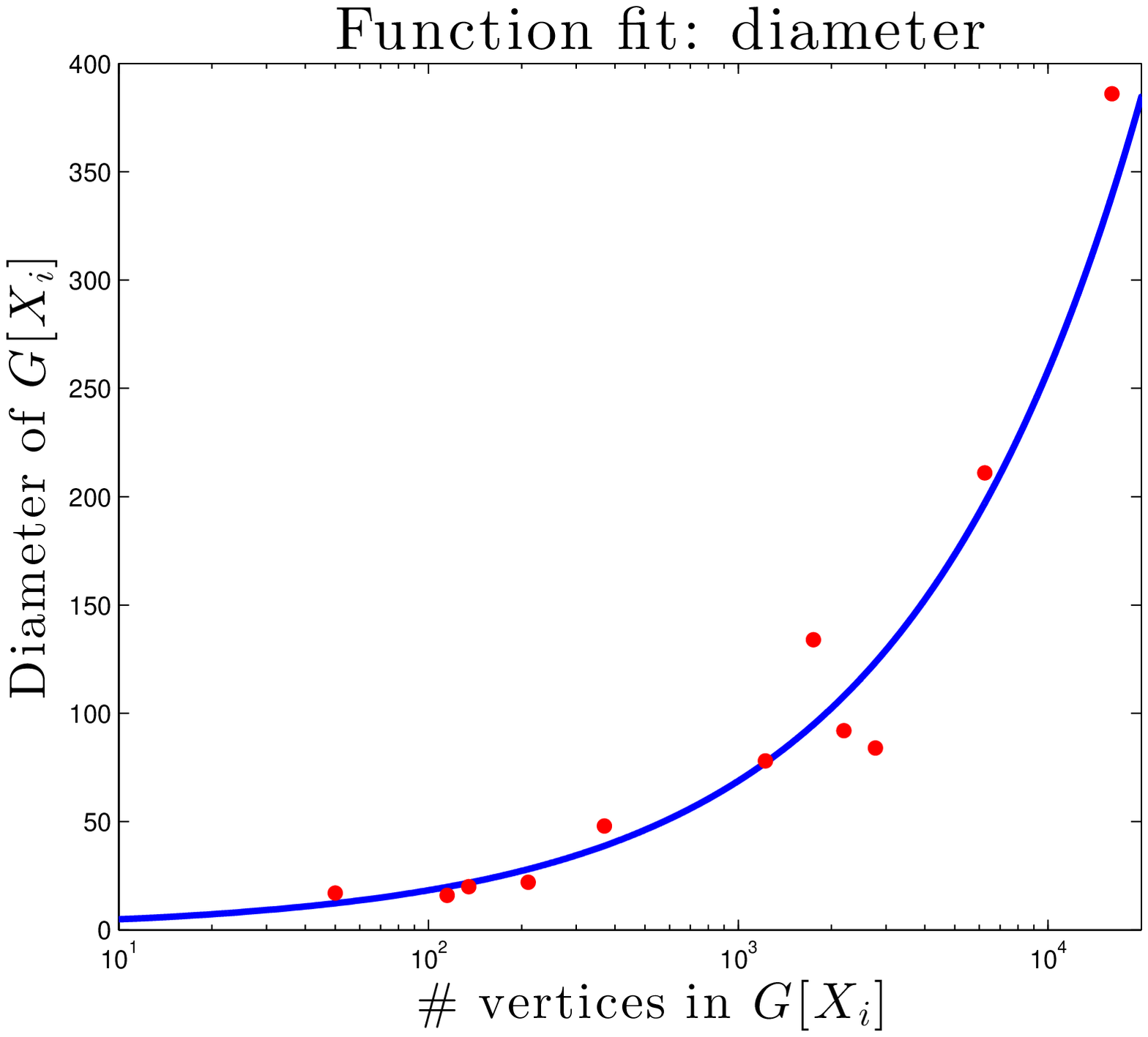}
\caption{} \label{fig:diamFit}
\end{subfigure}
\qquad
\begin{subfigure}[b]{0.23\textwidth}
\includegraphics[width=\linewidth]{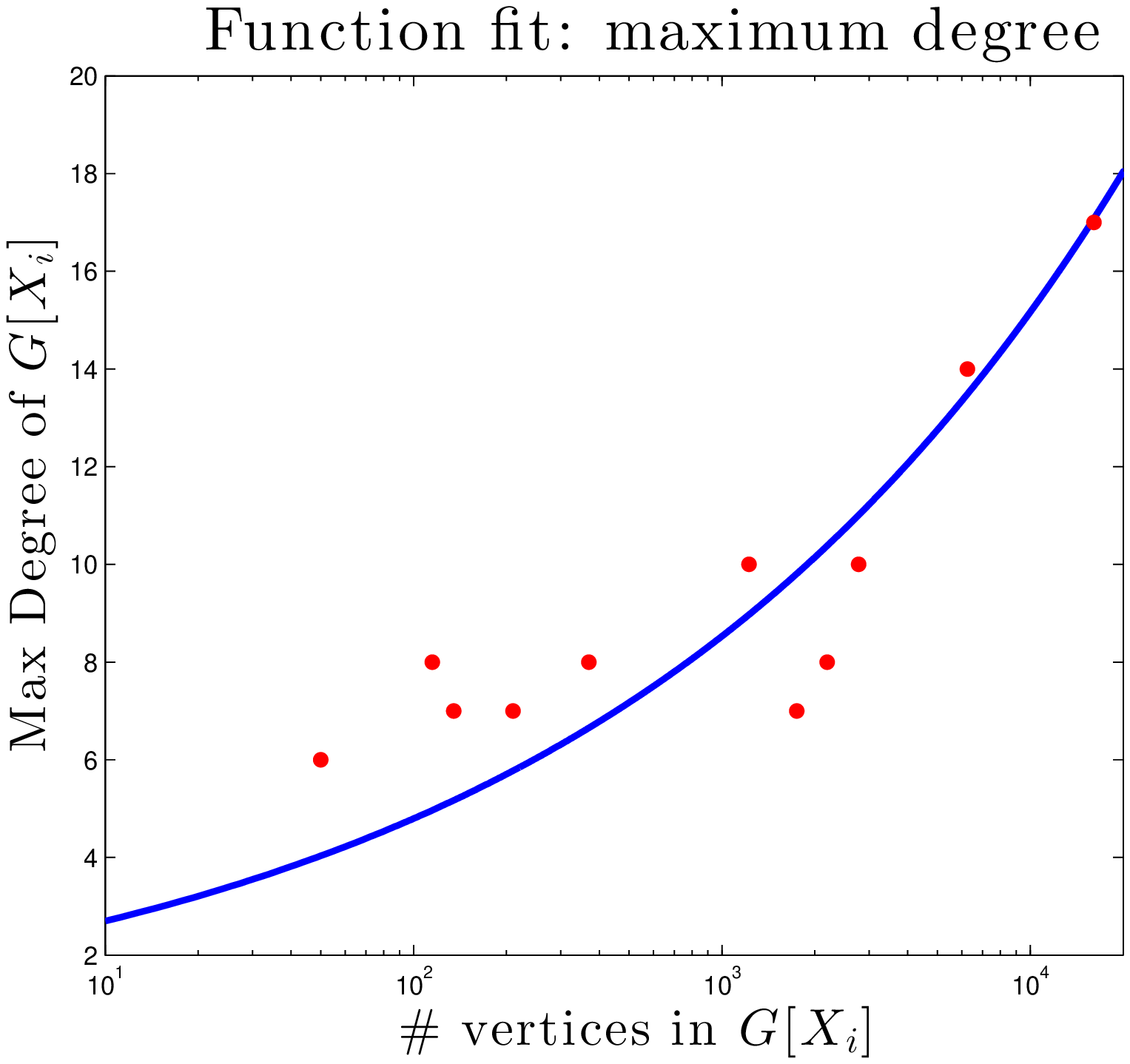}
\caption{} \label{fig:maxFit}
\end{subfigure}
\qquad
\begin{subfigure}[b]{0.25\textwidth}
\includegraphics[width=\linewidth]{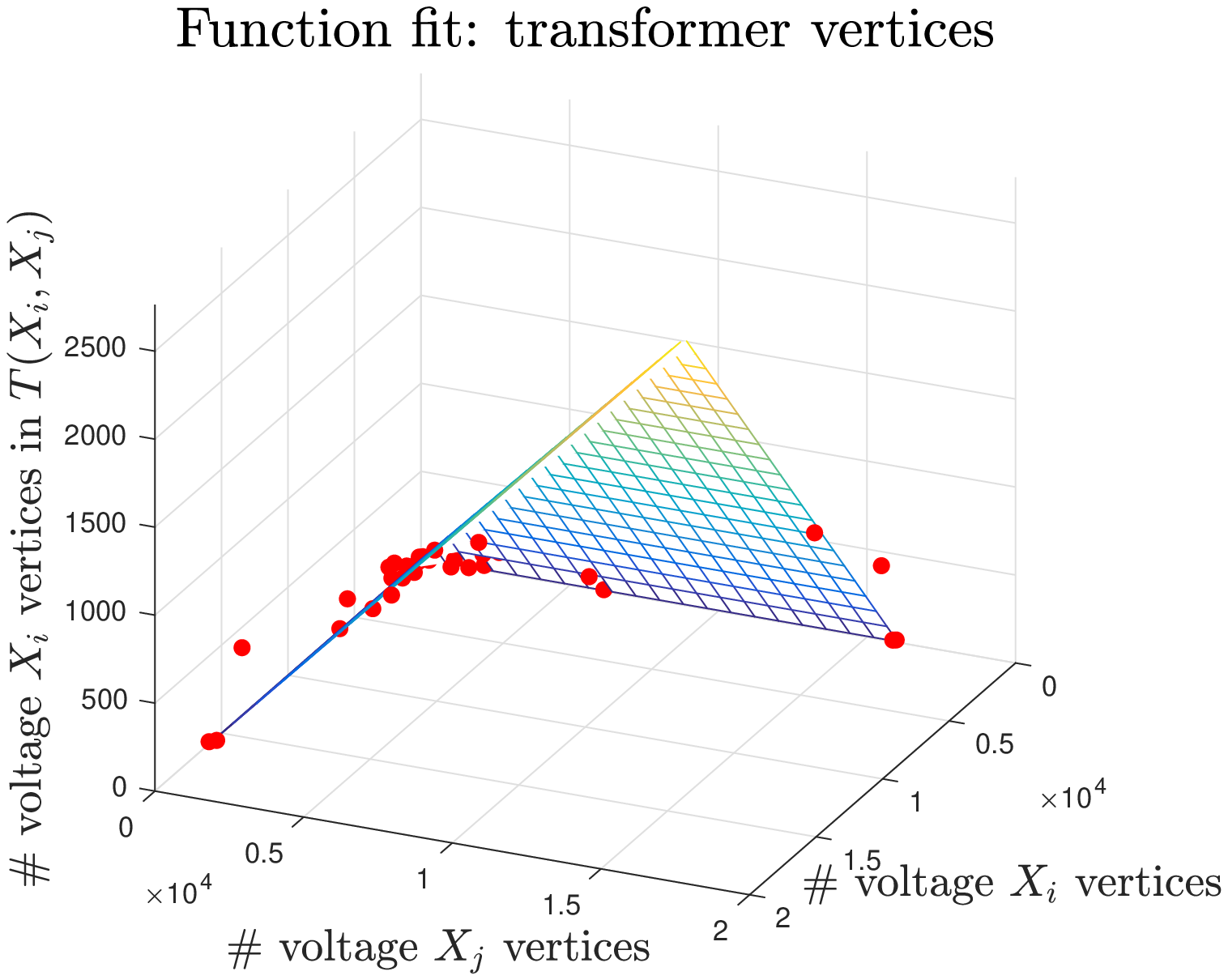}
\caption{} \label{fig:funFitFixed}
\end{subfigure}
\caption{Functions used to guide synthetic input generation vs. the data.}\label{fig:fitFun}
\end{figure}

\paragraph{Diameter, $\delta$} {We suggest the diameter of same-voltage subgraphs is on the order of $\sqrt{n}$, which Young et~al. \cite{stephen} also independently suggest in recent work. Indeed, for our data, the optimal mean-squared error minimizing function of the form
\[
f(n)=c\cdot n^{k},
\]
where $n$ is the number of vertices in the same-voltage subgraph and $f(n)$ is diameter yields $c\approx1.301$ and $k\approx 0.574$. Figure \ref{fig:diamFit} plots the fit.}

\paragraph{Same-voltage degree sequence, $\degseq^X$}
{To synthetically generate the Phase 1 degree sequences, we suggest a generalized log-normal degree distribution, where the number of degree $d$ vertices, satisfies
\[
n_d \propto \exp\left(-\left(\frac{\log{d}}{\alpha} \right)^\beta \right),
\]
for some parameters $\alpha,\beta$. As described in \cite{Kolda2014} and implemented in the supplemental graph generation software package \cite{feastpack}, one may conduct a parameter search to locate the optimal such $\alpha$ and $\beta$ given user-specified target values for average degree and maximum degree, denoted $\bar{d}$ and $d_{\max}$, respectively. Since average degree is twice the ratio of number of edges to vertices, one's choice of $\bar{d}$ as a function of $n$ reflects an assumption of how a graph's {\it edge density} varies (see \cite{leskovec2007graph} for more on this topic). In our data, we see no compelling trends of densification. Average degree is consistent across our 11 same voltage subgraphs, with mean $\bar{d}=2.425$, standard deviation $0.1846$ and coefficient of variation $7.6\%$. We suggest this value for $\bar{d}$ when generating log-normal degree distributions. As for maximum degree, research suggests that $d_{\max}$ is difficult to generalize as a function of $n$. For the case of networks with idealized power-law degree distributions, some work suggests $d_{\max} \sim n^{1/\gamma}$, where $\gamma$ depends on the power-law exponent, and may range between $1$ to $5$ (see \cite{newman2003structure}, which contains a survey of these results). Although the same-voltage subgraph degree distributions stray from idealized power-law degree distributions, we nonetheless find that $\gamma \approx 4$ seems to provide a good fit to our data. The optimal mean-squared error minimizing function of the form $g(n)= c\cdot n^{1/4}$ yields $c\approx 1.517$. Figure \ref{fig:maxFit} plots the fit. }

\begin{figure}[h!]
\centering
\includegraphics[scale=0.14]{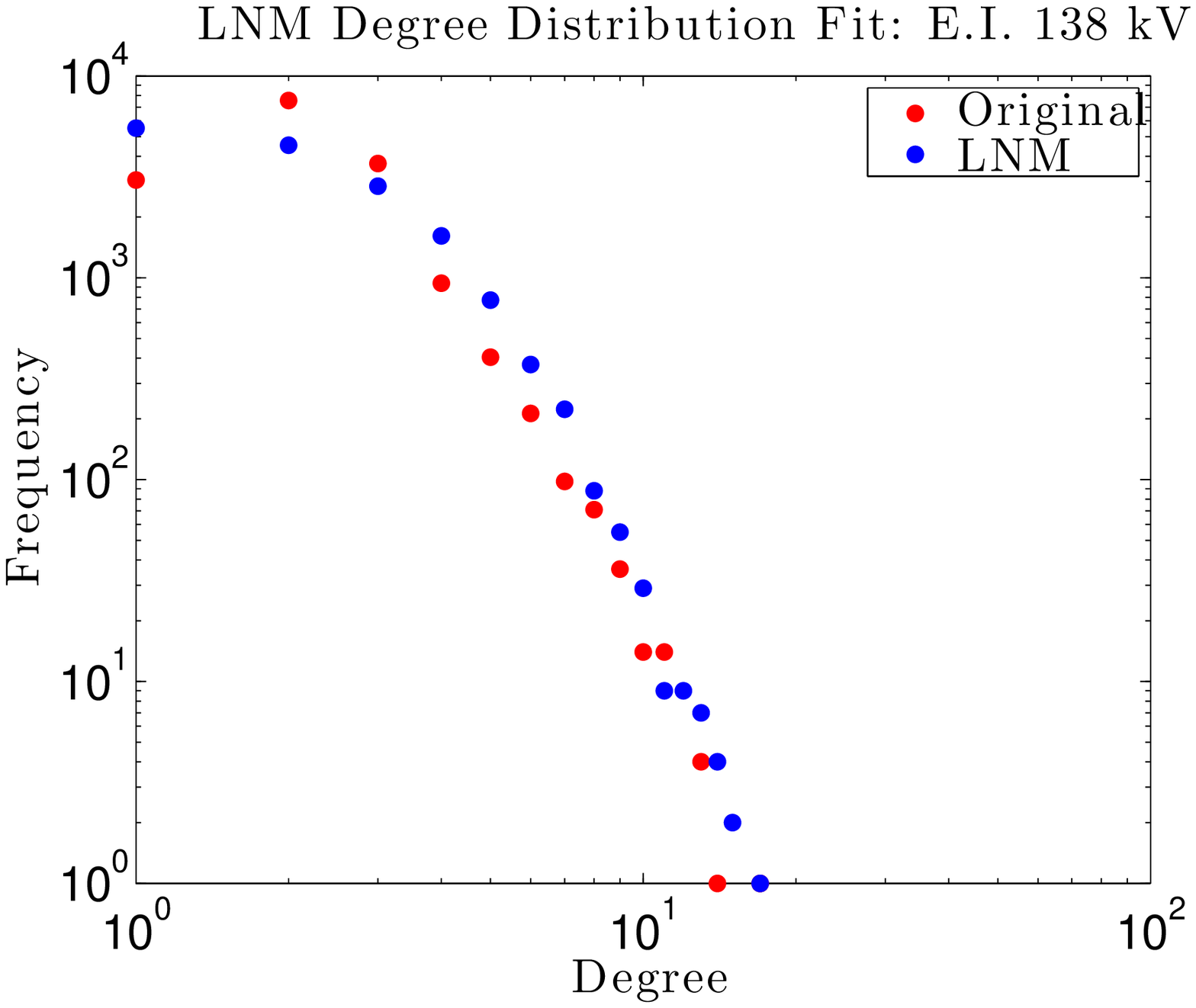}
\includegraphics[scale=0.14]{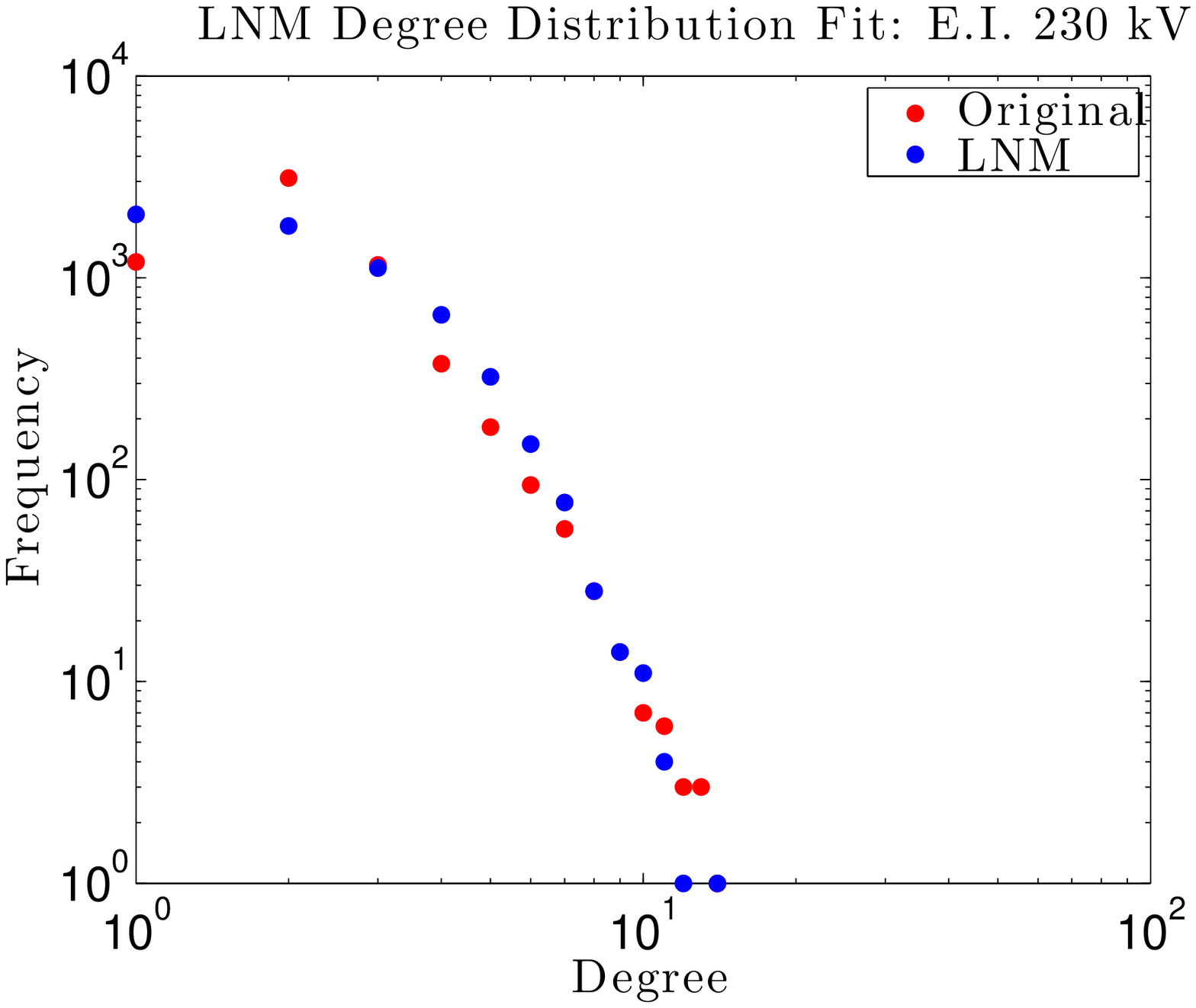}
\includegraphics[scale=0.14]{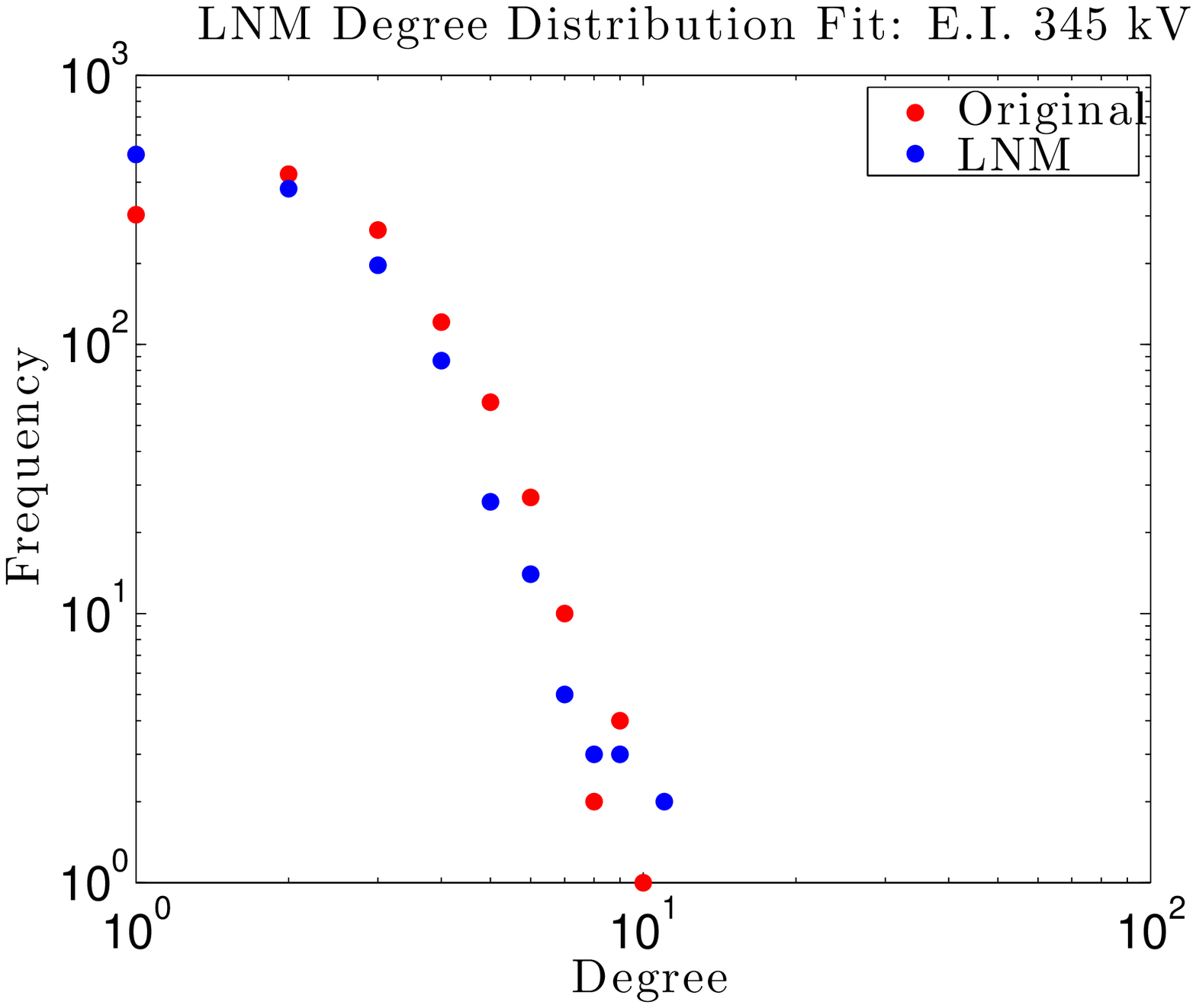}
\includegraphics[scale=0.14]{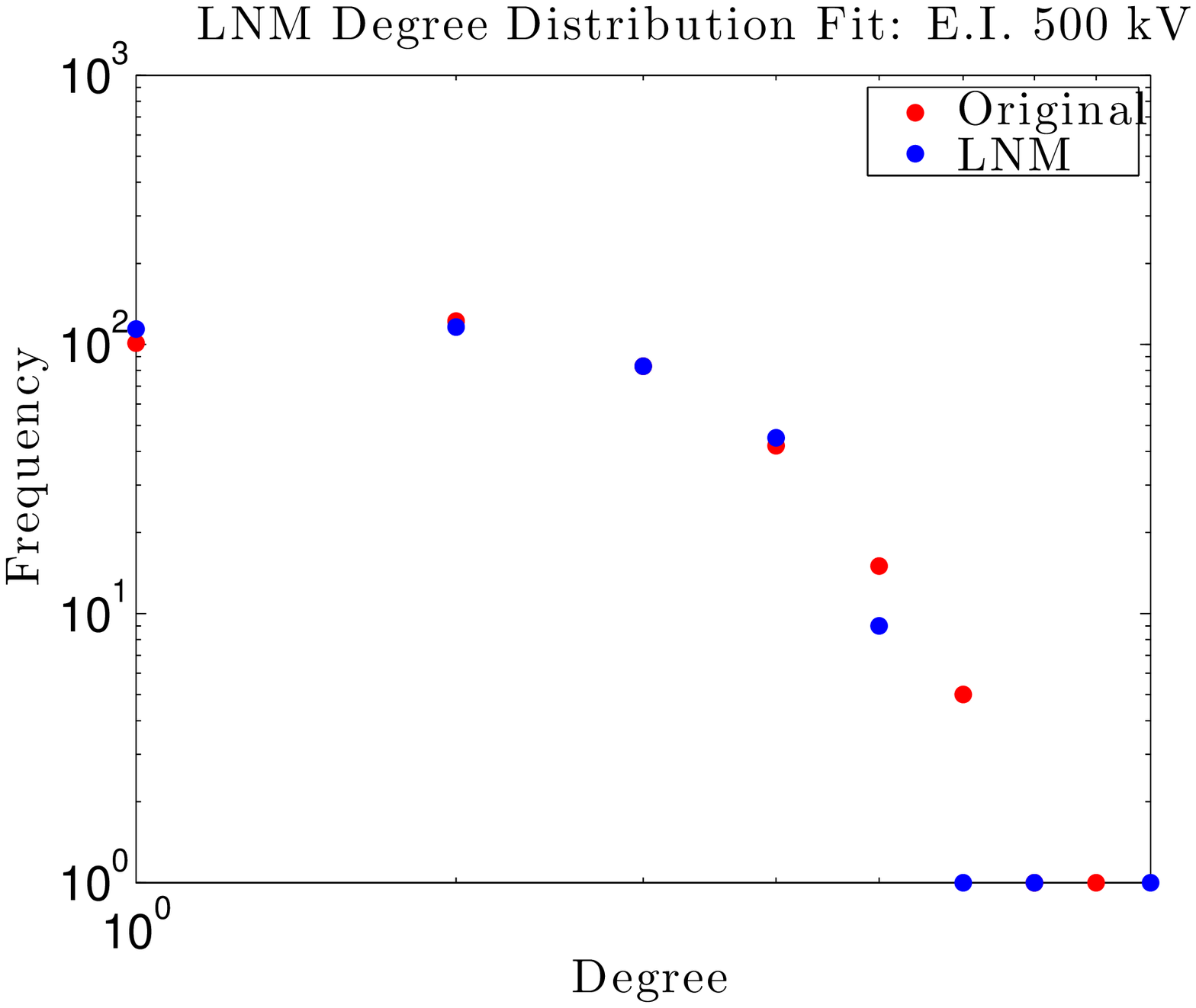}
\includegraphics[scale=0.14]{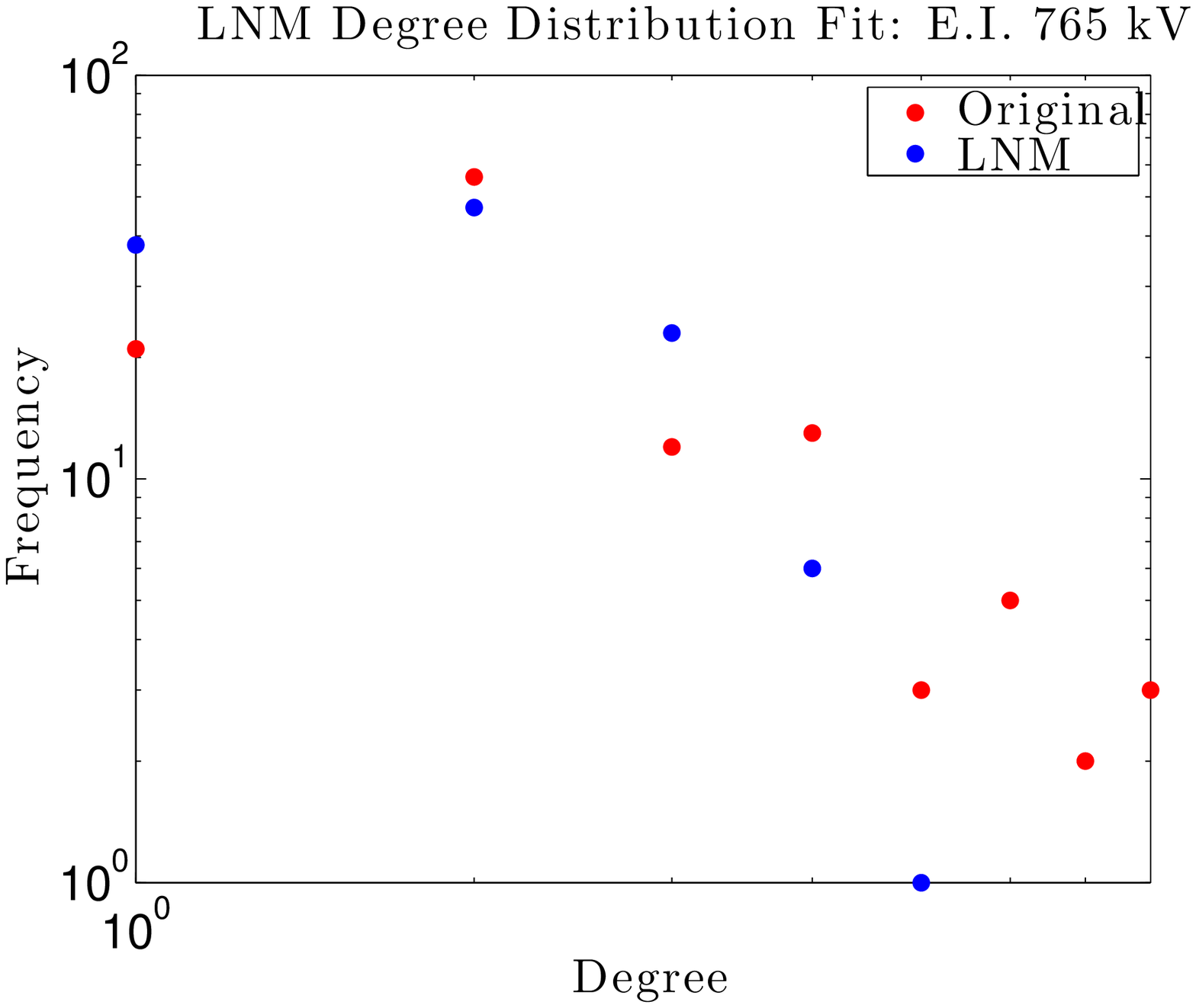}
\includegraphics[scale=0.14]{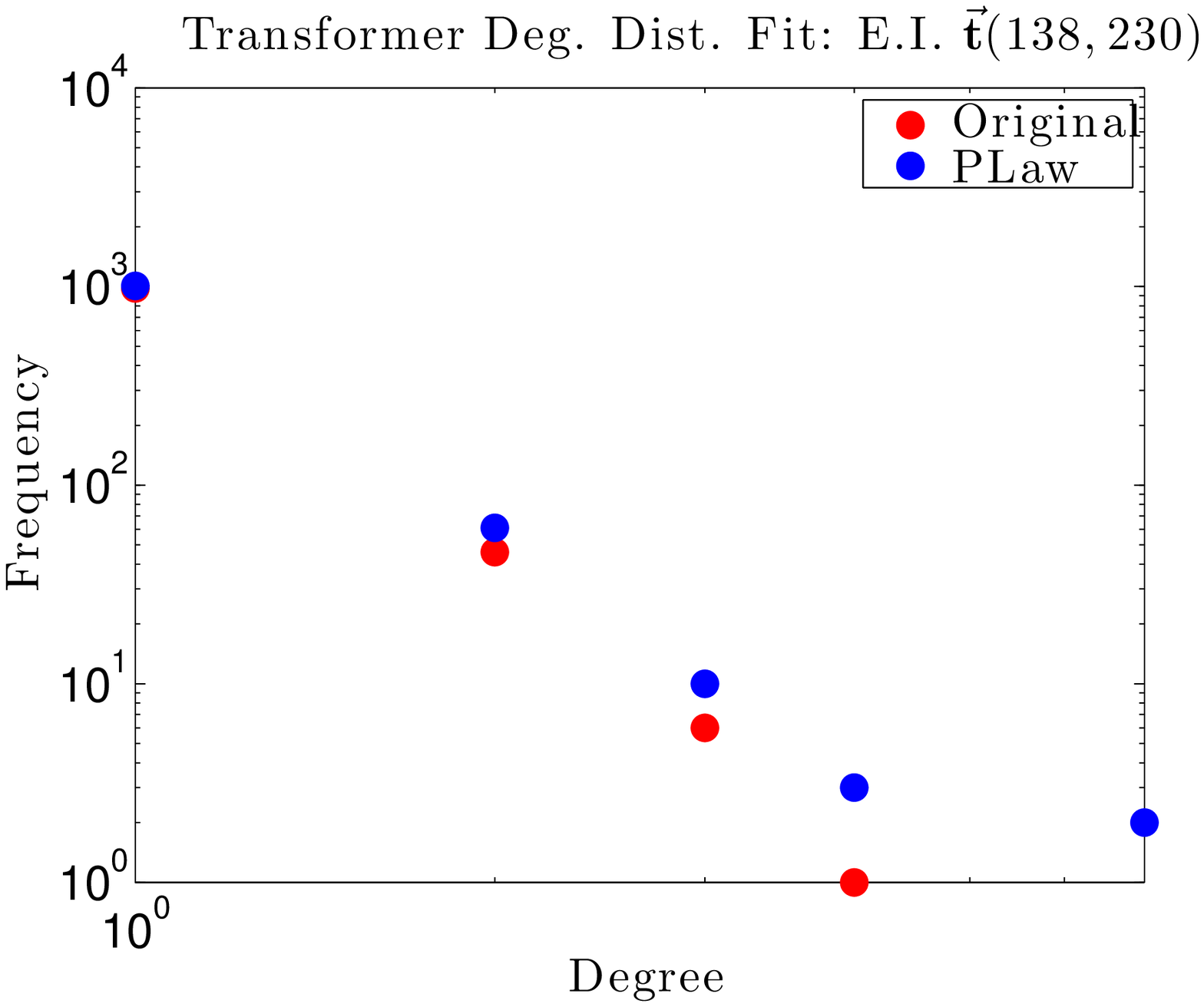}
\includegraphics[scale=0.14]{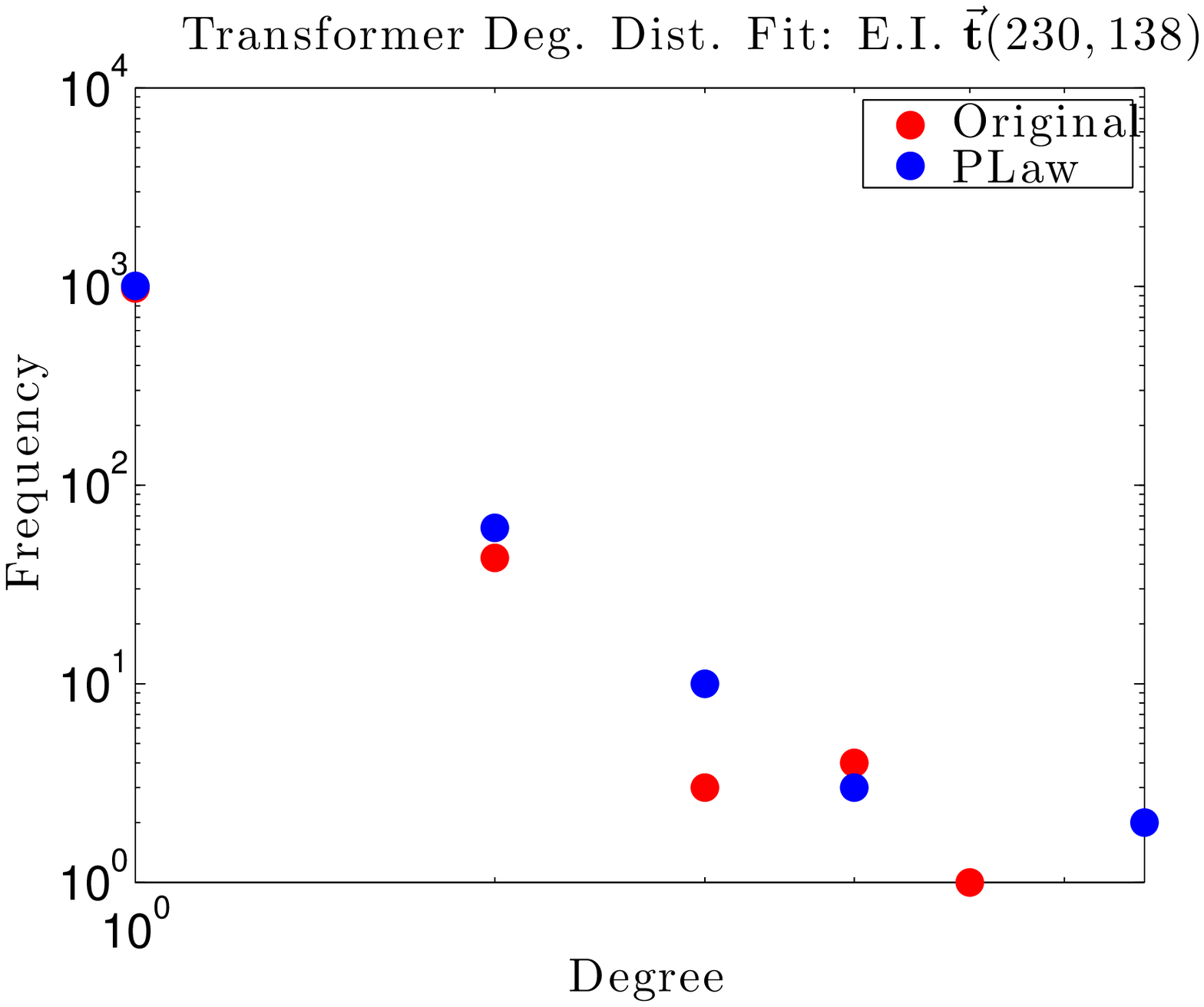}
\includegraphics[scale=0.14]{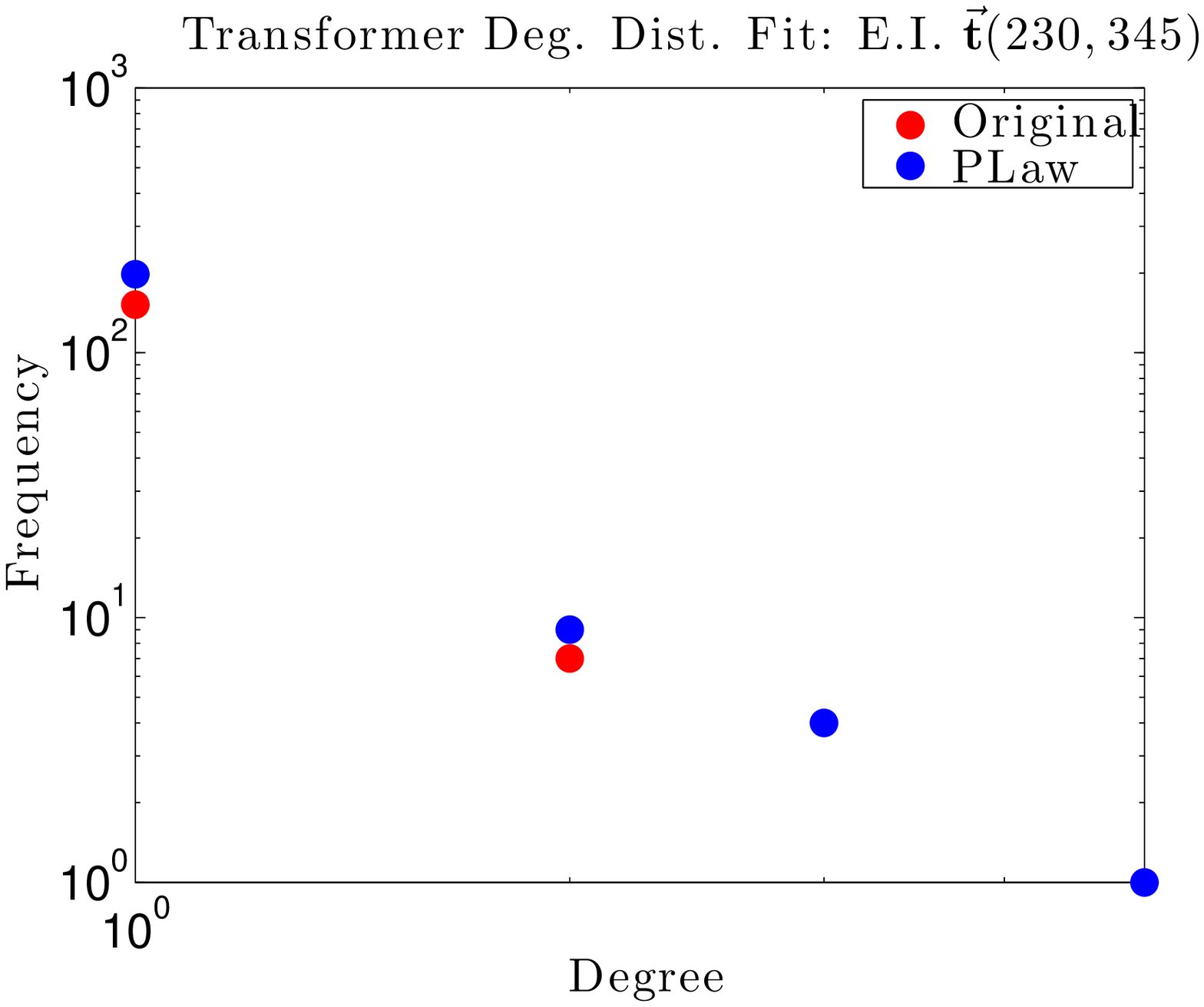}
\includegraphics[scale=0.14]{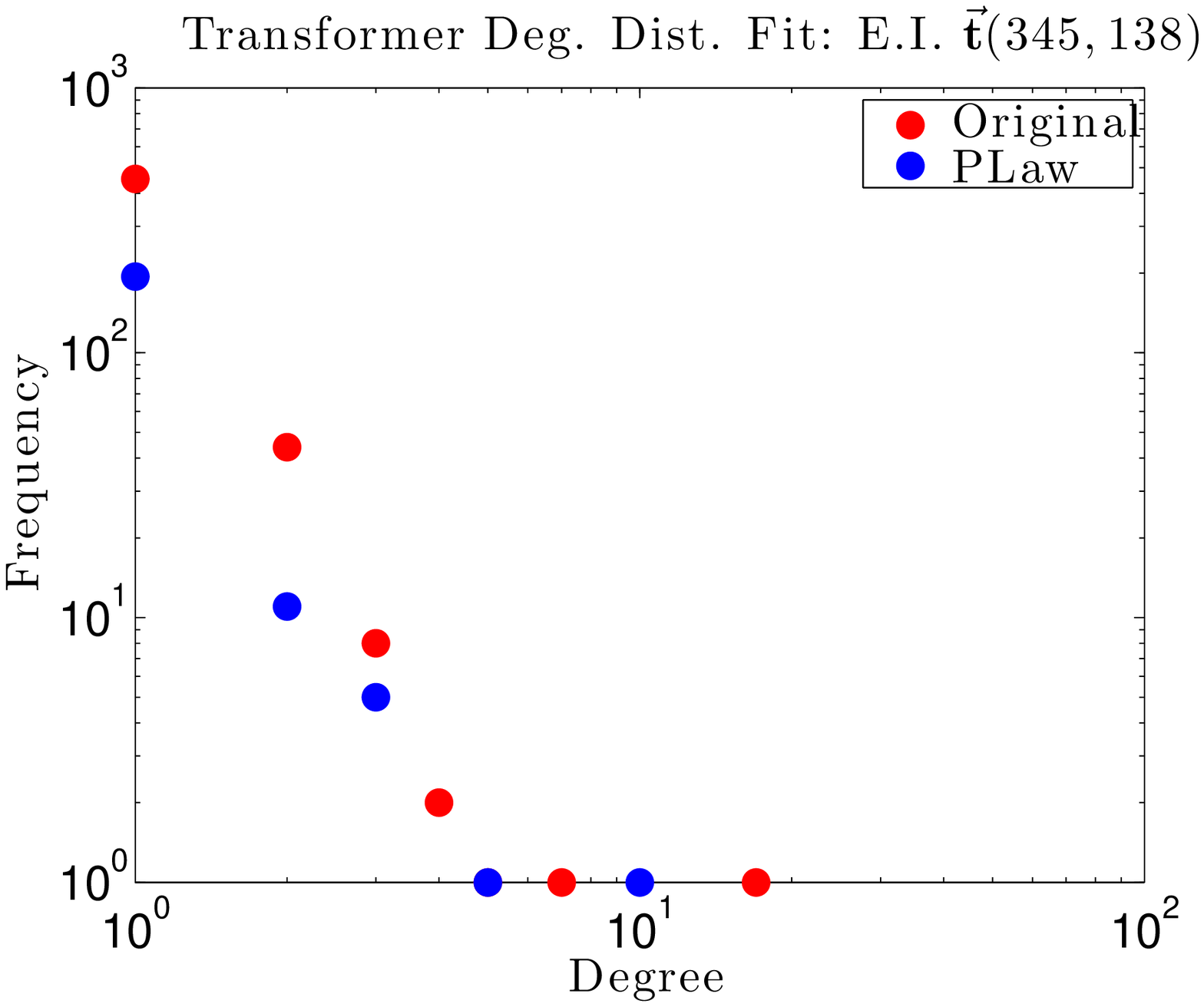}
\includegraphics[scale=0.14]{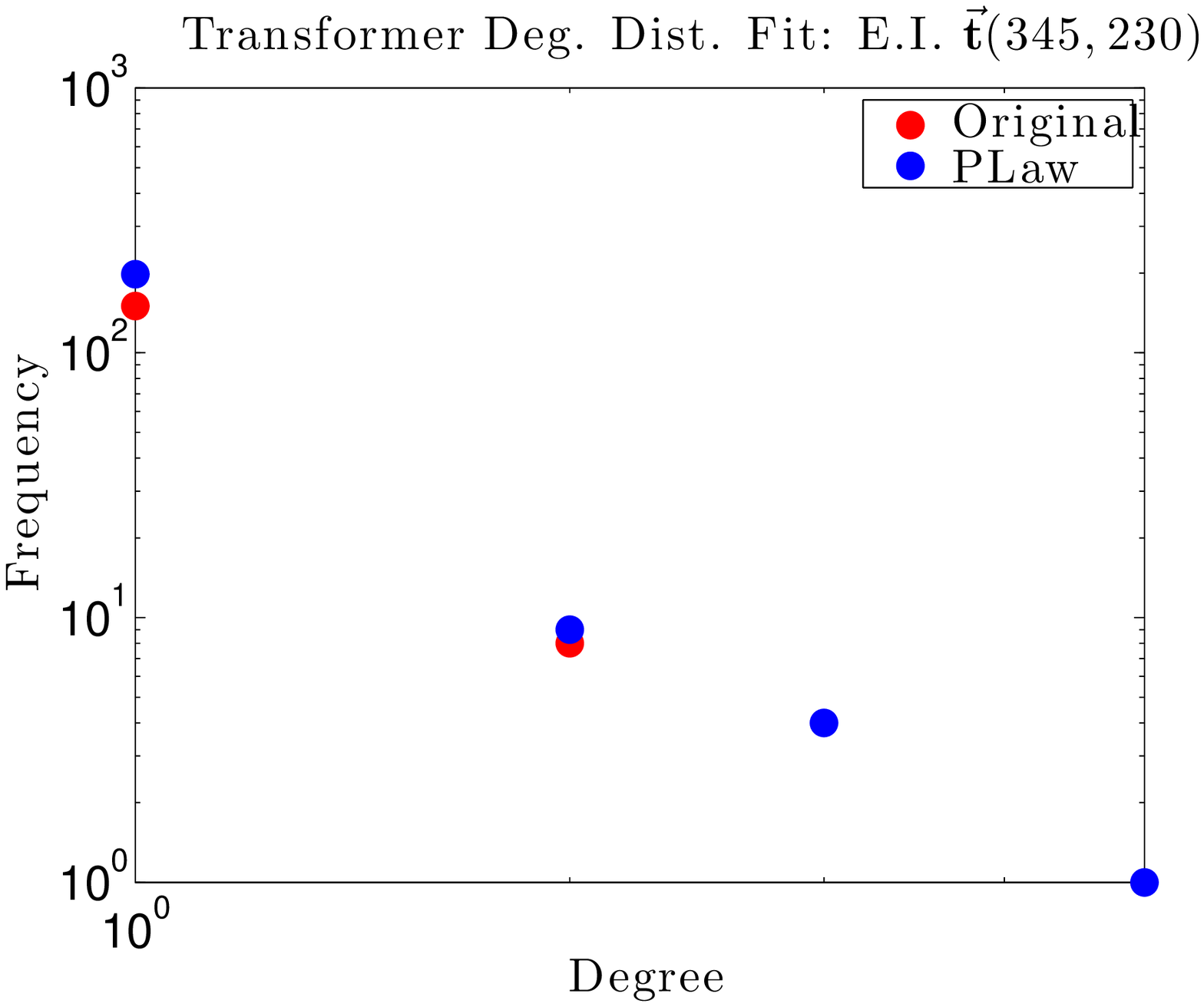}
\caption{{\it Top row:} Realizations of synthetic discrete lognormal degree distributions vs original distributions. {\it Bottom row:} Realizations of synthetic discrete power law transformer degree distributions vs original distributions.}\label{fig:ddLNM}
\end{figure}

{Estimating $\bar{d}$ and $d_{\max}$ with $f(n)$ and $g(n)$, we generate realizations of log-normal degree distributions using the aforementioned software \cite{feastpack} and plot the fit to the original data in Figure \ref{fig:ddLNM}. Finally, one may convert the outputted degree distribution for $G[X]$ to a degree sequence by assigning degrees to randomly chosen vertices from the pool of voltage $X$ vertices, yielding (along with the rounded prediction for diameter given by function $f$) the synthetic inputs for Phase 1. }

\subsection{{Synthetic Generation of Phase 2 Model Inputs}}
\paragraph{Transformer degree sequence, $\tdegseq[X,Y]$} {For any given transformer subgraph between voltage levels $X_i$ and $X_j$, recall there are two corresponding transformer degree sequences: $\tdegseq[X_i,X_j]$ and $\tdegseq[X_j,X_i]$. As a first step to synthetically generating these, one must specify how many vertices participate in each transformer graph, i.e. how many vertices of voltage $X_i$ are incident to a transformer edge having the other endpoint in voltage $X_j$ (and vice versa for $X_j$). While this can be invoked as a separate user-specified parameter, we approximate this based on the number of vertices in the same-voltage subgraphs.}

{More formally, given a set of $k$ voltage levels, $\mathcal{X}=\{X_1,\dots X_k\}$, let $n_i$ denote the number of voltage $X_i$ vertices and $t_i^j$ denote the number of voltage $X_i$ vertices which are incident (via a transformer) to a voltage $X_j$ vertex. We seek a function $h: \mathbb{R}^{+} \times \mathbb{R}^{+} \to \mathbb{R}^{+}$ such that $h(n_i,n_j)\approx t_i^j$. In order to guarantee this function outputs valid values for any valid input, it must also satisfy $h(n_i,n_j) \leq n_i$, so that the number of voltage $X_i$ vertices incident to a transformer doesn't exceed the total number of voltage $X_i$ vertices. We suggest a simple model satisfying these constraints,
\[
h(n_i,n_j)=c \cdot \min \{n_i, n_j\},
\]
and find that the optimal $c \approx 0.174$. Figure \ref{fig:funFitFixed} plots the fit. }

{Since transformer subgraphs are relatively small graphs consisting almost entirely of disjoint, small $k$-stars, where the number of $k$-stars decreases exponentially in $k$, their degree distributions are short and steep. We suggest generating simple power-law degree distributions, for some large power-law exponent $\gamma>4$. Based on fitting power-law exponents to the data and averaging, we use $\gamma \approx 4.15$. We plot realizations of such discrete power law degree distributions for some larger transformer subgraphs in the Eastern Interconnection in Figure \ref{fig:ddLNM}. Finally, for each pair of voltage levels, we suggest drawing a {\it single} power law degree distribution on $h(n_i,n_j)$ vertices. Recall that since transformer subgraphs are bipartite graphs, the degree sequences $\tdegseq[X,Y]$ and $\tdegseq[Y,X]$ must therefore sum to the same value for each pair of voltage levels. However, in practice two experimentally drawn degree distributions will not necessarily satisfy this constraint. Although there are possible workarounds\footnote{One workaround, mentioned in \cite{guillaume2006bipartite}, is to iteratively modify the outputted degree sequences until this constraint is met. In our case, one might choose to iteratively nullify the degree of randomly chosen vertices from the larger-sum sequence while redrawing degrees for null-degreed vertices in the smaller-sum sequence, until the sums are equal.}, we use this simpler approach. One may convert the outputted transformer degree distribution to the (different) degree sequences $\tdegseq[X,Y]$ and $\tdegseq[Y,X]$ by assigning degrees to randomly chosen vertices, among the pool of voltage $X$ and $Y$ vertices, respectively. Repeating this for all pairs of voltage levels yields the Phase 2 inputs. }



\subsection{Model performance under synthetically generated inputs}

\begin{figure}[t!]
\centering
\begin{subfigure}[b]{0.29\textwidth}
\includegraphics[width = \linewidth]{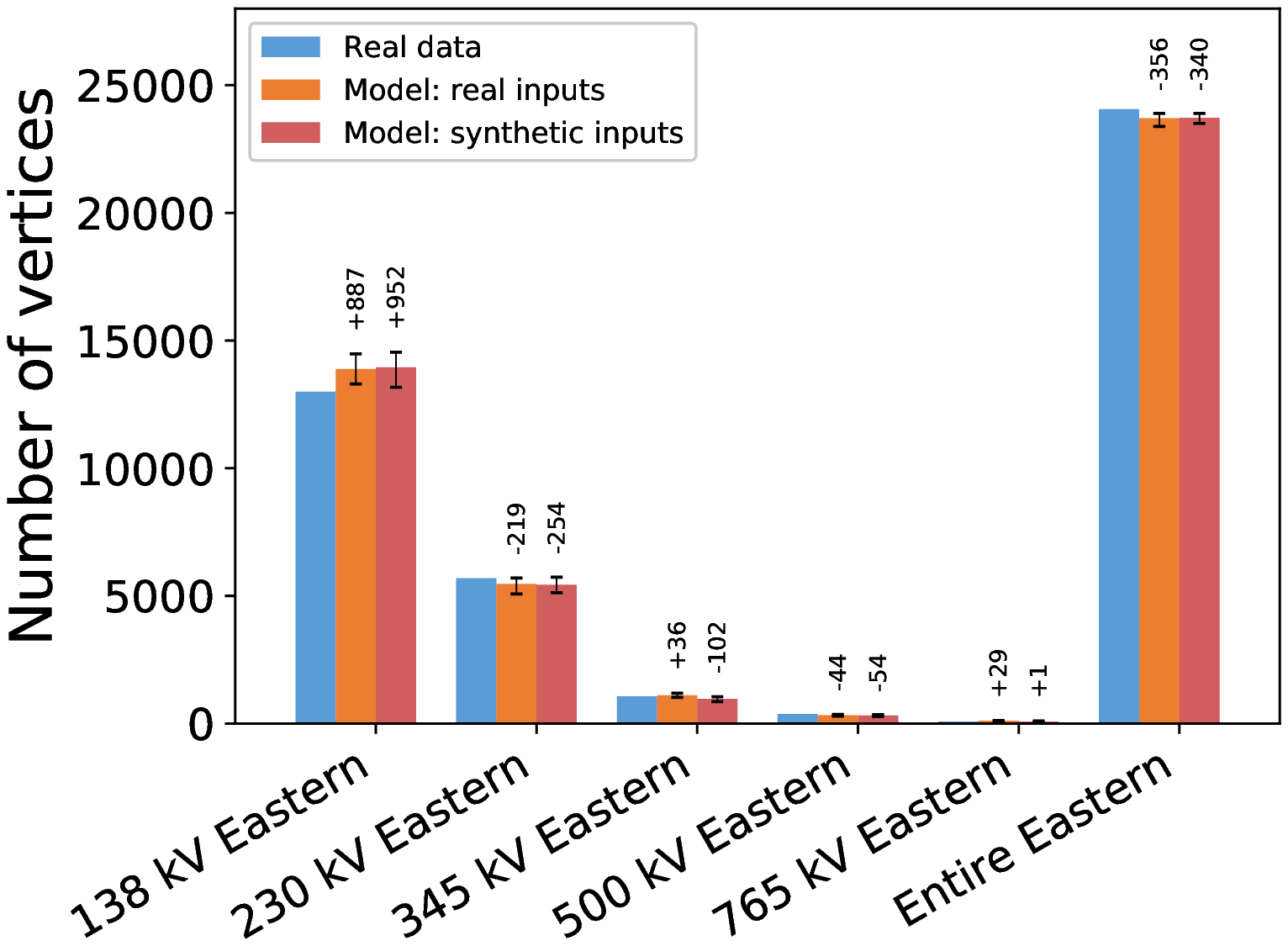}
\caption{}\label{fig:NumVertCompare-syn}
\end{subfigure}
\begin{subfigure}[b]{0.29\textwidth}
\includegraphics[width = \linewidth]{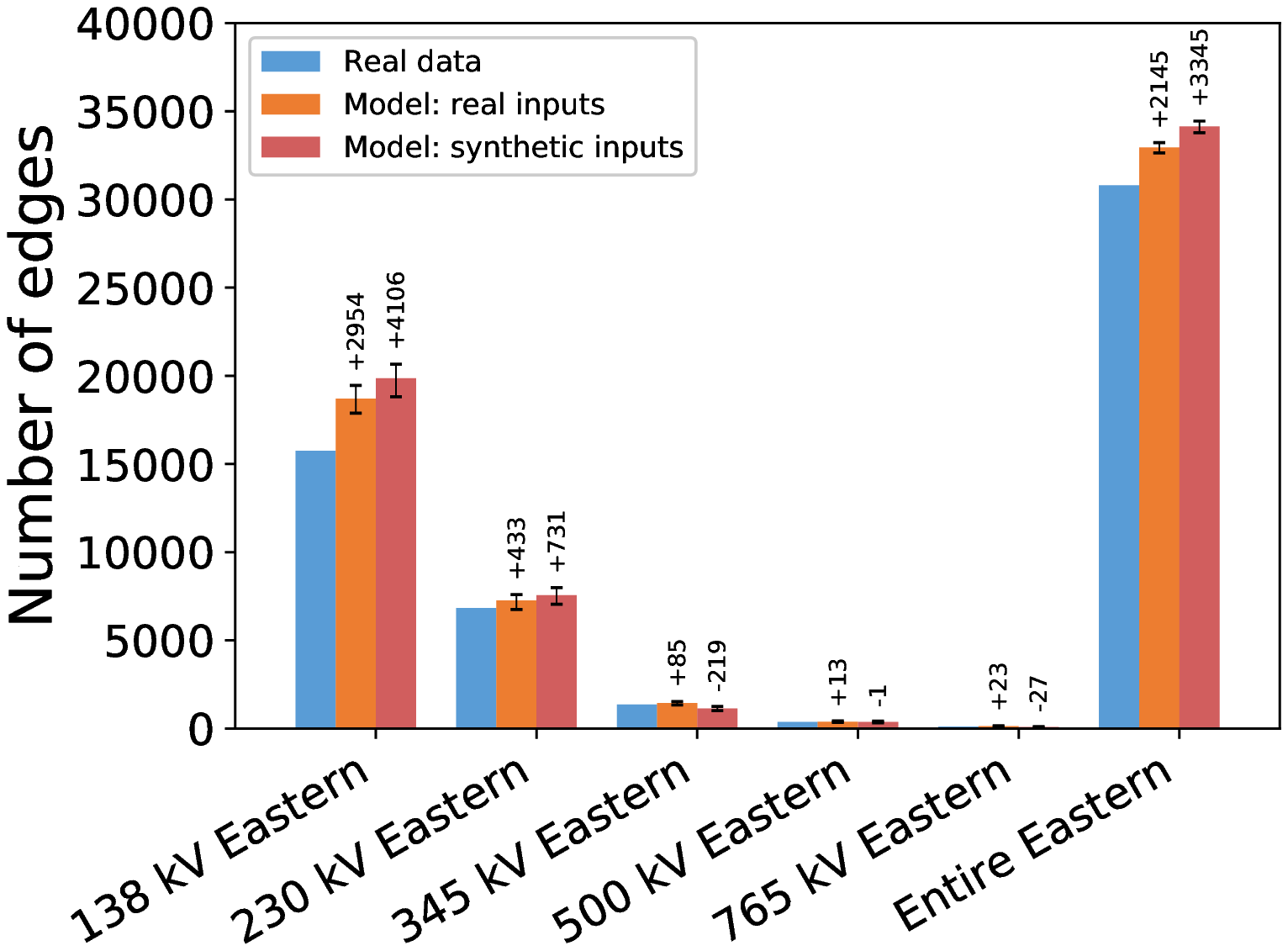}
\caption{}\label{fig:NumEdgeCompare-syn}
\end{subfigure}
\begin{subfigure}[b]{0.29\textwidth}
\includegraphics[width = \linewidth]{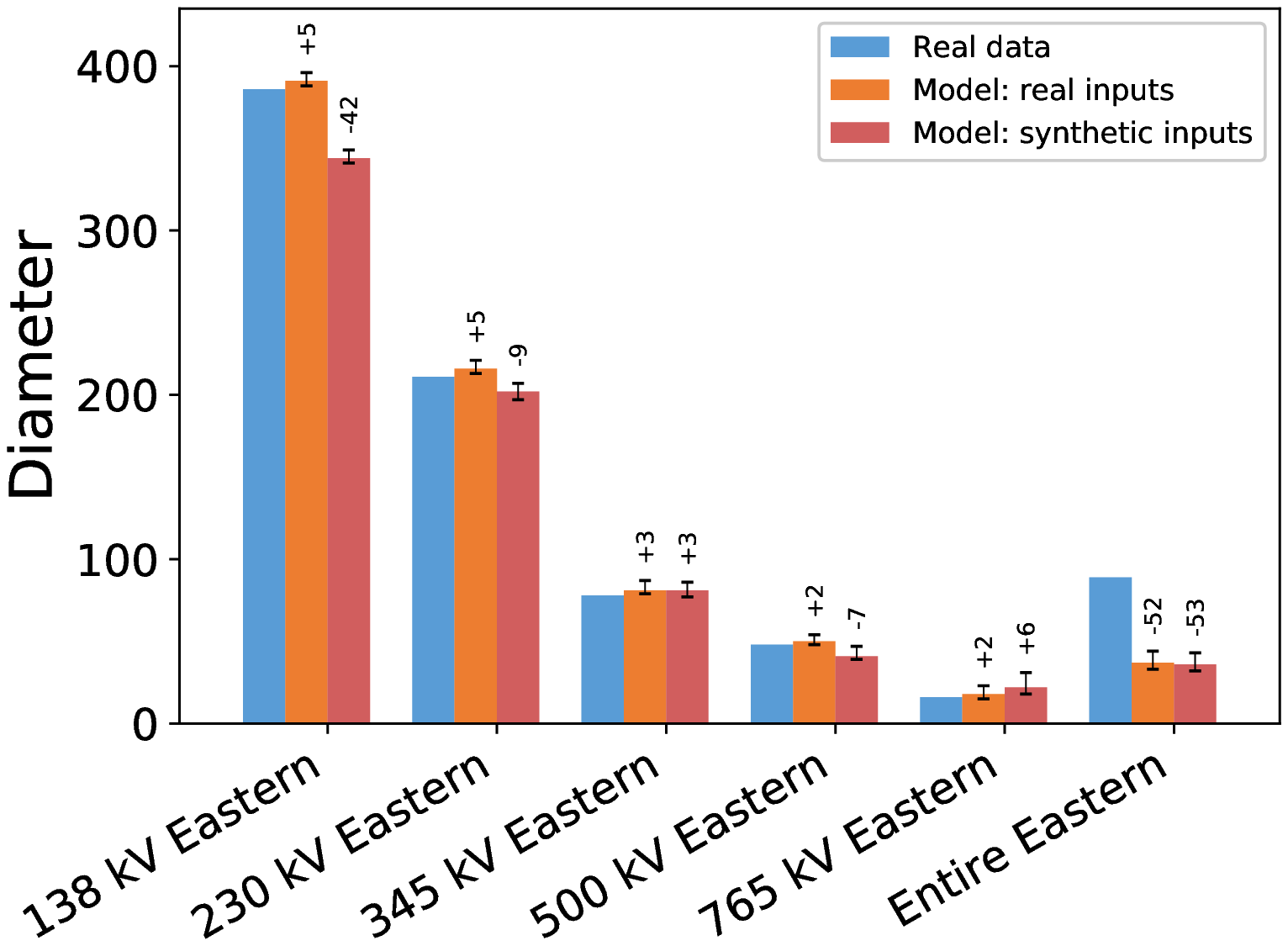}
\caption{}\label{fig:DiamCompare-syn}
\end{subfigure}
\\
\begin{subfigure}[b]{0.29\textwidth}
\includegraphics[width = \linewidth]{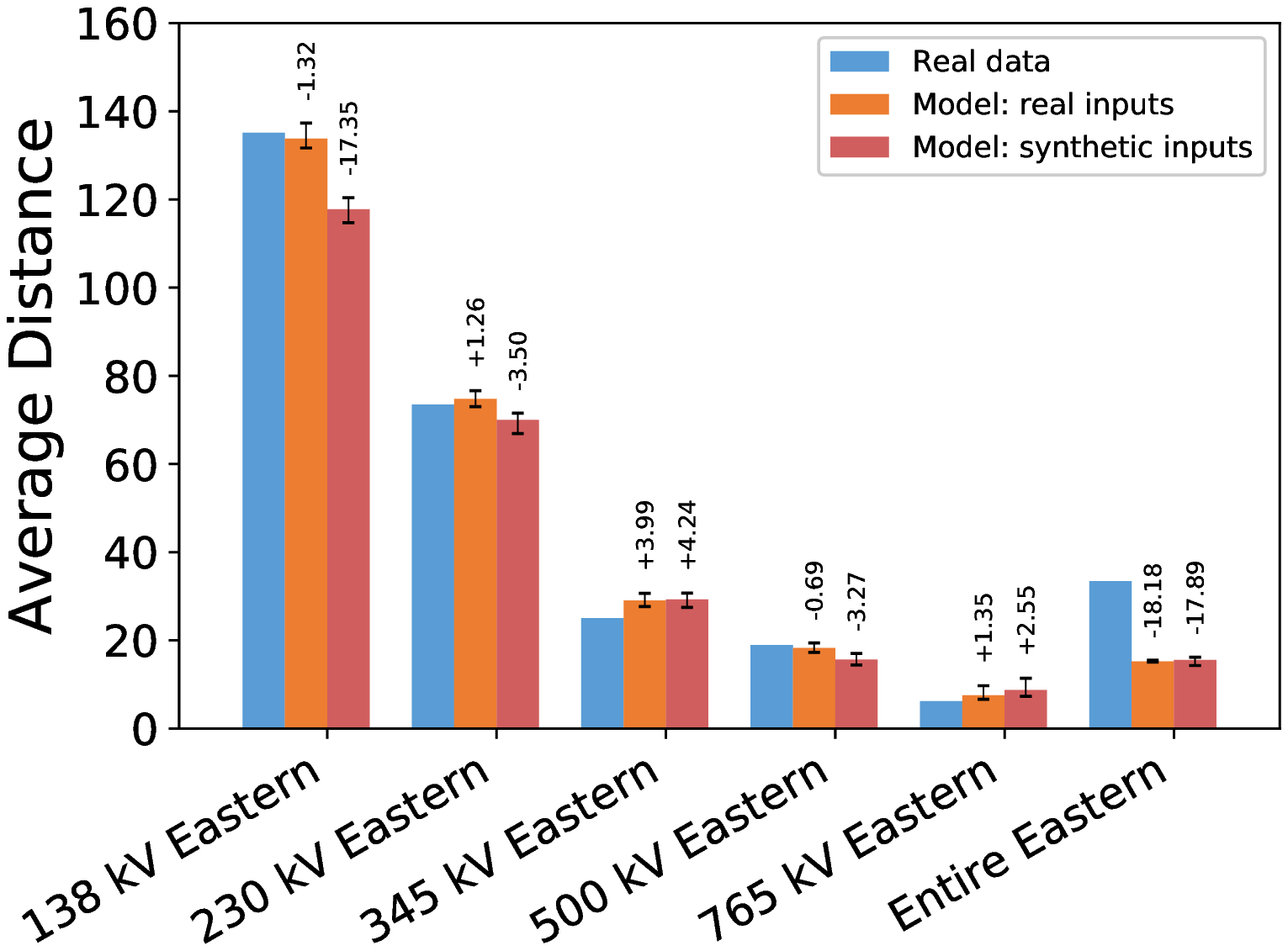}
\caption{}\label{fig:AvDistCompare-syn}
\end{subfigure}
\begin{subfigure}[b]{0.29\textwidth}
\includegraphics[width = \linewidth]{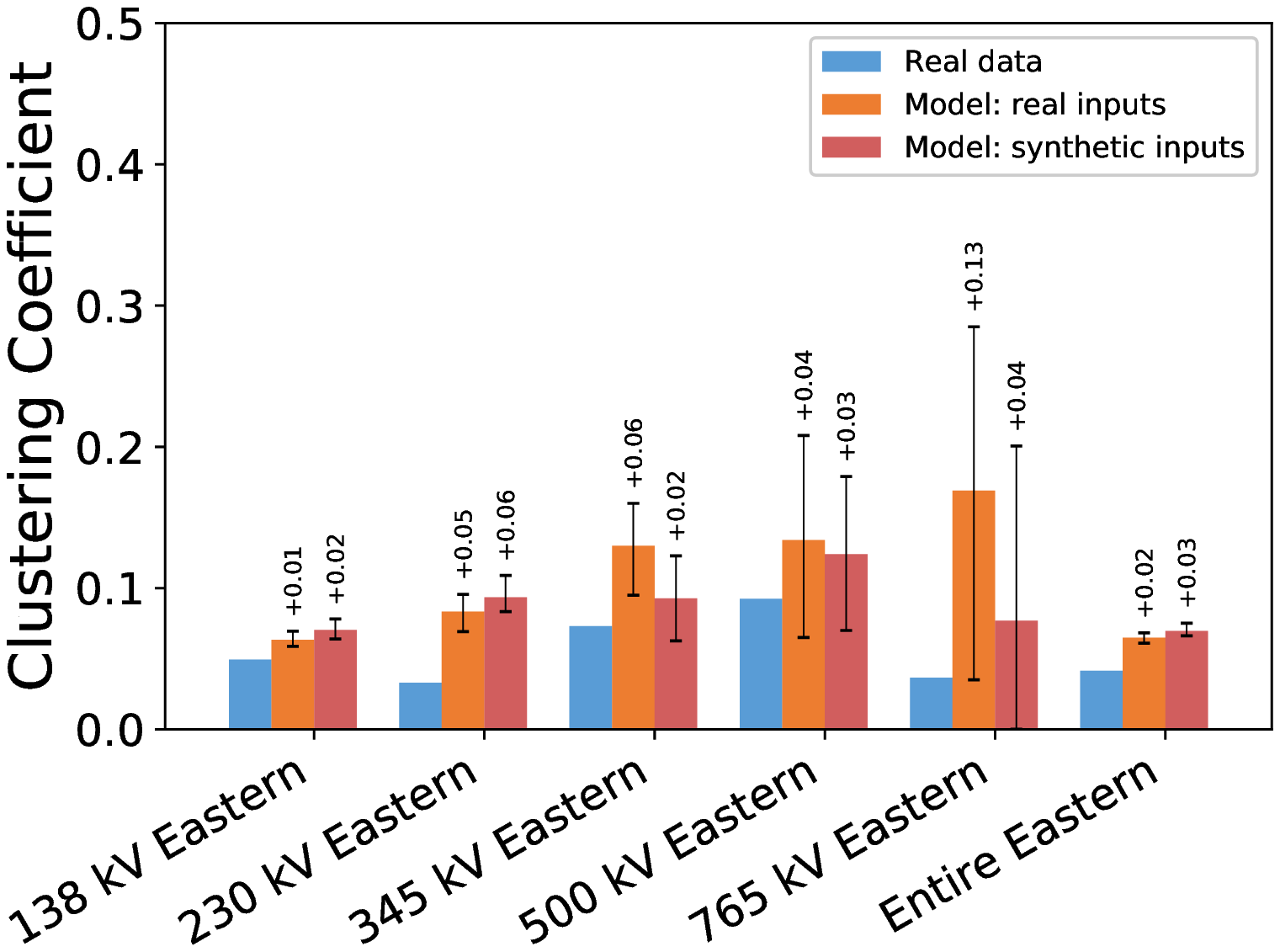}
\caption{}\label{fig:AvLCCCompare01-syn}
\end{subfigure}
\caption{Bar charts comparing measures in the same-voltage subgraphs.}
\label{fig:synResults}
\end{figure}

{We test our model's performance under instances of the Phase 1 and 2 inputs synthetically generated using the aforementioned guidelines. Recall that under these guidelines, the only information user-specified information that we extract from the real data is the number of vertices of each voltage level. Similar to our analysis in Section \ref{sec:results}, we generate 100 instances of each graph and collect the relevant measures for each run. Figure \ref{fig:synResults} compares the model's performance under these synthetically generated inputs to its performance under real inputs for both the same-voltage subgraph and entire aggregate graph of the Eastern Interconnection. Figures \ref{fig:NumVertCompare-syn}-\ref{fig:NumEdgeCompare-syn} show the model performs comparably for the number of vertices and edges in the largest connected component, with relatively marginal differences between output generated from real and synthetic inputs. In the case of diameter and average distance, Figures \ref{fig:DiamCompare-syn} and \ref{fig:AvDistCompare-syn} similarly show small differences, with the exception of the 138 kV subgraph: under synthetic inputs, the model output diameter is about 11\% less, while under real inputs, the model output is about 1\% more. As seen in Figure \ref{fig:diamFit}, this is by virtue of the guideline providing a slight underestimate of the input diameter for the 138 kV subgraph. Finally, as was the case under real inputs, the model's output under synthetic inputs has slightly larger clustering coefficients than in the data, although Figure \ref{fig:AvLCCCompare01-syn} shows the average values are still relatively small, around or below $0.1$. }

%% file: gpg-conclusion.tex
\section{Conclusion and future work} \label{sec:conc}

In order to account for the heterogeneity of real-world power grid networks, we designed a generative random graph model that distinguishes between nodes and edges of different type, according to their voltage level. We found that same-voltage subgraphs in real power grid graph data exhibited an unusual combination of structural properties which make modeling them accurately difficult. We proposed a two-phase model which generates both the same-voltage subgraphs as well as the transformer edges that connect them, according to tunable user-specified input. We found that this model either matched or outperformed the Chung-Lu model in nearly all categories tested, particularly with regard to the large diameter and average distance observed in the data.
The close match of these structural properties were also seemingly reflected in visualizations that, even at small scales, bore resemblance to the original graph data. Lastly, our model also produced graphs with comparable resiliency to single edge failures.

{Ultimately, our model may be used to generate synthetic graph data, test hypotheses and algorithms at different scales, and serve as a baseline model on top of which further electrical network properties, device models, and interdependencies to other networks, may be appended.  In particular, given that our model only requires desired degrees and diameter, users may easily generate power grid graphs using artificially generated inputs, under guidelines identical or similar to those we provided above. This feature is attractive in light of the limited availability of real power grid data: given that these guidelines don't require user-input beyond specifying the desired number of vertices of a given voltage level, our model may still be used in the complete absence of real data.}

{Many questions remain for future work. First, we note possible areas for improvement in our model and refinement in our analyses. Perhaps most notably, in Section \ref{sec:agg}, we observed the accuracy of the model in matching diameter declined for the aggregate graphs, which can likely be attributed to choosing leaf vertices of $k$-stars uniformly at random in the stars algorithm. Accordingly, one possible avenue to improving our model would be to devise a weighted sampling scheme for leaf selection, which we speculate might also produce more accurate structural match in the interconnection graphs. Beyond visualizations, we didn't investigate the structural properties of these interconnection networks; indeed, taking into account both vertex and edge-weights makes this analysis more complicated. With regard to resiliency, while we examined single edge failures, we didn't consider scenarios of multiple simultaneous or cascading edge failures, which would provide a more extensive perspective. And lastly, while our model generates static graphs, real-world physical transmission systems are evolving in time, and it would be advantageous to capture such changes. }

%% file: gpg-appendices.tex
\appendix
\newpage
\section{Appendices}

\subsection{Chung-Lu model description}

The Chung-Lu random graph model takes a desired degree sequence $\degseq=(d_1,\dots,d_n)$ as an input, where $d_i$ is the desired degree of vertex $i$.
Note that this degree sequence also stipulates a desired number of edges, $m=\frac{1}{2}\sum_i d_i$.
For every pair of vertices $i,j$, the probability of an edge is $\Pr(\{i,j\} \mbox{ is an edge})= \frac{d_i d_j}{2m}$.
In order to ensure that this probability is at most 1, one can further require that the input degree sequence $\degseq$ satisfies $\max_i d_i^2 < \sum_k d_k$.
In expectation, each vertex achieves its desired degree $d_i$, since
\[
\mathbb{E}(\mbox{degree of vertex } i)= \sum_j \Pr(\{i,j\} \mbox { is an edge})=\sum_j \frac{d_i d_j}{2m}=d_i.
\]
For more information on the Chung-Lu model, see the monograph \cite{Chung2006} by Chung.
Instead of flipping a weighted coin for all ${n \choose 2}=O(n^2)$ possible edges, an efficient implementation of the Chung-Lu model favored by the authors of \cite{Kolda2014}, known as ``fast Chung-Lu", instead chooses the endpoints of $m=\frac{1}{2}\sum_i d_i$ edges by sampling these endpoints proportionally to their desired degree.
More precisely, in fast Chung-Lu, $\Pr(i \mbox{ selected})=\frac{d_i}{2m}$.
The expected degree of a vertex is again its desired degree, since here too $\Pr(\{i,j\} \mbox{ is an edge})= 2m \cdot \Pr(i \mbox{ selected}) \cdot \Pr(j \mbox{ selected})=\frac{d_i d_j}{2m}$.
Since real world graphs are often large and {\em sparse}, with a number of edges $m$ linear in the number of vertices $n$, the fast Chung-Lu model may be preferred by practitioners; see also \cite{miller2011efficient} for another efficient implementation of Chung-Lu.
Note that in fast Chung-Lu, repeated edges or loops are possible since the endpoints of each edge are chosen independently.
In practice, these tend to be few and are simply discarded in post-processing.
For a comparison of the properties of the regular, fast, and other versions of the Chung-Lu model, see \cite{Winlaw2015}.

\subsection{Algorithms} \label{apen:algs}

\begin{algorithm}[h]
  \caption{Preprocessing stage for subgraph of voltage $X$.
  {\bf Input:} Desired degrees $\degseq=(d_1,\dots,d_n)$ and desired diameter $\delta$.
  {\bf Output:} Updated degree sequence, $\degseq'$, vertex-box sequence ${\bf v}=(v_1,\dots,v_n)$, diameter path vertices $D$, subdiameter path vertices $S$ }
  \label{alg:setup}
  \begin{algorithmic}[1]
\Procedure{Setup}{$\degseq, \delta$}
\State $\eta \gets |\{d \in \degseq: d>0\}|$
    \Comment{Desired number of non-isolated vertices}
\State $\delta \gets \mbox{round}(\delta-2\log{\frac{\eta}{\delta+1}})$
    \Comment{Diameter adjustment} \label{line:diamAdj}
\Statex
\LineComment{Inflate degree seq. until expected number of non-isolated vertices matches }
\While {$|\degseq|-\sum_{d_i \in \degseq}\exp(-d_i)\leq \eta$} \label{startLine:inflate}
    \State Randomly select nonzero $d \in \degseq$
        \Comment{Nonisolated vertex to be duplicated}
    \State $\degseq \gets (\degseq, d)$ \label{endLine:inflate}
\EndWhile
\Statex
\LineComment{Randomly distribute all non-isolated vertices into boxes }
\State $I_O \gets \{i: d_i\geq 1\}$
    \Comment{Non-isolated vertex indices}
\State $B \gets \{1,\dots,\delta+1\}$
    \Comment{Indices of $\delta+1$ boxes}
\LineComment{To be continued on next page...}
\algstore{setup}
  \end{algorithmic}
\end{algorithm}

\begin{algorithm}
\begin{algorithmic} [1]
\algrestore{setup}
\LineComment{Continued from previous page...}
\Statex
\LineComment{In expectation, each box should have sufficiently many vertices so
that any vertex can achieve the max desired degree}
\If {$\frac{\eta}{\delta+1}<\max(\degseq)$} \label{startLine:maxDeg}
    \State Randomly select $C \subseteq B$ with $|C|=\frac{\eta}{\max(\degseq)}$.\label{endLine:maxDeg}
        \Comment{\parbox[t]{0.35\linewidth}{If needed, make some boxes empty (except for diam path vertex)}}
\Else
    \State $C = B$
\EndIf
\State {\bf v} = \textsc{AssignBoxes}$(I_O, C)$
    \Comment{See \textsc{AssignBoxes} procedure below}
\Statex
\LineComment{Choose vertices for the diameter path and randomly (re)assign them to distinct boxes}
\If {$|\{i: d_i\geq 3\}|\geq \delta+1$}
    \Comment{\parbox[t]{0.5\linewidth}{Choose these from vertices of degree at least 3, if possible}}
    \State $I_{P}=\{i: d_i\geq 3\}$
\Else
    \State {$I_P=\{i: d_i\geq 2\}$}
\EndIf
\State Randomly choose $D \subseteq I_P$ with $|D|=\delta+1$
    \Comment{\parbox[t]{0.3\linewidth}{Selects diameter path vertices from pool}}
\ForEach {$i \in D$}
    \State Randomly choose $b \in B$
    \State $v_i \gets b, \ B \gets B\backslash b$
        \Comment{Assign each diameter path vertex to a distinct box}
\EndFor
\Statex
\LineComment{Choose vertices for the subdiameter path and randomly (re)assign
them to distinct boxes. Make subdiameter path as long as possible, up to diameter length}
\State $\alpha \gets \min(\delta+1, |I_P\backslash D|)-1$
    \Comment{Length of subdiameter path}
\State $\beta \gets \left\lfloor\frac{\delta+1}{2}\right\rfloor-\left\lfloor\frac{\alpha-1}{2}\right\rfloor$
    \Comment{Used to center subdiameter path below diameter path}
\State $B \gets \{\beta,\dots,\beta+\alpha\}$
    \Comment{Indices of boxes for the subdiameter path vertices}
\State Randomly choose $S \subseteq I_P \backslash D$ with $|S|=\alpha+1$
    \Comment{\parbox[t]{0.3\linewidth}{Select subdiameter path vertices from remaining pool}}
\ForEach{$i \in S$}
    \State Randomly choose $b \in B$
    \State $v_i \gets b, \ B \gets B\backslash b$
        \Comment{\parbox[t]{0.5\linewidth}{Assign each sub-diameter path vertex to a distinct box}}
\EndFor
\Statex \\
\Return {$(\degseq, {\bf v}, D, S)$}
\EndProcedure
\Statex
\Statex
\Procedure{AssignBoxes}{indices, boxes}
\ForEach {$i \in indices$}
\State Randomly choose $b \in boxes$
\State $v_i \gets b$
\EndFor
\State \Return ${\bf v}$
\EndProcedure
  \end{algorithmic}
\end{algorithm}

\clearpage

\begin{algorithm}[h]
  \caption{Generative model for subgraph of voltage $X$.
  {\bf Input:} Updated degree sequence output from Algorithm 1 $\degseq'=(d_1,\dots,d_n)$, vertex-box list ${\bf v}=(v_1,\dots,v_n)$, diameter path vertices $D$, subdiameter path vertices $S$.
  {\bf Output:} edge list $E$ }
  \label{alg:phase1}
  \begin{algorithmic}[1]
  \Procedure{CLC}{$\degseq', {\bf v}, D, S$}
\State $E \gets \varnothing$
\Statex
\LineComment{Make diameter path}
\For {$k=1,\dots,|D|-1$}
\State Find the (unique) vertices $i,j \in D$ such that $v_i=k$ and $v_j=k+1$.
\State $E\gets E \cup \{i,j\}$
\EndFor
\Statex
\LineComment{Make subdiameter path}
\For {$k=1,\dots,|S|-1$}
\State Find the (unique) vertices $i,j \in S$ such that $v_i=k$ and $v_j=k+1$.
\State $E\gets E \cup \{i,j\}$
\EndFor
\Statex
\LineComment{Create Chung-Lu graph on desired degrees of vertices in each box}
\For{$k=1,\dots,\max({\bf v})$}
\State $B_k\gets \{j: v_j=k\}$
    \Comment{All vertices in box $k$}
\State $m_k \gets \mbox{round}(\frac{1}{2}\sum_{i \in B_k} d_i)$
    \Comment{\parbox[t]{0.5\linewidth}{Desired number of edges in Chung-Lu graph for box $k$}}
\For{$c=1,\dots,m_k$}
\State Randomly select $i \in B_k$ proportional to $d_i/{2m_k}$
\State Randomly select $j \in B_k$ proportional to ${d_j}/{2m_k}$
\State $E\gets E \cup \{i,j\}$
    \Comment{\parbox[t]{0.5\linewidth}{Discard any loops or duplicate edges in post-processing}}
\EndFor
\EndFor \\
\Return $E$
  \EndProcedure
  \end{algorithmic}
\end{algorithm}


\begin{algorithm}[h!]
  \caption{Insert transformer edges between subgraphs of voltage $X$ and $Y$.
  {\bf Input}: Desired transformer degrees $\tdegseq[X,Y]=\tup{t[X,Y]_1,\dots,t[X,Y]_n}$  and $\tdegseq[Y,X]=\tup{t[Y,X]_1,\dots,t[Y,X]_n}$.
  {\bf Output:} edge list $E$ }
  \label{alg:phase2}
  \begin{algorithmic}[1]
  \Procedure{Stars}{$\tdegseq[X,Y], \tdegseq[Y,X]$}
\State $E \gets \varnothing, \  L^X \gets \varnothing, \ L^Y \gets \varnothing$
    \Comment{Edge list and ``leftover bins"}
\State $I^X_O \gets \{i: t[X,Y]_i=1\}, \  I^Y_O \gets \{i: t[Y,X]_i=1\}$
    \Comment{\parbox[t]{0.25\linewidth}{Deg. 1 ver. in $G[X],G[Y]$}}
\State $I^X_C \gets \{i: t[X,Y]_i\geq 2\}, \ I^Y_C \gets \{i: t[Y,X]_i \geq 2\}$
    \Comment{\parbox[t]{0.25\linewidth}{Degree at least 2 vertices in $G[X]$ and $G[Y]$}}
\Statex
\LineComment{To be continued on next page...}
\algstore{phase2}
\end{algorithmic}
\end{algorithm}

\begin{algorithm}
\begin{algorithmic} [1]
\algrestore{phase2}
\LineComment{Continued from the previous page...}
\Statex
\LineComment{Define procedure for $k$-stars centered at a vertex in $G[A]$ with $k$ leaves in $G[B]$}
\Procedure{$k$-Stars}{$A,B$}
\While {$I_C^A \not = \varnothing$}
\State Randomly select $i \in I^A_C$
\If { $|I_O^B| \geq t[A,B]_i $}
    \Comment{\parbox[t]{0.5\linewidth}{If there are enough leaf vertices from $Y$ for create a star centered at $i$}}
\For {$c=1,\dots,t[A,B]_i$}
\State Randomly select $j \in I_S^B$
\State $E \gets  E \cup \{i,j\}$
\State  $I_O^B \gets I_O^B \backslash j$
    \Comment{Remove leaf vertex from pool}
\EndFor
\State  $I_C^A \gets I_C^A \backslash i$
    \Comment{Remove center vertex from pool}
\Else
\State $L^A \gets i$
    \Comment{Put vertex in leftover bin for later}
\State  $I_C^A \gets I_C^A \backslash i$
\EndIf
\EndWhile
\EndProcedure
\Statex
\LineComment{Generate $k$-stars centered at vertices in both voltage $X$ and $Y$}
\State $\textsc{$k$-Stars}(X,Y)$
\State $\textsc{$k$-Stars}(Y,X)$
\Statex
\LineComment{Insert 1-stars (edges) on any remaining degree 1 vertices from $G[X]$ and $G[Y]$}
\While {$I_O^X \not = \varnothing$ and $I_S^Y \not = \varnothing$ }
\State Randomly select $i \in I_S^X$ and $j \in I_S^Y$
\State $E \gets E \cup \{i,j\}$
\State $I_O^X \gets I_O^X \backslash i, \ I_O^Y \gets I_O^Y \backslash j$
    \Comment{\parbox[t]{0.5\linewidth}{Remove endpoints of inserted edges from their pool}}
\EndWhile
\State $L^X \gets\left( L^X \cup I_O^X\right), \ L^Y \gets \left( L^Y \cup I_O^Y\right)$
\Statex
\LineComment{Create bipartite Chung-Lu graph on any leftover vertices}
\If {$L^X \not= \varnothing$}
\State $m \gets \sum_{i \in L^X} t[X,Y]_i$
    \Comment{Number of edges to be inserted} \label{startLine:bipartiteCL}
\For {$k=1,\dots,m$}
\State Randomly select $i \in L^X$ proportional to $t[X,Y]_i/m$
\State Randomly select $j \in L^Y$ proportional to $t[Y,X]_j/m$
\State $E \gets E \cup \{i,j\}$
    \Comment{Discard any duplicate edges in post-processing} \label{endLine:bipartiteCL}
\EndFor
\EndIf
\State \\
\Return $E$
  \EndProcedure
  \end{algorithmic}
\end{algorithm}

\clearpage

\begin{algorithm}[h]
 \caption{Generative model for entire power grid graph on $k$ voltage levels, $X_1,\dots,X_k$.
  {\bf Input:} Desired same-voltage degrees $\degseq^{X_1},\dots,\degseq^{X_k}$,
  same-voltage diameters $\delta^{X_1},\dots,\delta^{X_k}$ and  transformer
  degrees $\tdegseq[X_i,X_j]$ for each pair $i,j\in\{1,\dots,k\}$.
{\bf Output:} edge list $E$ }
  \label{alg:entire}
  \begin{algorithmic}[1]
\Procedure{CLCStars}{$\{\degseq^{X_i}\}, \{\delta^{X_i}\}, \{\tdegseq[X_i,X_j]\}$}
\Statex
\State $E \gets \varnothing$
\Statex
\LineComment{Create each same-voltage subgraph}
\For {$i=1,\dots,k$}
\State $(\degseq', {\bf v}, D,S)\gets \textsc{Setup}(\degseq^{X_i},\delta^{X_i})$
\State $E \gets E \cup \textsc{CLC}(\degseq',{\bf v},D,S)$
\EndFor
\Statex
\LineComment{Insert transformer edges between each pair of same-voltage subgraphs}
\For {$i=1,\dots,k$}
\For {$j=i+1,\dots,k$}
\State $E \gets E \cup  \textsc{Stars}(\tdegseq[X_i,X_j],\tdegseq[X_j,X_i])$
\EndFor
\EndFor
\Statex \\
\Return {$E$}
  \EndProcedure
  \end{algorithmic}
\end{algorithm}


\subsection{Remarks on algorithms} \label{apen:algRem}

Before discussing the algorithms constituting our generative model, we note our model is general in the sense that neither Phase 1 nor 2 require additional assumptions on its inputs except that they are valid and simultaneously realizable by some graph.
In the case of Phase 1, this means there must be sufficiently many vertices of degree at least 2 in order to generate the diameter path of desired length.
In the case of Phase 2, since the inputs $\tdegseq[X,Y]$ and $\tdegseq[Y,X]$ are in fact desired degree sequences associated with partitions of a bipartite graph, these sequences must sum to the same value.
To achieve this generality, Phase 1 and 2 contain adjustments that will only be made for extreme cases which appear to be rare or non-existent in our tested power grid graph data, but nonetheless possible in a graph.
Furthermore, the setup of Phase 1 also makes some subtle adjustments to better ensure that the isolated vertex count, diameter, and maximum degree stipulated by the model inputs are more accurately achieved in the output of the CLC model.
While none of these adjustments are central to the idea of the model, we explain and justify them below.

\paragraph{Algorithm \ref{alg:setup}}
\begin{itemize}
\item Line \ref{line:diamAdj}: {\it Adjusting diameter.}
While the diameter path created by the CLC model will have desired diameter length $\delta$, the diameter of the entire CLC graph may still be slightly larger.
This is because the creation of random Chung-Lu graphs within each box are likely to extend the diameter path in the first and last boxes.
Thus, we expect that the additional length added to the diameter path won't exceed twice the expected diameter for each Chung-Lu graph.
We estimate the diameter of each Chung-Lu graph by $\log\left(\mathbb{E}\left(\# \mbox{ vertices in each box}\right)\right)$.

\item Lines \ref{startLine:inflate}-\ref{endLine:inflate}: {\it Adjusting the expected number of isolated vertices.}
In a realization of the Chung-Lu model, not only will vertices with desired degree 0 be isolated with probability 1, but vertices with nonzero desired degree can also become isolated with nonzero probability.
Thus, the number of isolated vertices in an instance of the Chung-Lu model will exceed the number of vertices desiring degree 0.
To adjust for this, we slightly inflate the provided degree sequence by randomly duplicating vertices until the expected number of isolated vertices matches the number of vertices desiring degree 0.
Here, we use a result from \cite{Chung2006}, which states that the expected number of isolated vertices for a Chung-Lu graph with degree sequence ${\bf d}=(d_1,\dots,d_n)$ is
\[
\mathbb{E}(\# \mbox{ isolated vertices})=\sum_{i} \exp(-d_i) + O\left(\frac{\sum_id_i^2}{\sum_i d_i}\right)\approx \sum_{i} \exp(-d_i).
\]

\item Lines \ref{startLine:maxDeg}-\ref{endLine:maxDeg}: {\it Allowing maximum degree to be achieved in any box}.
In the CLC model, the expected number of vertices in each box, and thus in each Chung-Lu random graph, is $\frac{\# \mbox{ total vertices }}{\# \mbox{ boxes}}$. If the desired degree of a vertex exceeds the number of vertices in its box, that vertex will only be able to achieve a strictly smaller degree\footnote{Also, the desired degree sequence within each particular box still may not be graphical (e.g., the sum of desired degrees within each box may not be even).
However, the Chung-Lu model will still output a graph with an approximation of this degree sequence; for practical purposes, this has not posed issues for matching the overall desired degree sequence reasonably well.}.
We note that this scenario is extremely uncommon in our power grid data.
Nonetheless, in such cases, the CLC adjusts by randomly choosing a number of boxes to be empty and allocating vertices in the other boxes, so that the expected number of vertices in each non-empty box is at least the maximum desired degree.
\end{itemize}

\paragraph{Algorithm \ref{alg:phase2}}
\begin{itemize}
\item Lines \ref{startLine:bipartiteCL}-\ref{endLine:bipartiteCL}: {\it Sufficient condition for perfect transformer degree match}.
These lines describe the fast bipartite Chung-Lu model introduced in \cite{Aksoy2016}.
We apply this model only to any ``leftover" vertices which could not be allocated to create stars by the preceding procedure.
A sufficient condition for all vertices of voltage $X$ and $Y$ to be allocated to stars (in which case they are guaranteed to exactly achieve their desired degree) is that there are sufficiently many vertices of voltage $Y$ with transformer degree 1 so that every vertex in $X$ with transformer degree 2 or more can achieve its degree by being linked to these degree-1 vertices from $Y$, and vice versa.
Loosely speaking, this condition is likely to be satisfied when the transformer edges graph consists mainly of disjoint edges.
Stated formally, this condition is that the following inequalities hold
\[
\sum_{ i \ : \ t[X,Y]_i \geq 2} t[X,Y]_i \leq \left|\{ j \in Y: t[Y,X]_j=1\}\right|,\sum_{ i \ : \ t[Y,X]_i \geq 2} t[Y,X]_i \leq \left|\{ j \in X: t[X,Y]_j=1\}\right|.
\]

If the above condition is satisfied -- which is the case in our power grid graph data -- the leftover bins $L^X$ and $L^Y$ defined in Algorithm \ref{alg:phase2} will be empty and consequently lines \ref{startLine:bipartiteCL}-\ref{endLine:bipartiteCL} will never be executed by the model.
\end{itemize}

\subsection{Assortativity and spectral gap}\label{apen:eig}

\begin{figure}[h]
\centering
\begin{subfigure}[b]{0.32\textwidth}
\includegraphics[width=\linewidth]{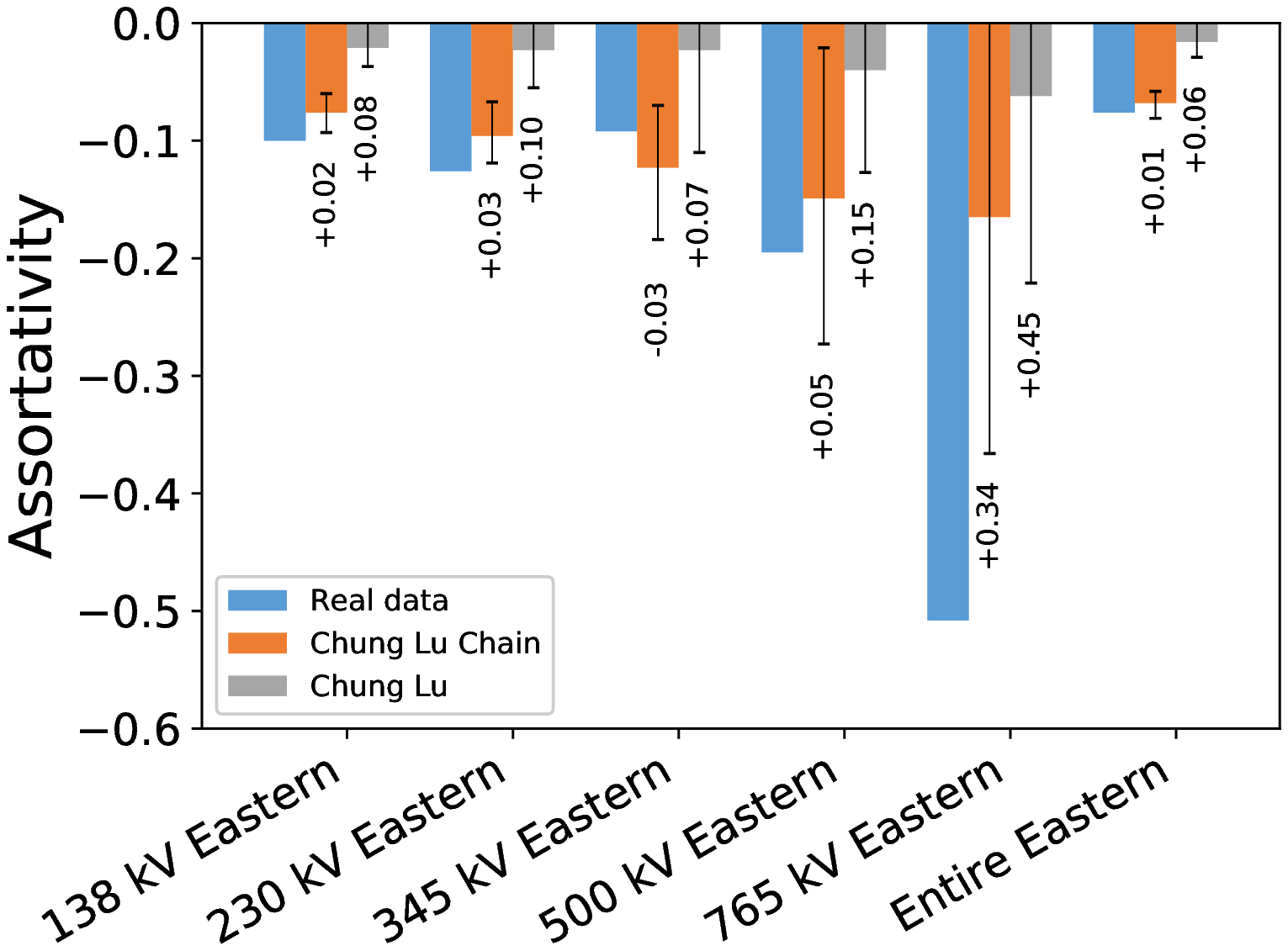}
\caption{}\label{fig:Assort-Eastern}
\end{subfigure}
\qquad
\begin{subfigure}[b]{0.32\textwidth}
\includegraphics[width=\linewidth]{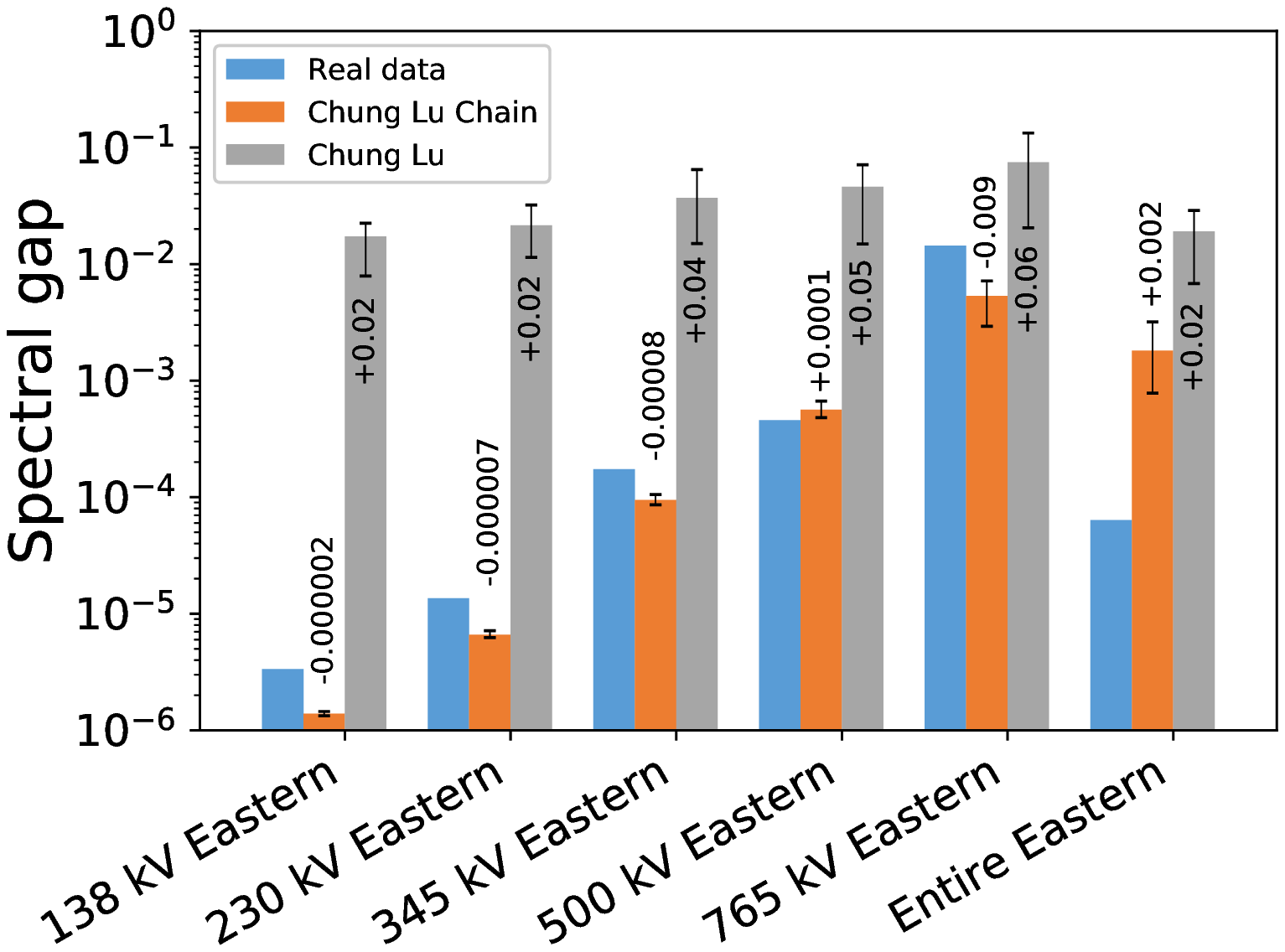}
\caption{}\label{fig:AC-Eastern}
\end{subfigure}
\caption{Bar charts comparison assortativity and spectral gap in the Eastern Interconnection networks.}
\label{fig:assort_ac}
\end{figure}

Below, we present supplemental results on our models performance with regard to {\it assortative mixing} and {\it spectral gap} for the Eastern Interconnection.
Bar plots comparing the real data to the graphs produced by our model are shown in Figure \ref{fig:assort_ac}.
As before, the bar heights are the average over 100 trials, and the error whiskers represent the minimum and maximum values observed.
These figures show that our CLC and CLC+Stars models achieve a closer match than the CL model in all cases.
The {\it assortativity coefficient} $r$ of a graph \cite{newman2002assortative} is the Pearson correlation coefficient of degree between pairs of adjacent vertices.
We note that $-1 \leq r \leq 1$, with small values of $r$ suggesting that high-degree vertices tend to link to low-degree vertices, while large values of $r$ suggest vertices tend to link with vertices of similar degree.
The {\it spectral gap} of a graph is the second smallest eigenvalue of the Laplacian matrix of the graph.
This eigenvalue is a numerical measure of the graphs connectivity, with smaller values suggesting a less-robust connectivity structure. The second Laplacian eigenvalue also corresponds to the Fiedler eigenvector \cite{fiedler1973algebraic} used in graph partitioning problems.
In order to facilitate more meaningful comparisons between networks of varying sizes, we consider the second smallest eigenvalue $\lambda_1$ of the {\it normalized} Laplacian matrix popularized by Chung \cite{fanSpectral}.
This quantity is intimately related to the Cheeger constant (also called {\it isoperimetric number} or {\it conductance}) of the graph, which measures the graphs ``bottleneckedness" or sparsest cut.
We note that $\lambda_1>0$ if and only if the graph is connected; thus, we restrict attention to the largest connected components.

